\documentclass{aa}  

\usepackage{graphicx}
\usepackage{txfonts}
\usepackage{amssymb}
\usepackage{amsmath}
\usepackage{graphicx}
\usepackage[colorlinks=true, allcolors=blue]{hyperref}
\usepackage{booktabs}
\usepackage{amsfonts}     
\usepackage[T1]{fontenc} 
\usepackage[utf8]{inputenc} 
\usepackage{multicol}
\usepackage{xcolor}
\usepackage[switch]{lineno}
\usepackage{gensymb}
\usepackage{soul}
\usepackage{rotating}
\usepackage{comment}
\usepackage{multirow}
\usepackage{longtable}
\usepackage{lscape}
\usepackage{color}
\usepackage{makecell}
\usepackage{placeins}  

\newcommand{\eqb}{\begin{eqnarray}}
\newcommand{\eqe}{\end{eqnarray}}

\defcitealias{Hovatta2021}{H21}
\defcitealias{Kouch2024_CGRaBSvIC}{K24}
\defcitealias{Kouch2025_WH2}{K25}
\defcitealias{Kouch2025_CAZ_catalog}{K26}

\begin{document}

\title{Association of the IceCube neutrinos with CAZ blazar light curves}

\subtitle{}

\author{
Pouya M. Kouch \inst{\ref{UTU},\ref{FINCA},\ref{MRO}} \thanks{\href{mailto:pouya.kouch@utu.fi}{pouya.kouch@utu.fi}} \orcid{0000-0002-9328-2750},
Talvikki Hovatta \inst{\ref{FINCA},\ref{MRO},\ref{ELE}} \orcid{0000-0002-2024-8199}, 
Elina Lindfors \inst{\ref{UTU},\ref{FINCA}} \orcid{0000-0002-9155-6199}, 
Ioannis Liodakis \inst{\ref{IA_FORTH_Crete}} \orcid{0000-0001-9200-4006}, 
Karri I.I. Koljonen \inst{\ref{NTNU}} \orcid{0000-0002-9677-1533}, 
and
Alessandro Paggi \inst{\ref{IA_FORTH_Crete}} \orcid{0000-0002-5646-2410} 
}

\institute{
Department of Physics and Astronomy, University of Turku, FI-20014 Turku, Finland \label{UTU}
\and
Finnish Centre for Astronomy with ESO (FINCA), Quantum, Vesilinnantie 5, University of Turku, FI-20014 Turku, Finland \label{FINCA}
\and
Aalto University Mets\"ahovi Radio Observatory, Mets\"ahovintie 114, FI-02540 Kylm\"al\"a, Finland \label{MRO}
\and
Aalto University Department of Electronics and Nanoengineering, PL~15500, FI-00076 Espoo, Finland\label{ELE}
\and
Institute of Astrophysics, Foundation for Research and Technology-Hellas, GR-71110 Heraklion, Greece \label{IA_FORTH_Crete}
\and
Institutt for Fysikk, Norwegian University of Science and Technology, H{\o}gskloreringen 5, Trondheim, 7491, Norway \label{NTNU}
}

\date{Received October 07, 2025; accepted February 18, 2026}


\abstract{The IceCube Neutrino Observatory has detected several hundred high-energy neutrinos from cosmic sources. Despite numerous studies searching for their origin, it is still not known which sources emit them. A few likely individual associations exist with active galactic nuclei (AGNs), mostly comprising blazars (AGNs with jets pointed toward Earth). Nonetheless, on a population level, blazar-neutrino correlation strengths are rather weak. This could mean that blazars as a population do not emit high-energy neutrinos or that the detection power of the tests is insufficient due to the strong atmospheric neutrino background. By assuming an increase in high-energy neutrino emission during major blazar flares, in our previous studies, we leveraged the neutrino arrival time to boost the detection power. Here, we utilize the same principle, while substantially increasing the number of blazars in the sample. We searched for the spatiotemporal correlation of 356 IceCube high-energy neutrinos with major optical flares of 3225 radio- and 3814 $\gamma$-ray-selected blazars. We found that despite the increase in data size, the number of likely spatiotemporal associations remained low and the overall correlation strengths weak. Two individual associations were shown to drive our strongest correlation, namely, the only $>$2$\sigma$ post-trial spatiotemporal correlation, occurring with the BL Lac objects of the radio-selected blazar sample. We estimated that $\lesssim$8\% of the detected cosmic neutrinos were emitted by blazars during major optical flares. As a complementary analysis, we compared the synchrotron peak frequency, redshift, Doppler factor, X-ray brightness, and optical variability of spatially neutrino-associated blazars to those of the general blazar population. We found that spatially neutrino-associated blazars have a Doppler factor and X-ray brightness that are higher than average.}


\titlerunning{Association of the IceCube neutrinos with CAZ blazar light curves}
\authorrunning{Kouch et al.}

\maketitle
%
\section{Introduction} \label{sec_intro}
At present, hundreds of high-energy (TeV-PeV) neutrinos from extraterrestrial sources have been detected and found to be in excess of the strong atmospheric neutrino background by various neutrino observatories, such as IceCube (e.g., \citealt{IC2021_history_ref}), ANTARES (e.g., \citealt{Albert2021_ANTARES}), and KM3NeT (e.g., \citealt{KM3NeT2025_100PeV_detection}). Further detections are expected by the next generation or future observatories, such as Baikal-GVD (e.g., \citealt{Baikal2018_status}), RNO-G (e.g., \citealt{Aguilar2021_RNO_Greenland}), GRAND (e.g., \citealt{Fang2017_GRAND}), P-ONE (e.g., \citealt{Malecki2024_P_ONE}), TRIDENT (e.g., \citealt{Ye2023_TRIDENT}), and HUNT (e.g., \citealt{Huang2024_HUNT}). Despite the hundreds of  detections that are available, the cosmic sources that produce high-energy neutrinos are still largely unknown (e.g., \citealt{IC2021_neut_review, Troitsky2021_neut_review}).

One of the main candidates for emitting high-energy (TeV--PeV) neutrinos (hereafter, neutrinos) are active galactic nuclei (AGNs; e.g., \citealt{Mannheim1989, Mucke2001, Hooper_Plant2023_leptonic_neut_production}). Powered by supermassive black holes, almost all AGNs display extreme nonthermal emission and $\sim$10\% possess highly collimated jets of relativistic plasma. AGN jets have the ability to accelerate hadrons to high-enough energies which could interact with photons to produce neutrinos (e.g., \citealt{Mannheim1993_hot_hadrons_in_jets}). In AGNs whose jets are viewed at small angles ($\lesssim$$10^{\circ}$; i.e., blazars), all electromagnetic and any potential neutrino emission from the jet becomes strongly Doppler boosted. This makes blazars some of the most luminous objects in the universe and prime targets for a population-based neutrino counterpart search. For recent reviews on blazars, see, for example, \cite{Bottcher2019}, \cite{Blandford2019}, and \cite{Hovatta_Lindfors2019}.

While the hottest neutrino spot in the sky is NGC~1068 (a weakly jetted Seyfert-II AGN, e.g., \citealt{Muxlow1996_NGC1068_jet, Padovani2024_NGC1068}) at a significance of 4.2$\sigma$ (\citealt{IC2022_NGC1068}), the only other few sources potentially associated with neutrinos have been blazars (e.g., TXS~0506+056, \citealt{IC2018_TXS0506}; PKS~1424+240, and GB6~J1542+6129, \citealt{IC2021_PKS1424_GB6J1542}). As such, many studies have focused on the connection of neutrinos with individual blazars (e.g., \citealt{Krauss2014, Kadler2016, Padovani2019_TXS0506_masqBLL, Righi2019, Franckowiak2020, Rodrigues2021_PKS1502_AM3_modeling, Padovani2022_PKS1424_masqBLL, Acharyya2023_MWL_assoc_w_0735, Kun2024, Blinov2025_PKS1502_repeating_pattern, Omeliukh2025_PKS0735_leptohadronic_modeling, KM3NeT2025_230213A_source_search, Kovalev2025_PKS1424_eye_of_sauran}) and blazar populations (e.g., \citealt{Aartsen2020, Plavin2020, Giommi2020, Plavin2021, Smith2021, Hovatta2021, Zhou2021, Bartos2021, Kun2022, Buson2022, Plavin2023, Novikova2023, Buson2023, Bellenghi2023, Suray2024, Albert2024_ANTARES_v_blz_assoc, Plavin2024, Kouch2024_CGRaBSvIC, Abbasi2024_stacking_w_MOJAVE, Abbasi2025_mm_blz_vs_neut, Lu2024}). Some of these population-based analyses hint toward a potential global connection between blazars and neutrinos, albeit rather inconclusively.

There is evidence that the most energetic neutrinos are not coming from continuously neutrino-emitting sources, but possibly from a source population that mainly and rarely emits the highest energy neutrinos (e.g., \citealt{Abbasi2024_no_corr_between_events_and_diffuse}). Blazars are suspected to be such a source population (e.g., \citealt{Padovani2024}). Yet, a confident population-based correlation between blazars and these highest energy neutrinos has not been established. This may be the result of: (1) only a few blazars emitting such neutrinos or (2) inadequate detection power due to suboptimal test statistics and observational limitations (i.e., low counts, low signalness,\footnote{Signalness is an estimate for the probability of a neutrino event being of cosmic origin.} and poor spatial resolution of the reconstructed neutrino events; e.g., \citealt{Liodakis2022_WH, Kouch2025_WH2}). To understand whether (1) is the reason and to confidently identify potential neutrino-emitting blazars, it is vital to diminish the effect of (2) by boosting the detection power. 

In our previous works (\citealt{Hovatta2021} and \citealt{Kouch2024_CGRaBSvIC}, hereafter ``H21'' and ``K24,'' respectively), we attempted to boost the detection power of our correlation analysis by leveraging the precise arrival time of neutrinos (e.g., \citealt{Abbasi2024_stacking_w_MOJAVE}). Since blazar emission is highly variable, with quiescent and flaring periods generally corresponding to lower and higher jet activity, respectively (e.g., \citealt{Lahteenmaki2003_radio_gamma_correlation, LeonTavares2011_radio_gamma_correlation}), we assumed that neutrino production would be enhanced when the jet is more active during major electromagnetic flares (e.g., \citealt{Oikonomou2019_indepth_blz_neut_connection, Kreter2020_fermi_flare_v_neut, Stathopoulos2022_xray_flare_v_neut}). Via this phenomenological assumption, we searched for temporal associations between individual high-energy neutrino arrival times and major, long-term flares experienced by spatially associated blazars in the radio (in both \citetalias{Hovatta2021} and \citetalias{Kouch2024_CGRaBSvIC}) and optical bands (in \citetalias{Kouch2024_CGRaBSvIC}). We obtained post-trial spatiotemporal correlations of $\sim$2$\sigma$ and 2.17$\sigma$ in \citetalias{Hovatta2021} and \citetalias{Kouch2024_CGRaBSvIC}, respectively.

In this paper, we use a similar neutrino sample as \citetalias{Kouch2024_CGRaBSvIC} (i.e., IceCat1+, but updated to include 2021--2023 neutrino events; see Sect. \ref{sec_data_neut}) and rely on the same phenomenological assumption as above (i.e., in blazars, major neutrino flares coincide with major electromagnetic flares). However, unlike the approach taken in \citetalias{Hovatta2021} and \citetalias{Kouch2024_CGRaBSvIC},  here we search for quasi-simultaneous coincidences between neutrino arrival times and major, short-term optical flares (some as short as a few days in duration). We focus on variability in the optical band because: (1) most blazars are variable in the optical band, with optical flaring being often correlated with flaring in higher energy bands (e.g., $\gamma$-rays; \citealt{Liodakis2019_optical_gamma_flare_correlation, deJaeger2023_optical_gamma_flare_correlation}) and (2) the optical band offers the largest number of well-sampled light curves, which is a necessary ingredient for performing temporal associations, owing to recent advances in all-sky surveys. Crucially, we can now enhance the detection power of our spatiotemporal analysis even further by: (1) using the largest sample of blazar optical light curves to date, compiled in our companion paper \cite{Kouch2025_CAZ_catalog}, hereafter referred to as K26 (e.g., see our simulations in \citealt{Liodakis2022_WH} which suggested that having more sources should result in a more significant signal if a spatiotemporal blazar-neutrino connection exists) and (2) using the most optimal test strategy as determined via our simulations in \cite{Kouch2025_WH2}, hereafter referred to as ``K25''. We note that the most optimal test strategy was also utilized in \citetalias{Kouch2024_CGRaBSvIC}.

In Sect. \ref{sec_data}, we describe the data used in this study. In Sect. \ref{sec_analysis}, the methodology of the spatiotemporal analysis and its most optimal test strategy are described. In Sect. \ref{sec_results}, the results are presented and discussed. Finally, we provide the summary and conclusions of our study in Sect. \ref{sec_conclusions}.

\section{Data} \label{sec_data}

\subsection{High-energy neutrino events} \label{sec_data_neut}
In this study, we use an updated version of the IceCat1+ neutrino sample, which we previously compiled in \citetalias{Kouch2024_CGRaBSvIC}. IceCat1+ consists of 267 neutrinos from the IceCube collaboration's IceCat-1 (\citealt{Abbasi2023_IceCat1}), which is the first catalog of muon-track high-energy neutrino events selected based on real-time alert criteria (recorded between May 2011 and December 2020). IceCat1+ additionally contains 16 muon-track events from another IceCube collaboration study (\citealt{Abbasi2022_extra_neut}), where events recorded between May 2009 (when the telescope was still in partial configuration) and December 2018 were selected based on non-real-time criteria. In this paper, we additionally include 73 events between January 2021 and October 2023, which were added to IceCat-1 after its initial publication (v4.0\footnote{\href{https://dataverse.harvard.edu/file.xhtml?fileId=7502710}{https://dataverse.harvard.edu/file.xhtml?fileId=7502710}}). Therefore, this ``updated IceCat1+'' comprises 356 events in total and is given as an electronic table in this paper. This table provides the arrival time, likely sky localization, kinetic energy estimate, and signalness ($\mathcal{S}$) of the 356 events.

While the arrival times are known precisely (on the order of nanoseconds), the sky localizations are not. Using the reported asymmetric right ascension (RA) and declination (Dec.) errors as one-sided 2$\sigma$ Gaussian error estimates, we can trace an ellipsoid corresponding to the $\gtrsim$90\%-likelihood error region of each event in the sky (see Fig. 1 of \citetalias{Kouch2025_WH2}). The area of such an error region ($\Omega$) is calculated as follows:

\begin{equation} \label{eqn_omega}
    \Omega = \frac{\pi}{4}\left(\alpha^+\cdot\delta^++\alpha^-\cdot\delta^++\alpha^-\cdot\delta^-+\alpha^+\cdot\delta^-\right)
,\end{equation}
where $\alpha^{+/-}$ and $\delta^{+/-}$ represent the asymmetric $2\sigma$ error bars in the RA and Dec. directions, respectively. The median $\Omega$ for the updated IceCat1+ is $\widetilde{\Omega}=6.73$~deg$^2$, which is rather large and results in numerous spatial associations. 

Notably, the median signalness for the updated IceCat1+ is $\widetilde{\mathcal{S}}=0.421$, implying that more than half of its neutrinos are of atmospheric origin. These atmospheric neutrinos act as noise when searching for a global blazar-neutrino association. In Sect. \ref{sec_analysis}, we describe a test strategy that involves weighing down neutrinos with large $\Omega$ and small $\mathcal{S}$ to optimally boost the detection power of the blazar-neutrino search.

\subsection{Blazars, their light curves, and their flares} \label{sec_data_blz}
In this study, we used two blazar-dominated AGN samples: (1) a statistically complete subsample of the Radio Fundamental Catalog containing 3225 sources (hereafter denoted as RFC$^\dagger$; see Sect. \ref{sec_data_blz_RFC}) and (2) the Fourth Catalog of Active Galactic Nuclei present in the Large Array Telescope (4LAC) containing 3814 sources (see Sect. \ref{sec_data_blz_4LAC}). Both of these are substantially larger than the blazar samples we used in our previous studies. In \citetalias{Hovatta2021} we used several samples with a combined count of 1795; whereas in \citetalias{Kouch2024_CGRaBSvIC}, we only focused on the CGRaBS sample (the best from \citetalias{Hovatta2021}) with a count of 1157.

In \citetalias{Kouch2025_CAZ_catalog}, we compiled the largest blazar catalog to date, consisting of 7918 unique sources from the RFC$^\dagger$, 4LAC, third catalog of HSP (3HSP), Candidate Gamma-Ray Blazar Survey (CGRaBS), and Fifth Roma-BZCAT Multifrequency Catalogue of Blazars (5BZC) catalogs. Notably, the RFC$^\dagger$ and 4LAC samples are both fully contained in the CAZ catalog. Together they constitute 5880 sources, making up the majority ($\sim$74\%) of the CAZ catalog. In this study, we use these two subsamples of the CAZ catalog instead of its entirety in order to investigate whether radio- or $\gamma$-ray-bright blazars in particular would be more likely to be neutrino emitters. In \citetalias{Kouch2025_CAZ_catalog}, we also tabulated, when available, the physical parameters of the CAZ sources, such as source type, synchrotron peak frequency ($\nu_\mathrm{sy}$), redshift ($z$), radio variability Doppler factor ($D_\mathrm{var}$), and average X-ray flux density ($S_\text{x-ray}$). We provide their X-ray light curves and spectra in another companion paper Paggi et al. (in prep.). In the CAZ catalog, the source types are: flat spectrum radio quasar (FSRQ; denoted as Q), BL Lac object (BLL; denoted as B), host-galaxy dominated BLL (denoted as G), blazar of unknown type (denoted as U), or non-blazar AGN (denoted as A). We additionally divide the sources into classes based on their $\nu_\mathrm{sy}$ following the low-, intermediate-, high-, and extremely high-synchrotron peaked scheme (abbreviated as LSP, ISP, HSP, and EHSP, respectively). LSPs have $\nu_\mathrm{sy}<10^{14}$~Hz, ISPs $10^{14} \leq \nu_\mathrm{sy}<10^{15}$~Hz, HSPs $10^{15} \leq \nu_\mathrm{sy}<10^{17}$~Hz, and EHSPs $\nu_\mathrm{sy} \geq 10^{17}$~Hz. We note that FSRQs are generally LSPs, but BLLs can range from LSPs to EHSPs and are, thus, further divided into: LBLs, IBLs, HBLs, and EHBLs. In Table \ref{table_blazar_info}, we summarize the number and fraction of available parameters for the RFC$^\dagger$ and 4LAC samples.

In \citetalias{Kouch2025_CAZ_catalog}, we additionally constructed decade-long, high-cadence optical light curves for most of the 7918 CAZ sources using the Catalina Real-time Transient Survey (CRTS), the Asteroid Terrestrial-impact Last Alert System (ATLAS), and the \textit{Zwicky} Transient Facility (ZTF) all-sky light curves, which are collectively referred to as the CAZ light curves. Their absolute flux density levels are not always indicative of the flux density of the jet. The host-galaxy and non-jet components of the blazar can contaminate the observed flux densities. Likewise, nearby bright sources can cause flux density contamination as a result of the unsupervised data reduction procedures of the all-sky surveys (e.g., automatic forced-photometry). Therefore, we used the relative changes in flux density rather than the absolute values to trace the intrinsic changes in the activity level of a blazar jet. As such, we identified the variability (see Sect. \ref{sec_data_blz_CAZ_LCs}), periods of enhanced emission (see Sect. \ref{sec_data_blz_BB95}), and prominent flaring periods (see Sect. \ref{sec_data_blz_BBHOP} and \ref{sec_data_blz_BB95_and_BBHOP}) of the CAZ light curves. These light-curve and variability data, along with some of the physical parameters of the blazars, were used in the spatiotemporal analysis of this study. Crucially, only sources with CAZ light curves deemed variable were considered in the spatiotemporal analysis. However, in Sect. \ref{sec_results_candidate_blazar_properties}, we describe our  complementary, spatial-only analysis using all blazars regardless of their variability.

\begin{table*} 
\centering 
\caption{Summary of the count and fraction of available parameters for the RFC$^\dagger$ and 4LAC sources as well as their variable subsamples.}
\label{table_blazar_info}
\begin{tabular}{l|rrrr} 
\hline\hline 
Parameter & RFC$^\dagger$ & Variable RFC$^\dagger$ & 4LAC & Variable 4LAC \\ 
(1) & (2) & (3) & (4) & (5) \\ 
\hline 
Total  & 3225* & 1170* (36.3\% of 3225) & 3814* & 1917* (50.3\% of 3814) \\
\hline
Overlap & 1159 (35.9\%) & 723 (61.8\%) & 1159 (30.4\%) & 723 (37.7\%) \\
\hline
Type (any)                   & 3225 (100\%)  & 1170 (100\%) & 3814 (100\%)  & 1917 (100\%) \\
.. Q (FSRQ)                  & 1773 (55.0\%) & 716 (61.2\%) & 792 (20.8\%)  & 504 (26.3\%) \\
.. B (BLL)                   & 296 (9.2\%)   & 212 (18.1\%) & 1458 (38.2\%) & 857 (44.7\%) \\
.. G (galaxy dominated)      & 4 (0.1\%)     & 2 (0.2\%)    & 0 (0.0\%)     & 0 (0.0\%)    \\
.. U (unclassified blazar)   & 340 (10.5\%)  & 102 (8.7\%)  & 1493 (39.1)   & 529 (27.6\%) \\
.. A (non-blazar AGN)        & 812 (25.2\%)  & 138 (11.8\%) & 71 (1.9\%)    & 27 (1.4\%)   \\
\hline
$\nu_\mathrm{sy}$ class (any)  & 1046 (32.4\%) & 681 (58.2\%) & 3085 (80.9\%) & 1689 (88.1\%) \\
.. LSP  & 992 (30.8\%)  & 646 (55.2\%) & 1699 (44.5\%) & 1016 (53.0\%) \\
.. ISP  & 36 (1.1\%)    & 25 (2.1\%)   & 536 (14.1\%)  & 304 (15.9\%)  \\
.. HSP  & 18 (0.6\%)    & 10 (0.9\%)   & 710 (18.6\%)  & 336 (17.5\%)  \\
.. EHSP & 0 (0.0\%)     & 0 (0.0\%)    & 140 (3.7\%)   & 33 (1.7\%)    \\
\hline
$z$              & 2581 (80.0\%) & 1089 (93.1\%) & 3158 (82.8\%) & 1775 (92.6\%) \\
\hline
$D_\mathrm{var}$ & 792 (24.6\%)  & 512 (43.8\%)  & 584 (15.3\%)  & 456 (23.8\%) \\
\hline
$S_\text{X-ray}$ & 2038 (63.2\%) & 882 (75.4\%)  & 2821 (74.0\%) & 1491 (77.8\%) \\
\hline
\end{tabular}
\tablefoot{Columns (1)--(5) show the parameter for which the available values are counted, the count (and fraction) of available values for RFC$^\dagger$ sources, RFC$^\dagger$ sources which are variable, 4LAC sources, and 4LAC sources which are variable, respectively. Counts of the overlapping sources between RFC$^\dagger$ and 4LAC are given in ``Overlap''. Unless specified, all fractions (given as percentages) are with respect to the total count of each column (denoted using *).}
\end{table*}

\subsubsection{RFC$^\dagger$ sample} \label{sec_data_blz_RFC}
The Radio Fundamental Catalog (RFC\footnote{\href{https://astrogeo.org/sol/rfc/}{https://astrogeo.org/sol/rfc/}}) contains tens of thousands of precisely localized compact radio sources detected via very long baseline interferometry (VLBI). A subsample of it, limited to $S_\text{X-band}^{\text{VLBI}} \geq 150$~mJy, can be considered statistically complete as a result of observations by the VCS (Very Long Baseline Array calibrator surveys; e.g., \citealt{kovalev2007_vcs5, gordon2016_vcsii}) and Australian LBA (Australian Long Baseline Array; e.g., \citealt{petrov2019_lba}). Since flux-limited VLBI samples are generally blazar-dominated (e.g., \citealt{readhead1978_flux_limited_sample_mostly_blz}), we began compiling the CAZ catalog using this subsample of RFC\footnote{The \texttt{rfc\_2023d} version of RFC was used in \citetalias{Kouch2025_CAZ_catalog}.}. The sources in the CAZ catalog were checked to be: (1) AGN; and (2) unique within $\pm$0.001$^\circ$ (see \citetalias{Kouch2025_CAZ_catalog} for justification).

Of the 7918 CAZ sources, 3225: (1) are in RFC; (2) are categorized as AGNs or candidates; and (3) fulfill $S_\text{X-band}^{\text{VLBI}} \geq 150$~mJy. These 3225 sources form the statistically complete RFC sample (i.e., RFC$^\dagger$), whose parameter availability is summarized in Table \ref{table_blazar_info} column (1). We note that RFC$^\dagger$ is dominated by FSRQs and LBLs, as expected from a radio flux-limited sample. There are 1159 sources in RFC$^\dagger$ that are also found in the 4LAC sample.

\subsubsection{4LAC sample} \label{sec_data_blz_4LAC}
4LAC is the second blazar-dominated sample we use in this study, which is also contained in its entirety in the CAZ catalog (see \citetalias{Kouch2025_CAZ_catalog}). We adopted all 3814 low- and high-latitude sources in the third 4LAC data-release (4LAC-DR3\footnote{\href{https://fermi.gsfc.nasa.gov/ssc/data/access/lat/4LACDR3/}{https://fermi.gsfc.nasa.gov/ssc/data/access/lat/4LACDR3/}}) from August 2022, which is based on 12 years of data from the \textit{Fermi} Large Array Telescope (LAT; \citealt{Atwood2009_FermiLAT}). Crucially, we used the radio counterpart coordinates of the 4LAC sources instead of the less reliably constrained $\gamma$-ray coordinates (\citealt{Ajello2022_4lac_dr3}).

In Table \ref{table_blazar_info} column (4), we show the number and fraction of available parameters for the 3814 sources of the 4LAC sample. Notably, as expected from a $\gamma$-ray-selected sample, 4LAC contains more BLLs than FSRQs when compared to RFC$^\dagger$. Additionally, unlike RFC$^\dagger$, a significant portion of the 4LAC sample comprises non-LSPs, although LSPs still make up the majority of the 4LAC sources. Lastly, as mentioned previously, the 4LAC and RFC$^\dagger$ samples have 1159 sources in common.

\subsubsection{CAZ light curves and variability} \label{sec_data_blz_CAZ_LCs}
In \citetalias{Kouch2025_CAZ_catalog} we compiled as long and as high cadence optical light curves as possible for the 7918 sources in the CAZ catalog. We extracted multiband light curves from the CRTS\footnote{\href{http://crts.caltech.edu/}{http://crts.caltech.edu/}} (e.g., \citealt{drake2009}), ATLAS\footnote{\href{https://fallingstar-data.com/}{https://fallingstar-data.com/}} (e.g., \citealt{tonry2018}), and ZTF\footnote{\href{https://ztf.caltech.edu/}{https://ztf.caltech.edu/}} (e.g., \citealt{bellm2019}) all-sky surveys. This was done by performing forced-photometry on the coordinates of the CAZ sources. We additionally took data from the Tuorla blazar monitoring program\footnote{\href{https://tuorlablazar.utu.fi/}{https://tuorlablazar.utu.fi/}} (e.g., \citealt{nilsson2018}) and the Katzman Automatic Imaging Telescope (KAIT\footnote{\href{https://w.astro.berkeley.edu/bait/kait.html}{https://w.astro.berkeley.edu/bait/kait.html}}) monitoring program (e.g., \citealt{filippenko2001, Cohen2014_KAIT_blazars}), when possible.

To ensure maximal cadence for the CAZ light curves, we merged the filters of the different surveys\footnote{There are 11 filter+survey combinations in total: one from CRTS (no filer, but closest is V-band), five from ATLAS (o, c, r, g, and i), three from ZTF (r, g, and i), one from Tuorla (R-band), and one from KAIT (V-band).} by translating them in flux density space. This was done, in order of decreasing priority, by minimizing the difference between simultaneous flux density measurements of two filters (within $\pm$1, or then $\pm2$ days), or minimizing the difference between the average flux density of two filters (within their overlap periods, or then within a $\pm$50-day extension). The procedure to obtain the final CAZ light curves is described in more detail in Sect. 3 of \citetalias{Kouch2025_CAZ_catalog}. We note that all CAZ light curves are publicly available. In total, there are 7722 CAZ light curves with $\ge$10 data points and $\ge$30~d duration. 

In \citetalias{Kouch2025_CAZ_catalog}, we characterized the variability of the CAZ light curves, and then determined periods of enhanced emission and prominent flaring periods in the variable ones. These periods are used in this study to search for temporal coincidences with neutrinos. In short, we focused on the CRTS and ZTF data, while ignoring ATLAS data due to its noisiness, to determine whether a light curve is confidently variable. We quantified their variability using fractional variability ($F_\mathrm{var}$; e.g., \citealt{sokolovsky2017}) and used white dwarf light curves as control. A CAZ light curve is considered variable if either its CRTS $F_\mathrm{var}$ or its ZTF $F_\mathrm{var}$ reach certain critical thresholds. For more details on the variability quantification, we refer the reader to Sect. 4.1 of \citetalias{Kouch2025_CAZ_catalog}. Of the 7722 dense and long enough CAZ light curves, 2798 (36.2\%) were determined to be confidently variable. Notably, this does not mean that the other 4924 (63.8\%) light curves are confidently non-variable, but that their variability is comparable to that of white dwarfs within the predefined thresholds. Since a typical CAZ light curve can be rather noisy, the threshold to confidently ``detect'' variability is rather high.

Since for the main analysis of this study we only performed spatiotemporal correlation tests (requiring flares; see Sect. \ref{sec_analysis}), here we solely focus on variable blazars whose flares can be confidently identified. This inadvertently excludes 63.7\% of RFC$^\dagger$ and 49.7\% of 4LAC blazars from our analysis\footnote{This exclusion rate was set at a similar fraction (i.e., 70\%) in the simulations of \citetalias{Kouch2025_WH2}.}. We provide the parameter availability of the variable RFC$^\dagger$ and 4LAC samples in Table \ref{table_blazar_info}, columns (3) and (5), respectively. Nonetheless, we note that, in Sect. \ref{sec_results_candidate_blazar_properties}, we describe our complementary, spatial-only analysis, which uses all RFC$^\dagger$ and 4LAC blazars regardless of their variability.

\subsubsection{Periods of enhanced emission (BB95 periods)} \label{sec_data_blz_BB95}
In \citetalias{Kouch2025_CAZ_catalog}, to characterize periods of enhanced emission and prominent flaring (Sect. \ref{sec_data_blz_BBHOP}), we applied the Bayesian block (BB; \citealt{scargle2013}) algorithm to the CAZ light curves. Using inherently non-variable white dwarf light curves, which were constructed identically to blazar light curves, we calibrated the BB algorithm to partially account for noise-driven variability. Furthermore, when constructing the BB light curves, we marked extended periods ($>$60~d) without data as seasonal gaps\footnote{Since optical observations are only possible during dark skies, long-term optical light curves often exhibit extended periods of gaps in the data due to the seasonal inability to observe the sources.}.

Subsequently, we defined periods of enhanced emission as those BB whose averaged brightness exceeds the 95th percentile flux density level, referred to as ``BB95.'' Such strict flux density condition ensures that BB95 periods exclusively trace the most extreme optical periods of a variable light curve. As shown in Fig. 7 of \citetalias{Kouch2025_CAZ_catalog}, BB95 durations have an extended range, with typical values of $\sim$2--13~d. Unsurprisingly, almost all variable CAZ light curves result in some BB95: 1161 (99.2\%) of 1170 variable RFC$^\dagger$ sources, and 1906 (99.4\%) of 1917 variable 4LAC sources. The few light curves that do not result in any BB95 have such large flux density error bars that none of their binned flux densities reach the 95th percentile level. 

While BB95 are guaranteed to characterize the most extreme activity periods of a blazar, they suffer from two main caveats: (1) they are easily affected by extreme outliers and (2) they miss most flaring periods since only a handful of flares reach the most extreme flux densities. As an alternative to BB95, in the following subsection, we describe how prominent flaring periods are identified in variable CAZ light curves.

\subsubsection{Prominent flaring periods (BBHOP flares)} \label{sec_data_blz_BBHOP}
In \citetalias{Kouch2025_CAZ_catalog}, to determine prominent flaring periods, we initially applied the BB algorithm to the CAZ light curves. Then we searched for BB whose average flux density corresponds to a local maximum or minimum (i.e., ``hills'' and ``valleys'' in BB flux density, respectively). Subsequently, using a hopping algorithm (e.g., HOP, \citealt{Eisenstein1998_HOP}), we determined valley-hill-valley groups without going through any seasonal gap. These BB groups are referred to as ``BBHOP'' flares (e.g., \citealt{Meyer2019_BBHOP}). To ensure that major flaring periods are characterized in their entirety and that small flux density variations do not result in premature BBHOP flares, we iteratively merged all pairs of neighboring BBHOP flares if: (1) their rise or fall time was short ($<$2~d); or (2) their rise or fall flux density significance change was small ($<$3$\sigma$). For more details, see Sect. 4.3 of \citetalias{Kouch2025_CAZ_catalog}.

Merged BBHOP flares (hereafter, simply referred to as BBHOP flares) characterize most flaring periods in reasonably high-cadence portions of a light curve, even when the flux rise is relatively small. However, we were interested in the most prominent flaring periods, that is, those with the most prominent rise in flux. Therefore, we defined prominent BBHOP flares as those that: (1) occurred in a variable CAZ light curve; (2) had a peak BB brighter than the 75th percentile flux density; (3) had $>$3$\sigma$ rise or fall flux density significance change; and (4) did not consist of three BBs only with just one data point in their peak BB. These conditions resulted in 25194 prominent BBHOP flares across 2639 variable CAZ light curves. In terms of the blazar samples used in this study, 1108 (94.7\%) of 1170 variable RFC$^\dagger$ sources and 1872 (97.7\%) of 1917 variable 4LAC sources have at least one prominent BBHOP flare. Lastly, we note that prominent BBHOP flares have a wide range of total durations, with typical values of $\sim$30--130~d.

While prominent BBHOP flares characterize most major flaring periods, they suffer from a couple of caveats. Firstly, the efficiency of the BBHOP algorithm to characterize flares depends strongly on the cadence and continuity of the data. When the cadence drops or when there are seasonal gaps in the data, it is not possible to reliably identify local minima and maxima in flux density. Secondly, outliers can result in BBHOP flares. Unfortunately, low cadence, seasonal gaps, and noisy data are prevalent in the CAZ light curves, despite our attempts to mitigate them (for more details, see \citetalias{Kouch2025_CAZ_catalog}).

\subsubsection{BB95 periods as the peak of BBHOP flares} \label{sec_data_blz_BB95_and_BBHOP}
Since both BB95 and prominent BBHOP flares have strengths and weaknesses for characterizing outburst periods in blazars, we opt to use both of them. Additionally, we consider them simultaneously to identify the peak of the most extreme flares. So, as a third temporal metric in our spatiotemporal correlation tests, we look for BB95 periods which form the peak of a prominent BBHOP flare. Notably, this combined metric does not probe for BB95 periods which occur in the rise or decay segments of a BBHOP flare, but only those which coincide with its peak. For example, in the association of the blazar CAZJ0207+0950 with the neutrino IC190629A (Fig. \ref{fig_CAZJ0207+0950_neut_MJD56579.909}), both the BB95 and BBHOP metrics are met separately, but the combined metric is not met. By identifying the peaks of the most prominent flaring periods via two criteria, we reduce the chance of misidentifying noise-driven scatter as intrinsic. The obvious downside of this combined metric is its need for a high cadence.

\section{Analysis} \label{sec_analysis}
In this work, we searched for spatiotemporal correlation between optical light curves of blazars and IceCube neutrinos. The spatiotemporal test we adopted is similar to the one we used in \citetalias{Hovatta2021} and \citetalias{Kouch2024_CGRaBSvIC}. Here, we use blazar flares identified in variable CAZ optical light curves in \citetalias{Kouch2025_CAZ_catalog}, which limits the blazar sample of the spatiotemporal search of this paper to a subset of optically variable blazars only.

As a first step, we identified blazars that are spatially associated with the neutrino events. We did this by searching for blazar coordinates that fall within the enlarged (by 1$^\circ$ added in quadrature) $\gtrsim$90\%-likelihood error region of the neutrinos. The enlargement was introduced to account for potential unknown systematic errors in the spatial reconstruction of the events (e.g., \citealt{aartsen2013_sys_err}; see Sect. 3.3 of \citetalias{Kouch2024_CGRaBSvIC} for more details). Additionally, previous works (e.g., \citealt{Plavin2020}, \citetalias{Hovatta2021}, \citetalias{Kouch2024_CGRaBSvIC}, \citealt{abbasi2023_plavin_test}) have shown that such an enlargement leads to a hint of a correlation with blazars, since several spatiotemporally associated blazars narrowly miss the published $\gtrsim$90\%-likelihood error region of some well-reconstructed neutrino events. We emphasize that our spatiotemporal method, specifically with a top-hat-like weighting scheme, is robust to such error region enlargements. In fact, our simulations in \citetalias{Kouch2025_WH2} (see Sect. 4.2.6 therein) showed that enlarging the neutrino error regions yields conservative estimates for the strength of the underlying blazar-neutrino correlation. Therefore, here we only performed our spatiotemporal tests using the enlarged $\gtrsim$90\%-likelihood error regions.

Once the spatially associated blazars are identified, we can search their CAZ light curves to determine whether they are temporally associated with their spatially associated neutrino. Based on the simple phenomenological assumption that an enhancement in neutrino production is accompanied by an enhancement in electromagnetic emission, we defined a temporal association based on the following Boolean condition: If the neutrino arrival time coincides with a CAZ optical flare, there is a temporal association; otherwise, there is no temporal association. This assumes simultaneity between neutrino and optical flares (within a few days, considering data limitations). In support of this scenario, there are theoretical models that predict optical flares lagging behind neutrino flares by $\lesssim$1~d (e.g., \citealt{Rodrigues2025_modeling_of_TXS}). However, we caution that there are also theoretical models which predict potential time delays of months to years between neutrino flares and their electromagnetic counterparts (e.g., \citealt{Podlesnyi2025}).

Using the CAZ light curves, we define a temporal association if the neutrino arrived within $\pm$2~d of a BB95 (Sect. \ref{sec_data_blz_BB95}), during a prominent BBHOP flare (Sect. \ref{sec_data_blz_BBHOP}), or within $\pm$2~d of a BB95, which is the peak of a prominent BBHOP flare (Sect. \ref{sec_data_blz_BB95_and_BBHOP}). Since the CAZ light curves suffer from numerous limitations (e.g., uneven sampling, seasonal gaps, outliers, potentially poorly color-corrected simultaneous data), quasi-simultaneity cannot be defined exactly and requires an error margin. Throughout \citetalias{Kouch2025_CAZ_catalog}, we used $\pm$2~d as the lower limit for such an error margin; namely, we used it when merging the filters onto each other (Sect. \ref{sec_data_blz_CAZ_LCs}) and when merging neighboring BBHOP flares together (Sect. \ref{sec_data_blz_BBHOP}). This $\pm$2~d margin is especially necessary when checking for quasi-simultaneity with BB95 periods due to their rather short durations (e.g., as short as $\sim$1~d); however, it is not necessary for BBHOP flares because they typically constitute tens of BBs and are $\sim$30--130~d in duration.

We then counted the spatiotemporal associations to construct our test-statistic (TS) parameter. To obtain the global strength of the blazar-neutrino spatiotemporal correlation, we determined the probability of obtaining the observed TS parameter by chance via a Monte Carlo simulation. We randomized the blazar RA coordinates\footnote{Randomization in declination is avoided since the reliability of the reconstruction of the neutrino events is strongly declination dependent.} and then repeated the above process to count the number of random spatiotemporal associations, resulting in a random TS parameter. This randomization was repeated $10^5$ times such that we obtain $10^5$ random TS parameters. Subsequently, using the following equation, we can obtain the global strength of the blazar-neutrino spatiotemporal correlation:

\begin{equation} \label{eqn_pval}
    p = \frac{M+1}{N+1}
,\end{equation}
where $M$ is the number of random TS parameters greater than or equal to the observed TS parameter, and $N$ is the total number of random TS parameters (\citealt{davison_hinkley_1997}); in this work, $N=10^5$. This process is visualized in Fig. 3 of \citetalias{Kouch2024_CGRaBSvIC}. Although this correlation strength is obtained through the uniform randomization of blazars in the sky, we verified that the results do not change if we exclude blazars within $\pm$10$^\circ$ from the Galactic plane.

In \citetalias{Hovatta2021} and \citetalias{Kouch2024_CGRaBSvIC}, we constructed the TS parameters via both averaging and counting techniques. However, in this work, we only focus on the counting technique since our extensive simulations in \citetalias{Kouch2025_WH2} showed that counting offers a greater detection power than averaging. Additionally, \citetalias{Kouch2025_WH2} showed that employing a weighting scheme to account for poorly reconstructed neutrino events is better than setting neutrino sample cuts (e.g., $\mathcal{S}$ or $\Omega$ cuts). Among the tested weighting schemes, the top-hat scheme performed better than the Gaussian one. Therefore, before constructing the TS parameter, we weigh them down using a top-hat weight ($W_\mathrm{T}$; first introduced in \citetalias{Kouch2024_CGRaBSvIC}). So, the global TS parameter is equal to $\sum_{i=1}^{K}W_{\mathrm{T},\, i}$ where $W_{\mathrm{T},\, i}$ is the top-hat weight of the $i^\mathrm{\, th}$ spatiotemporal association and $K$ is the total number of spatiotemporal associations globally.

The top-hat weight is only dependent on the properties of the neutrino (i.e., $\mathcal{S}$ and $\Omega$) and is calculated as follows:
\begin{equation} \label{eqn_W_T}
    W_\mathrm{T} =
    \begin{cases}
      0 & \text{if $d_{\mathrm{BN},\phi} > R_\phi$ (no spatial association);}\\
      \mathcal{S} & \text{if $d_{\mathrm{BN},\phi} \le R_\phi$ and $\Omega \le \widetilde{\Omega}$};\\
      \mathcal{S} \cdot \widetilde{\Omega} \ / \ \Omega & \text{if $d_{\mathrm{BN},\phi} \le R_\phi$ and $\Omega > \widetilde{\Omega}$.}
    \end{cases}
\end{equation}
Here, $\widetilde{\Omega}$ is the global median of $\Omega$ (see Eqn. \ref{eqn_omega}), $R_\phi$ the distance between the center and the edge of the error region at a given phase angle, $\phi$, and $d_{\mathrm{BN},\phi}$ is the distance between the blazar and the center of the error region along $\phi$ (see Fig. 1 of \citetalias{Kouch2025_WH2}). As noted in Sect. \ref{sec_data_neut}, $\widetilde{\Omega}=6.73~\mathrm{deg^2}$ for the published error regions. However, here, we use the enlarged error regions, whose larger $\Omega$ values result in $\widetilde{\Omega}=10.36~\mathrm{deg^2}$. When applied to all 356 enlarged neutrino events of the updated IceCat1+, the maximum, median, and minimum of $W_\mathrm{T}$ are 0.997, 0.316, and 0.003, respectively.

Notably, $W_\mathrm{T}$ reduces the effect of poorly localized events (i.e., those with $\Omega > \widetilde{\Omega}$) on the statistics, while preventing the well-localized events (i.e., those with $\Omega \ll \widetilde{\Omega}$) from dominating the statistics. This makes the top-hat weighting scheme superior to Gaussian weighting schemes, which are typically dominated by associations made with the most highly localized events. This is a caveat for the Gaussian weights since mid-range events (i.e., those with $\mathcal{S} \sim \widetilde{\mathcal{S}}$ and $\Omega \sim \widetilde{\Omega}$) are more numerous than the most highly localized events, even after accounting for their smaller $\mathcal{S}$ (see Sect. 4.2.5 of \citetalias{Kouch2025_WH2}). The advantage of $W_\mathrm{T}$ comes from its ability to recover more signal from the mid-range events. 

We note that due to the statistical nature of our analysis, not all of the identified associations are expected to be real. Following our previous works (\citetalias{Hovatta2021}, \citetalias{Kouch2024_CGRaBSvIC}, and \citetalias{Kouch2025_WH2}), the global TS parameter is constructed by counting the (weighted) number of unique spatiotemporally associated blazar-neutrino pairs. This allows for: (1) one blazar to be associated with multiple neutrinos; and (2) one neutrino to be associated with multiple flaring blazars. Notably, the latter is physically impossible. Yet, it enhances the statistical power of the test, because a neutrino associated with multiple flaring blazars is more likely to have come from a blazar than a neutrino associated with only one flaring blazar. Therefore, our global TS parameters should be taken as statistical measures rather than real counts. As a sanity check, we also test our spatiotemporal analysis by only considering one flaring blazar association for each neutrino event. In this approach, the TS parameter is calculated by counting the number of neutrino events which have at least one spatiotemporal association. This approach does not suffer from the second limitation mentioned above. We find that the correlation strengths remain largely identical to those obtained via the original formulation.

\section{Results and discussion} \label{sec_results}

\begin{figure*}
    \centering
    \includegraphics[width=18cm]{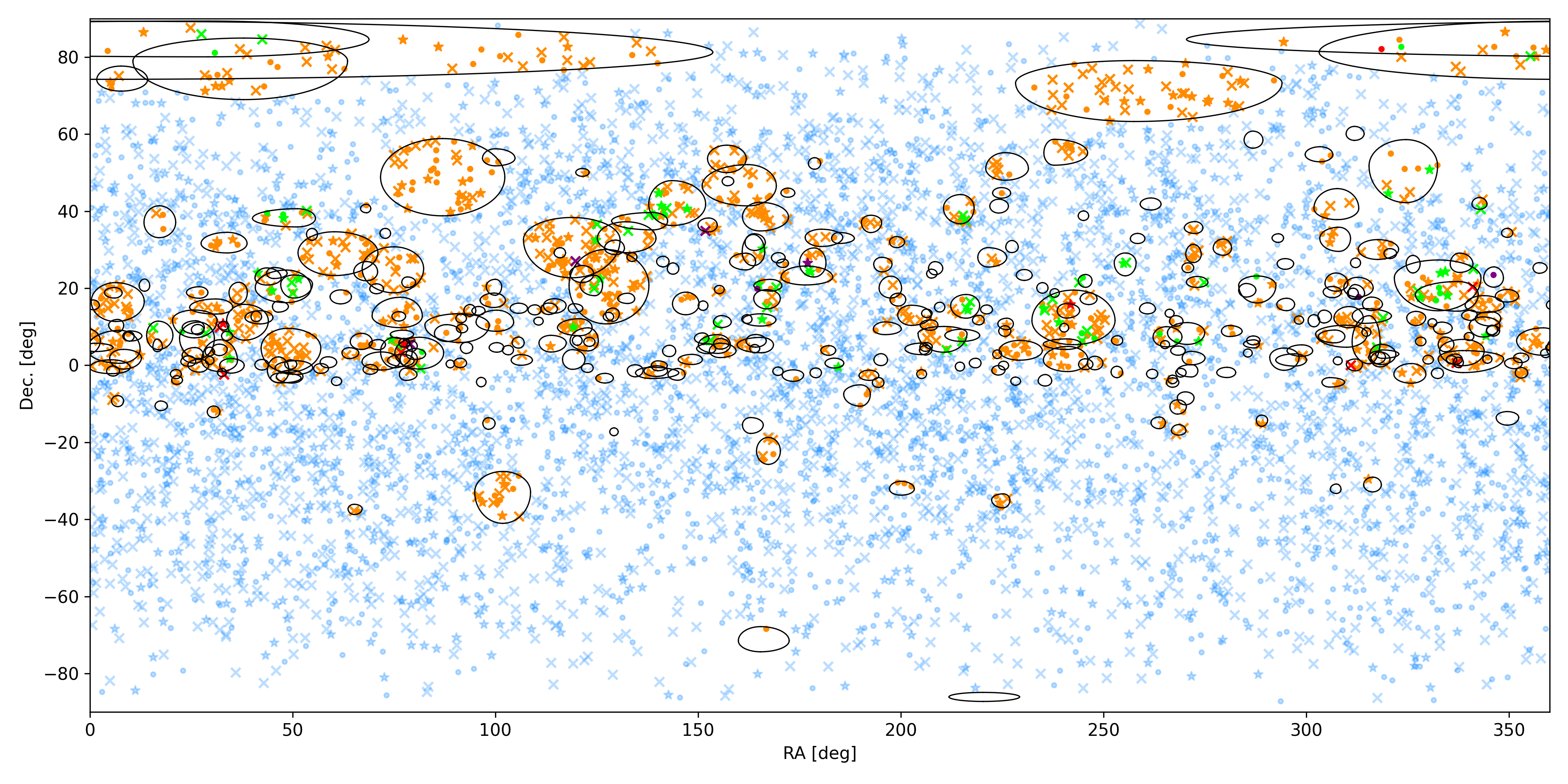}
    \caption{Sky distribution of the 356 updated IceCat1+ neutrinos and the 5880 blazars of the RFC$^\dagger$ and 4LAC samples. The black ellipses show the enlarged $\gtrsim$90\%-likelihood error region of the neutrinos. The circles (symbol count: 2066), crosses (2655), and stars (1159) display blazars present only in RFC$^\dagger$, only in 4LAC, and in both RFC$^\dagger$ and 4LAC, respectively. The faded blue markings (4822) represent uncorrelated blazars, the orange (955) markings refer to spatially associated blazars, the purple markings (7) refer to those blazars which are spatiotemporally associated with at least one neutrino via BB95, the green markings (86) refer to  blazars that are spatiotemporally associated with at least one neutrino via prominent BBHOP flares, and the red markings (10) refer to those blazars which are spatiotemporally associated with at least one neutrino via both metrics. We note that the ten red markings include (but are not limited to) the five associations arising from BB95 at the peak of BBHOP flares (see Sect. \ref{sec_data_blz_BB95_and_BBHOP}). In total, there are $7+10=17$ and $86+10=96$ blazars with at least one spatiotemporal association using the BB95 and prominent BBHOP metrics, respectively.}
    \label{fig_all_assoc_sky_map}
\end{figure*}

\begin{table*} 
\centering 
\caption{Results of the spatiotemporal correlation tests between the updated IceCat1+ neutrinos and blazars of the RFC$^\dagger$ and 4LAC samples.}
\label{table_results}
\begin{tabular}{c|ccc|c|c|c} 
\hline\hline 
Blz & Type & \# & Var \# & BB95 & Prominent BBHOP & BB95 as BBHOP peak\\ 
(1) & (2) & (3) & (4) & (5) & (6) & (7) \\ 
\hline
\multirow{3}{*}{\makecell{RFC$^\dagger$}}
 & (all) & 3225 & 1170 & 8 $\rightarrow$ 2.26 ($p$=$0.1215$) & 55 $\rightarrow$ 6.19 ($p$=$0.8000$) & 3 $\rightarrow$ 1.30 ($p$=$0.0206$) \\
 & FSRQ  & 1773 & 716  & 3 $\rightarrow$ 0.81 ($p$=$0.4861$) & 32 $\rightarrow$ 3.01 ($p$=$0.9000$) & 0 $\rightarrow$ 0.00 ($p$=$1.0000$) \\
 & BLL   & 296  & 212  & 3 $\rightarrow$ 1.31 ($p$=$0.0072$) & 18 $\rightarrow$ 2.91 ($p$=$0.0984$) & 2 $\rightarrow$ 1.30 ($p$=$0.0005$*) \\
\hline
\multirow{3}{*}{\makecell{4LAC}}
 & (all) & 3814 & 1917 & 14 $\rightarrow$ 3.44 ($p$=$0.2038$) & 85 $\rightarrow$ 11.15 ($p$=$0.8695$) & 4 $\rightarrow$ 1.37 ($p$=$0.1086$) \\
 & FSRQ  & 792  & 504  & 1  $\rightarrow$ 0.11 ($p$=$0.9008$) & 21 $\rightarrow$ 2.18  ($p$=$0.9150$) & 0 $\rightarrow$ 0.00 ($p$=$1.0000$) \\
 & BLL   & 1458 & 857  & 11 $\rightarrow$ 2.63 ($p$=$0.0471$) & 53 $\rightarrow$ 7.09  ($p$=$0.4708$) & 4 $\rightarrow$ 1.37 ($p$=$0.0129$) \\
\hline
\end{tabular}
\tablefoot{Columns (1)--(4) represent the main blazar sample, its subsample being tested against the updated IceCat1+, the number of blazars within the subsample, and the number of variable blazars within the subsample, respectively. Columns (5)--(7) represent the spatiotemporal correlation results using the BB95 (Sect. \ref{sec_data_blz_BB95}), the prominent BBHOP (Sect. \ref{sec_data_blz_BBHOP}), and the BB95 as the peak of BBHOP metrics (Sect. \ref{sec_data_blz_BB95_and_BBHOP}), respectively. The arrows connect the unweighted TS (i.e., the count of spatiotemporal associations via each metric) to the weighted TS, and ``$p$'' gives the pre-trial $p$-value of the weighted spatiotemporal correlation. Only one of the 18 $p$-values (marked with *) remains significant at the 2$\sigma$ level after trial correction.}
\end{table*}

\begin{table*} 
\centering 
\caption{List of spatiotemporal associations with $W_\mathrm{T}>0.316$ between the updated IceCat1+ neutrinos and blazars of RFC$^\dagger$ and 4LAC.}
\label{table_assoc}
\begin{tabular}{ccccccccccccccc}
\hline\hline 
CAZ J2000 & IC ID & MJD & $\mathcal{S}$ & $\Omega$ & $W_\mathrm{T}$ & {\rotatebox[origin=c]{90}{\,RFC$^\dagger$\,}} & {\rotatebox[origin=c]{90}{\,4LAC\,}} & {\rotatebox[origin=c]{90}{\,BB95\,}} & {\rotatebox[origin=c]{90}{\,BBHOP\,}} & T & $\nu_\mathrm{sy}$ & $z$ & $D_\mathrm{var}$ & $S_\text{X-ray}$ \\ 
(1) & (2) & (3) & (4) & (5) & (6) & (7) & (8) & (9) & (10) & (11) & (12) & (13) & (14) & (15) \\ 
\hline 
J0211+1051 & 131014A & 56579.9 & 0.67 & 4.5 & 0.665 & Y & Y & Y* & Y* & B & 14.2 & 0.2 & 8.4 & 5.6e-12 \\ 
J0207+0950 & 131014A & 56579.9 & 0.67 & 4.5 & 0.665 & - & Y & Y & Y & U & 13.4 & 0.6 & - & 1.7e-12 \\ 
J0509+0541 & 170922A & 58018.9 & 0.63 & 4.9 & 0.631 & Y & Y & Y* & Y* & B & 14.6 & 0.3 & 14.7 & 3.5e-12 \\ 
J0212$-$0221 & 230724A & 60149.1 & 0.53 & 3.6 & 0.526 & - & Y & Y & Y & B & 14.7 & 0.2 & - & 7.9e-12 \\ 
J2251+4030 & 131108A & 56604.6 & 0.50 & 8.9 & 0.502 & - & Y & - & Y & B & - & 0.2 & - & 5.0e-13 \\ 
J1103+1158 & 200109A & 58858.0 & 0.77 & 20.6 & 0.387 & Y & Y & - & Y & Q & 14.1 & 0.9 & - & 1.5e-12 \\ 
J1615+2130 & 190413B & 58586.7 & 0.38 & 8.8 & 0.383 & - & Y & - & Y & U & 13.1 & 1.6 & - & - \\ 
J1619+2247 & 190413B & 58586.7 & 0.38 & 8.8 & 0.383 & Y & - & - & Y & Q & - & 2.0 & 9.4 & - \\ 
J2304+2331 & 100608X & 55355.5 & 0.72 & 20.2 & 0.370 & Y & - & Y & - & Q & - & 1.2 & - & - \\ 
J1018+1036 & 220317A & 59655.1 & 0.36 & 10.6 & 0.356 & - & Y & - & Y & U & - & 0.7 & - & - \\ 
J1819+2132 & 220822A & 59813.9 & 0.38 & 11.8 & 0.335 & - & Y & - & Y & B & 14.3 & 0.4 & - & 3.3e-13 \\ 
J1058+1951 & 130408A & 56390.2 & 0.53 & 16.4 & 0.331 & Y & - & Y & - & Q & - & 1.1 & 1.3 & 2.2e-12 \\ 
J1117+2014 & 130408A & 56390.2 & 0.53 & 16.4 & 0.331 & - & Y & - & Y & B & 16.2 & 0.1 & - & 6.8e-12 \\ 
J1059+2057 & 130408A & 56390.2 & 0.53 & 16.4 & 0.331 & Y & Y & - & Y & Q & - & 0.4 & 13.6 & 1.0e-12 \\ 
$\vdots$ & & & & & & & & & & & & & & \\
\hline
\end{tabular}
\tablefoot{Column (1) gives the CAZ J2000 name of the spatiotemporally associated blazar; (2) IceCube ID of the spatiotemporally associated neutrino; (3) arrival MJD of the neutrino; (4) signalness of the neutrino; (5) area of the enlarged $\gtrsim$90\% error region of the neutrino; (6) top-hat weight of the neutrino; (7) presence of the blazar in the RFC$^\dagger$ sample; (8) presence of the blazar in the 4LAC sample; (9) temporal association via a BB95 period; (10) temporal association via a prominent BBHOP flare; (11) blazar type from the CAZ catalog; (12) synchrotron peak frequency in log(Hz); (13) redshift; (14) radio variability Doppler factor; and (15) median flux density in the X-ray band (in erg cm$^{-2}$ s$^{-1}$). In columns (7)--(10), ``Y'' means ``present''. In columns (9) and (10), ``*'' indicates an association based on the BB95 as the peak of BBHOP metric (i.e., when the spatially associated neutrino falls within $\pm$2~d of the peak of a prominent BBHOP such that the peak itself is a BB95; see Sect. \ref{sec_data_blz_BB95_and_BBHOP}). For brevity, this table only shows blazar-neutrino spatiotemporal associations with $W_\mathrm{T}>0.316$ (i.e., those with higher than median weight), hiding 96 low-weight associations. The full list of the spatial and spatiotemporal associations is given as an electronic table.}
\end{table*}

We show the results of the spatiotemporal correlation analysis between the CAZ optical light curves and the IceCube neutrinos in Table \ref{table_results}. We list all individual spatiotemporal associations with $W_\mathrm{T}>0.316$ (i.e., greater than the median weight) in Table \ref{table_assoc}. The full list of the spatial and spatiotemporal blazar-neutrino associations is given as an electronic table. The associations are visualized in an all-sky map in Fig. \ref{fig_all_assoc_sky_map}.

As described in Sect. \ref{sec_data_blz}, the test is performed on two samples of blazars: RFC$^\dagger$ and 4LAC. Since FSRQs and BLLs are thought to be physically distinct (see Sect. \ref{sec_intro}), several studies investigated their correlation with neutrinos separately (e.g., \citealt{Aartsen2017_IceCube_stacking_blz_subpop}). As such, in this paper, we additionally tested the spatiotemporal correlation of neutrinos with FSRQs and BLLs separately. This results in a total of six different blazar subsamples being tested, hence, six rows in Table \ref{table_results}. As described in Sects. \ref{sec_data_blz_BB95}, \ref{sec_data_blz_BBHOP}, and \ref{sec_data_blz_BB95_and_BBHOP}, the temporal association is tested via three metrics: (1) BB95 periods; (2) prominent BBHOP flares; and (3) BB95 periods forming the peak of a prominent BBHOP flare. Thus, altogether we have 18 $p$-values.

Overall, we find that the total number of high-weight blazar-neutrino spatiotemporal associations is low. As a result, blazars and neutrinos exhibit a weaker than 2$\sigma$ (pre-trial) spatiotemporal correlation in 14 out of the 18 tested scenarios. Of the four scenarios crossing the 2$\sigma$ threshold, three are weaker than 3$\sigma$, while one reaches 3.48$\sigma$ (pre-trial). Crucially, the strongest correlations are driven by only two high-weight individual associations, namely: (1) the IBL CAZJ0211+1051 (4FGL~J0211.2+1051) with the neutrino IC131014A ($W_\mathrm{T}$=$0.665$); and (2) the IBL CAZJ0509+0541 (TXS~0506+056) with the well-known IC170922A ($W_\mathrm{T}$=$0.631$). 

As we consider 18 test scenarios, we correct for the look elsewhere effect by multiplying all of the $p$-values with 18, following the Bonferroni trial correction method. This results in only one $>$2$\sigma$ post-trial significance, arising between neutrinos and the BB95 periods forming the peak of BBHOP flares of RFC$^\dagger$ BLLs. The exact post-trial $p$-value for this scenario is 0.0005$\times$18$=$0.009, which is equivalent to 2.61$\sigma$. As a sanity check, we also performed the trial correction using the less conservative method of \cite{Benjamini_Hochberg1995}, which led to identical results as the Bonferroni. We note that we choose not to use the harmonic mean $p$-value of \cite{Wilson2019_harmonic_mean_pval} to calculate a combined post-trial $p$-value as done in \citetalias{Kouch2024_CGRaBSvIC}, because here we test various subpopulations of blazars with vastly different sample sizes unlike in \citetalias{Kouch2024_CGRaBSvIC}. In the following subsections we discuss and interpret our results further.

\subsection{Effect of seasonal gaps} \label{sec_results_observational_gaps}
The low number of spatiotemporal associations could be due to a major data limitation present in the CAZ light curves: frequent and long seasonal gaps. Of the 605 spatial associations between neutrinos and variable blazars, in 333 (55.0\%) the neutrino arrives in a part of the light curve which is without data. Either the neutrino arrives outside of the total range of the light curve (in 79 cases) or within a seasonal gap (in 172 cases in the CRTS portion, and 82 in the rest of the light curve).

While it is not known which of these spatially neutrino-associated blazars would have been flaring at the time of the neutrino arrival, we can estimate the fraction. For the 272 (45.0\% of 605) spatial associations for which we have data available, in 110 (39.7\% of 272) of them, the neutrino coincides with a BB95 period or a prominent BBHOP flare (8 with BB95 only, 92 with BBHOP only, and 10 with both BB95 and BBHOP simultaneously). Unsurprisingly, this coincidence rate is similar to the combined average duty cycle of prominent BBHOP flares ($\sim$34\%; see Fig. 12 of \citetalias{Kouch2025_CAZ_catalog}) and that of BB95 ($\sim$5\%; see Fig. 14 of of \citetalias{Kouch2025_CAZ_catalog}). If we assume such a flare coincidence rate for all variable light curves globally, we estimate that around $333 \times 0.397 \approx 132$ potential spatiotemporal associations are missed because of the seasonal gaps. This is a notable caveat for our spatiotemporal analysis because it prevents us from identifying a potential temporal association in around $132/605 \approx 22\%$ of the spatial associations.

\subsection{BB95 periods versus prominent BBHOP flares} \label{sec_results_BB95_v_BBHOP}
As seen from Table \ref{table_results}, using BB95 periods as a proxy for temporal associations leads to substantially stronger, albeit fewer, correlations than using prominent BBHOP flares. As described in Sect. \ref{sec_data_blz_BB95}, BB95 periods are exclusive to the top 5\% flux densities and can be as short as a few days. On the other hand, as described in Sect. \ref{sec_data_blz_BBHOP}, BBHOP flares trace a major outburst, whose peak reaches the 75th percentile flux density, from its inception to its termination regardless of the flux density at the start and end. BBHOP flares can last as long as a few hundred days at a time.

A stronger association when using BB95 suggests that neutrinos may be preferably emitted during the most extreme optical outbursts rather than more generally between the start and end of major flares. Nonetheless, we caution that this result is sensitive to how reliably BB95 periods and BBHOP flares are identified. As mentioned in Sects. \ref{sec_data_blz_BB95} and \ref{sec_data_blz_BBHOP}, both of these identifications suffer from caveats, with noise-driven variations affecting the reliability of the BBHOP flares more.

Interestingly, the strongest correlation strengths are found upon minimizing the effect of the noise-driven variations. This is done by temporally associating neutrinos with BB95 periods which form the peak of a prominent BBHOP flare (see Sect. \ref{sec_data_blz_BB95_and_BBHOP}). Since this metric requires a relatively high cadence to be identified, they are generally rare in the CAZ light curves. Still, two of the five blazar-neutrino associations emerging from this metric are with the two most high-weight spatiotemporal associations, which greatly enhances the correlation strengths obtained via this temporal metric. This reinforces the above interpretation that neutrinos preferably arrive at the peak of the largest outburst periods from blazars.

\subsection{FSRQs versus BLLs} \label{sec_results_FSRQ_vs_BLL}
When testing blazar-neutrino correlations with blazars divided into FSRQs and BLLs, Table \ref{table_results} shows that BLLs are more strongly spatiotemporally correlated with the updated IceCat1+ neutrinos than FSRQs. In both RFC$^\dagger$ and 4LAC samples, and via all the temporal metrics, the correlation strength substantially increases when going from all blazars to BLLs, while a decrease is seen when going from all blazars to FSRQs. This generally contrasts the results that suggest FSRQs are more strongly correlated with neutrinos (e.g., \citealt{Moretti2025_FSRQ_corr_w_neut}).

Naively, our result would suggest that BLLs are better neutrino emitters than FSRQs. However, this generalization should be considered cautiously, because the correlation is driven by only two individual high-weight associations, namely: (1) the IBL CAZJ0211+1051 with IC131014A; and (2) the IBL CAZJ0509+0541 with IC170922A. First, it is possible that we are missing other potential spatiotemporal associations due to the presence of long and frequent seasonal gaps in the CAZ light curves (see Sect. \ref{sec_results_observational_gaps}). With more data, it is possible that this apparent BLL exclusivity disappears. Second, these correlations could arise because of some specific properties present in these individual sources, possibly unique to them rather than their entire class. Third, this generalization is strongly dependent on the reliability of the source classifications. If these blazars are misidentified, they can falsely favor the misidentified class as the more likely neutrino emitter because of their high weight. In this regard, the BLL-classification of some neutrino-emitting candidate blazars have been questioned in the past (e.g., that of TXS~0506+056, \citealt{Padovani2019_TXS0506_masqBLL}).

Interestingly, we find that the apparent BLL-neutrino correlation is stronger when using RFC$^\dagger$ than 4LAC. This is unexpected if BLLs as a whole were the more likely neutrino emitters, since the BLL-to-FSRQ ratio in RFC$^\dagger$ is $\sim$0.2 while $\sim$1.8 in 4LAC. Instead, this suggests that a subpopulation of BLLs that are present in RFC$^\dagger$ are the more likely neutrino emitters. Given that RFC$^\dagger$ is a radio-flux-limited sample, this subpopulation consists of radio-bright BLLs. Indeed, such sources have been suggested to be transitional between BLLs and FSRQs (e.g., \citealt{Ghisellini2011_transitional_BLL_FSRQ}). Such transitional sources could be intrinsically FSRQs which can easily get misclassified as BLLs due to their emission lines becoming outshined by a strong jet continuum. These so-called ``masquerading BLLs'' have been linked to neutrino emission (e.g., \citealt{Padovani2019_TXS0506_masqBLL, Padovani2022_PKS1424_masqBLL, Sahakyan2023_PKS0735_masq_BLL, Rodrigues2024_spectra_of_neut_blz}), which is partly inline with our finding that radio- and $\gamma$-ray-bright IBLs are observationally the most likely candidates for neutrino emission.

\subsection{Fraction of cosmic neutrinos emitted by blazars} \label{sec_results_frac_of_neutrinos_coming_from_blazars}
In \citetalias{Kouch2025_WH2} we found that even if only 20\% of neutrinos were emitted by blazars at the time of a major flare, our most optimal spatiotemporal test setup (counted and top-hat weighted; used in this study) should confidently detect the correlation. The simulations had $\sim$4000 blazars with a 30\% flaring rate, comparable to the RFC$^\dagger$ and 4LAC tests here. Of the 1000 simulations we ran, in 87.9\% the blazar-neutrino spatiotemporal correlation was detected at least at the $\sim$4$\sigma$ level, and in 97.8\% at the 3$\sigma$ level. The mean of the 1000 $p$-values was 0.0004. To compare the observed significances to the simulation ones, we focus on $p$-values obtained via the BB95 temporal metric. This is because the BB95 temporal metric uses the extreme tail of the flux density distribution to identify flares, similar to how flares were defined in the simulations. Thus, the observed $p$-value for all RFC$^\dagger$ blazars is 0.1246 (i.e., 1.54$\sigma$) and for all 4LAC blazars 0.2038 (i.e., 1.27$\sigma$). Clearly, fewer than 20\% of the updated IceCat1+ cosmic neutrinos were emitted during major optical flares of blazars.

There are two differences between the tests here and those of \citetalias{Kouch2025_WH2}: (1) the effect of seasonal gaps was not simulated in \citetalias{Kouch2025_WH2}, whereas here the light curves have a seasonal gap rate of 55\% leading to the omission of $\sim$22\% of spatial associations as potential spatiotemporal ones (see Sect. \ref{sec_results_observational_gaps}); and (2) in \citetalias{Kouch2025_WH2} the original IceCat1+ neutrino sample (283 events) was used, but here we use the updated version (356 events; larger by $\sim$26\%). While the effect of (1) may be critical and needs to be accounted for, we checked that the effect of (2) is minimal by repeating the simulations of \citetalias{Kouch2025_WH2} using the updated IceCat1+.

To roughly estimate the fraction of the updated IceCat1+ cosmic neutrinos emitted by blazars during major optical flares, here we run similar simulations to those of \citetalias{Kouch2025_WH2} while taking into account the seasonal gaps. Within a simulation run, we randomly generate one set of blazars with a sample size of 4000 and a variability rate of 30\%, which are generally comparable to those of the 4LAC and RFC$^\dagger$ samples. The random generation process is described in detail in Sect. 3.1.1 of \citetalias{Kouch2025_WH2}. In short, we generate 4000 random coordinates, assign each a random $F_\mathrm{var}$ following the observed $F_\mathrm{var}$ distribution, and finally assign each a random flux density distribution similar to those observed, which is then used to randomly generate flux densities with respect to all neutrinos in the updated IceCat1+. At the end, some of these 4000 random blazars can be spatiotemporally associated with a neutrino by chance, if they happen to fall within the error region of the neutrino, have a critically large $F_\mathrm{var}$, and have a critically large flux density associated with the neutrino. In addition to these, we simulate a number of blazars to be spatiotemporally associated with neutrinos from the updated IceCat1+ (selected randomly based on their signalness). By design, these blazars are randomly centered on the most likely arrival direction of their respective neutrino event with a Gaussian deviation equivalent to the neutrino event error bars, have a critically large $F_\mathrm{var}$, and a critically large flux density associated with the neutrino (see Sect. 3.1.2 of \citetalias{Kouch2025_WH2}). Since the exact number of spatiotemporally neutrino-associated blazars depends on the signalness of the neutrinos, it can vary from one simulation run to another. This number also depends on the fraction of cosmic neutrinos assumed to be coming from blazar flares. In the ``realistic'' scenario of \citetalias{Kouch2025_WH2}, we assumed a 20\% fraction. However, here, we gradually increase this fraction from 0\% to 20\%.

After each simulation run, we estimate the spatiotemporal correlation strength between the generated sample of blazars and the updated IceCat1+. We focus on the counted, top-hat-weighted scenario, as used in the real tests of this study. We simulate the effect of the 55\% light curve gap fraction by randomly excluding 50\% of spatiotemporal associations from the $p$-value calculations. Next, we repeat this while incrementally increasing the fraction of blazar-flare-emitted cosmic neutrinos from the updated IceCat1+ sample by 1\% from 0\% to 20\%. These simulations are repeated 500 times for each incremental fraction. We note that the Monte Carlo number for each of the 500 repetitions is $10^4$, meaning that the smallest obtainable $p$-value is 0.0001 (i.e., 3.9$\sigma$; similar to \citetalias{Kouch2025_WH2}).

\begin{figure}
    \centering
    \includegraphics[width=0.49\textwidth]{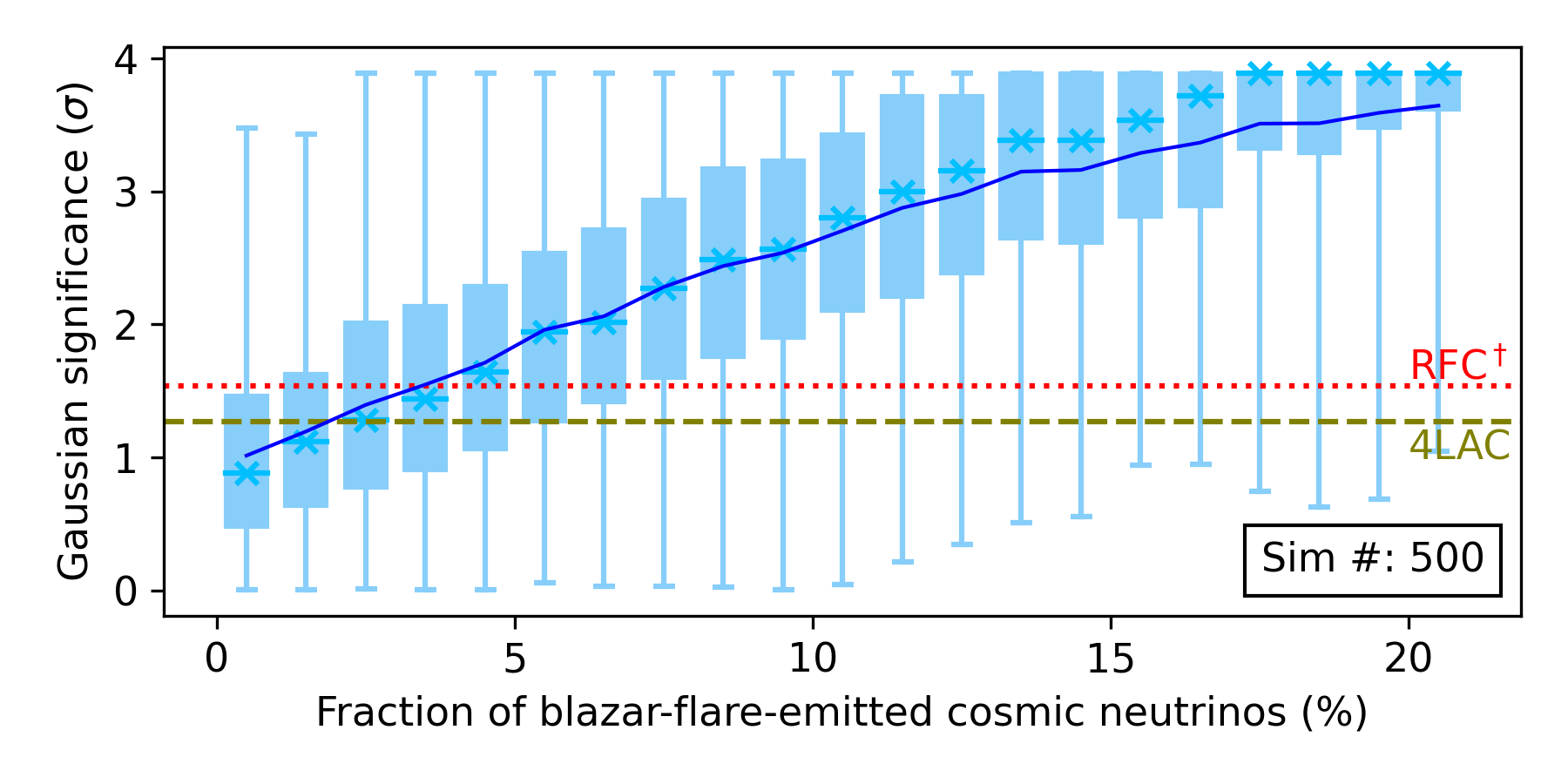}
    \caption{Spatiotemporal correlation significance in Gaussian $\sigma$ against fraction of blazar-flare-emitted cosmic neutrinos from the updated IceCat1+ sample in \%. The blue box plots represent the distribution of the significances for 500 simulations. The running solid (dark blue) line denotes the mean of the significances. The median is shown using a blue cross in the middle of the boxes, while the boxes themselves stretch from the 25th to 75th percentiles. The lines extended out of the boxes show the minimum and maximum significances obtained globally. The dotted red and the dashed olive horizontal lines show the correlation significances between the updated IceCat1+ neutrinos and all blazars of the RFC$^\dagger$ and 4LAC samples, respectively, via the BB95 temporal metric (see Sect. \ref{sec_results_frac_of_neutrinos_coming_from_blazars}). The significance in each simulation is limited to a maximum of 3.9$\sigma$ (i.e., $p=0.0001$). In these simulations we used a light curve gap fraction of 50\%.}
    \label{fig_distr_blz_emitted_neut_frac}
\end{figure}

In Fig. \ref{fig_distr_blz_emitted_neut_frac}, we plot the correlation strength of the simulations in Gaussian $\sigma$ against the fraction of the updated IceCat1+ cosmic neutrinos emitted during major blazar flares in the optical band. Notably, when this fraction is $\sim$20\%, the typical correlation strengths are $\sim$4$\sigma$, similar to those from \citetalias{Kouch2025_WH2}. This suggests that excluding even as much as 50\% of the spatiotemporal associations from the $p$-value calculations does not seem to affect the $p$-values by much. Therefore, while the presence of seasonal gaps can substantially reduce the total number of spatiotemporal associations, it only modestly weakens the overall spatiotemporal correlation strengths. Nevertheless, we caution that this conclusion may only be valid within the assumed 0\%--20\% fraction of the cosmic neutrinos from blazar flares.

In Fig. \ref{fig_distr_blz_emitted_neut_frac}, where the gap fraction is taken as 50\%, we additionally show the observed correlation strengths for all RFC$^\dagger$ and 4LAC blazars via the most comparable temporal metric (i.e., BB95). These are 1.54$\sigma$ and 1.27$\sigma$, respectively. As made evident from Fig. \ref{fig_distr_blz_emitted_neut_frac}, 1.54$\sigma$ spatiotemporal correlation strengths are most often obtained when $\sim$2--8\% of the updated IceCat1+ cosmic neutrinos are emitted during major blazar flares. Based on the 4LAC tests, this fraction would be even smaller. This fraction would also be smaller if the light curve gap fraction were smaller than 50\%. For example, for a light curve gap fraction of 20\%, the observed fraction of the blazar-flare-emitted cosmic neutrinos would be $\lesssim$5\%. Thus, we can confidently conclude that the fraction of the updated IceCat1+ cosmic neutrinos emitted during major optical flares of blazars is $\lesssim$8\%. Remarkably, this is well in agreement with the previous estimates for the fraction of neutrinos coming from blazars (e.g., \citealt{Aartsen2017_blz_maximal_contr, Huber2019_blz_maximal_contr, Oikonomou2022_frac_of_neutrinos_by_blazars}). Interestingly, even among other neutrino-emitting source candidates, like Seyfert-II AGN, neutrino emission appears to be dominated by only a few sources rather than the whole population (e.g., \citealt{Saurenhaus2025_diffuse_neutr_from_Seyferts}).

If indeed $\lesssim$8\% of the updated IceCat1+ cosmic neutrinos arise during blazar optical flares, it is then expected that only $\lesssim$12 of the neutrinos were emitted during bright optical flares. Due to the previously discussed light curve limitations, we unfortunately cannot confidently identify most of their potential emitters (apart from CAZJ0211+1051 and CAZJ0509+0541).

\begin{figure*}
    \centering
    \includegraphics[width=18cm]{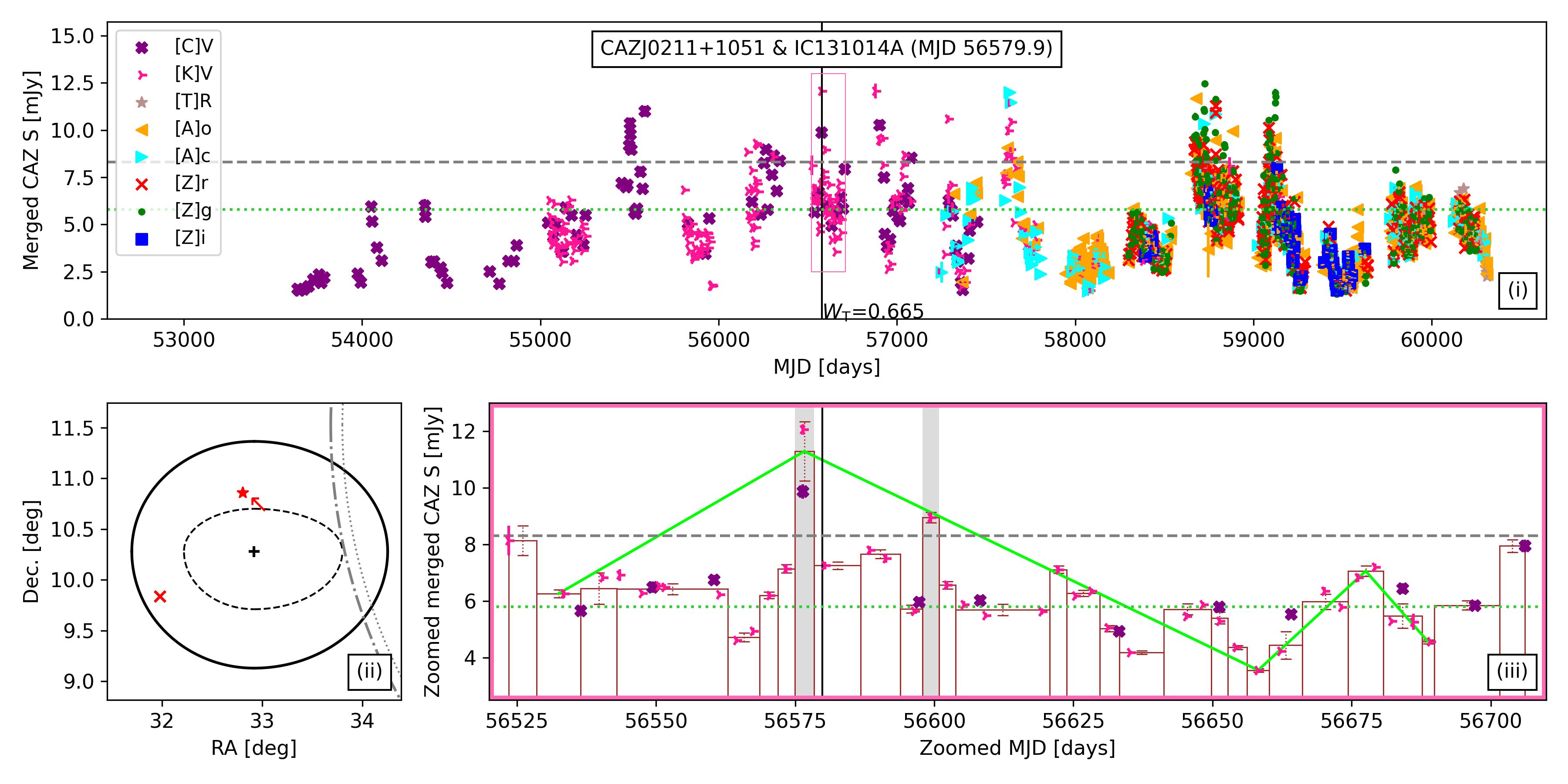}
    \caption{Light curve and sky map of the blazar CAZJ0211+1051, which is spatiotemporally associated with the neutrino IC131014A ($W_\mathrm{T}=0.665$). Plot (i) shows the entire CAZ light curve of the blazar along with the arrival time of the associated neutrino (shown using a solid, vertical, black line). The horizontal dotted green line shows the 75th percentile flux density, and the horizontal dashed grey line the 95th percentile. In the plot legend, ``[C]V'' refers to the CRTS data points (obtained without filter), ``[K]V'' to the V-band of KAIT, ``[T]R'' to R-band of Tuorla, ``[A]o'' to o-filter of ATLAS, ``[A]c'' to c-filter of ATLAS, ``[Z]r'' to r-filter of ZTF, ``[Z]g'' to g-filter of ZTF, and ``[Z]i'' to i-filter of ZTF. Plot (ii) gives the sky map centered around the main associated neutrino, whose enlarged error region edge is drawn using a solid black line and its published error region using a dashed black line. All other neutrino events are plotted in grey (enlarged error regions using dash-dotted lines and published ones using dotted lines). The location of the main blazar within the main associated neutrino is shown with an arrow. Similar to Fig. \ref{fig_all_assoc_sky_map}, blazars which are only in RFC$^\dagger$ are shown as circles, those only in 4LAC as crosses, and those in both RFC$^\dagger$ and 4LAC as stars. Likewise, all blazars which are uncorrelated to the main associated neutrino are marked in faded blue, in orange if only spatially associated, in grey if spatiotemporally associated via BB95, in green if spatiotemporally associated via prominent BBHOP flares, and in red if via both BB95 periods and prominent BBHOP flares simultaneously (not to be confused with the BB95 at the peak of a BBHOP metric described in Sect. \ref{sec_data_blz_BB95_and_BBHOP}). CAZJ0211+1051 is 0.17$^\circ$ outside of the published error region of IC131014A. Plot (iii) shows a zoomed-in version of plot (i) around the arrival time of the associated neutrino. The zoom is demarcated in plot (i) using a pink box. The reddish brown solid lines in the background show the BB, the grey vertical bands show the BB95, and the green solid lines forming a wedge show the prominent BBHOP flares (connecting the start of the flare to its peak and subsequently to its end).}
    \label{fig_CAZJ0211+1051_neut_MJD56579.909}
\end{figure*}

\begin{figure*}
    \centering
    \includegraphics[width=18cm]{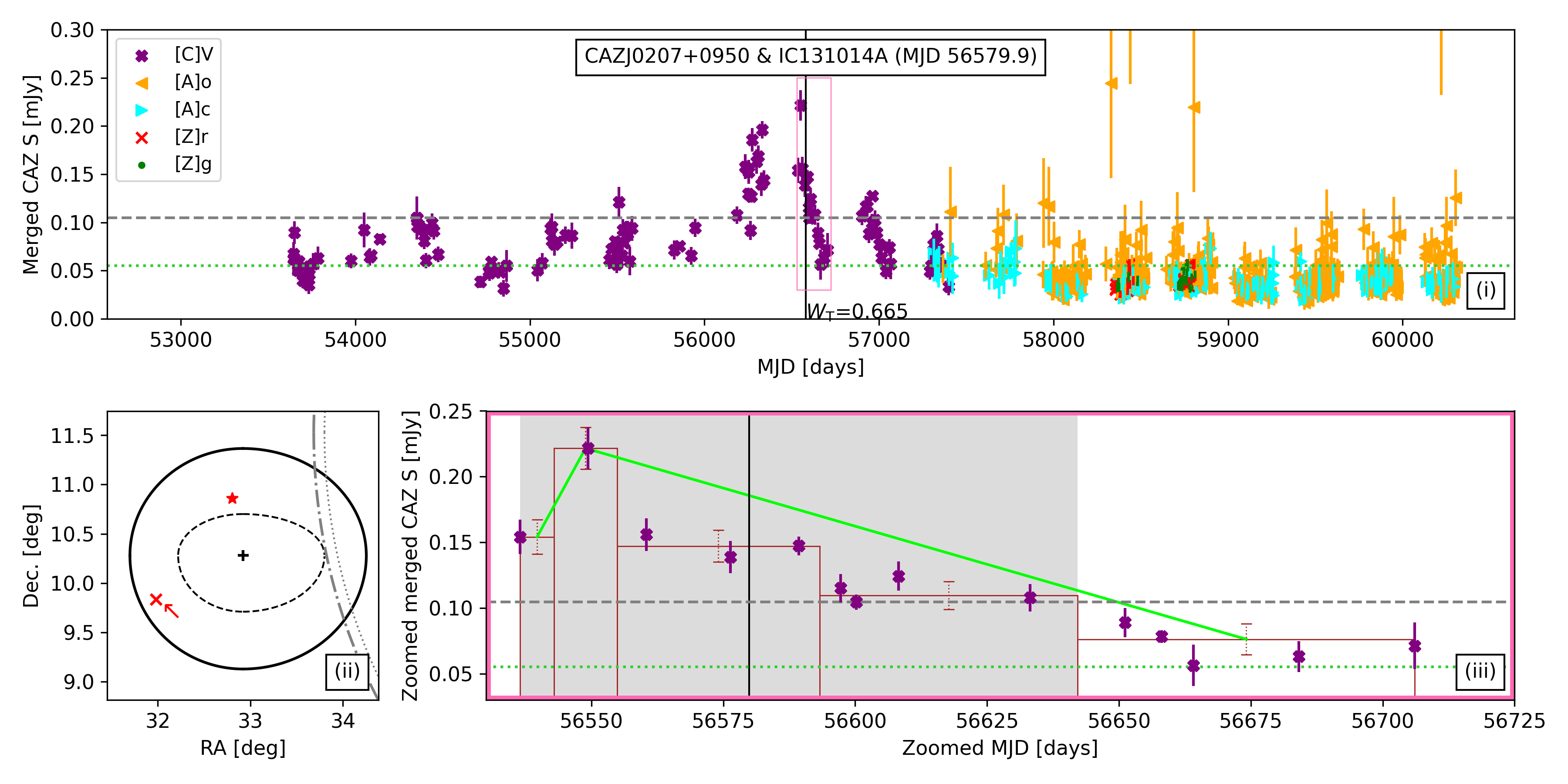}
    \caption{Light curve and sky map of the blazar CAZJ0207+0950 which is spatiotemporally associated with the neutrino IC131014A ($W_\mathrm{T}=0.665$). For plot details, see description of Fig. \ref{fig_CAZJ0211+1051_neut_MJD56579.909}. We note that the y-axis in plot (i) is truncated to omit a few outliers. Spatially, CAZJ0207+0950 is 0.37$^\circ$ outside of the published error region of IC131014A.}
    \label{fig_CAZJ0207+0950_neut_MJD56579.909}
\end{figure*}

\begin{figure*}
    \centering
    \includegraphics[width=18cm]{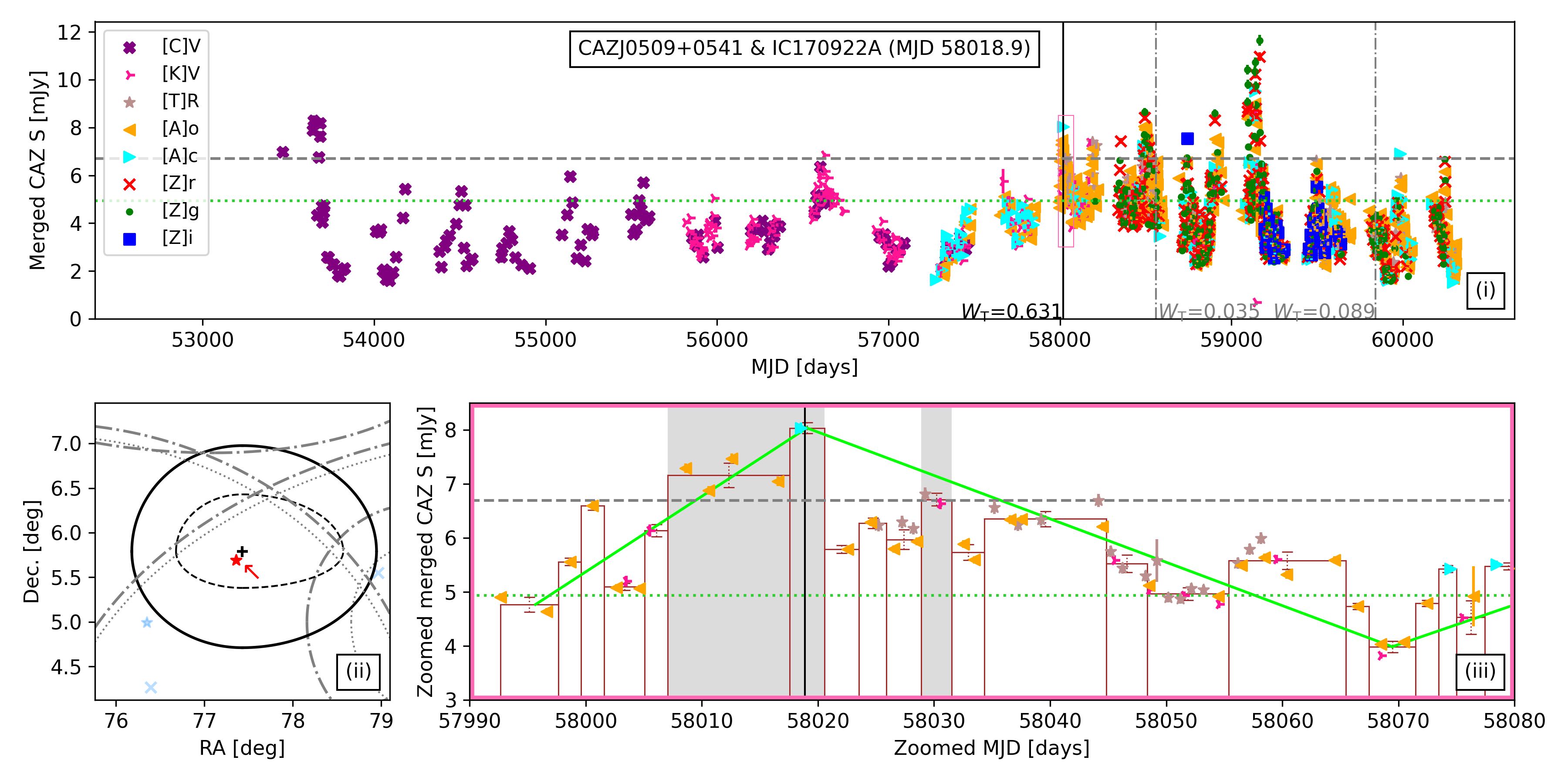}
    \caption{Light curve and sky map of the blazar CAZJ0509+0541 (TXS~0506+056) which is spatiotemporally associated with the neutrino IC170922A ($W_\mathrm{T}=0.631$). For plot details, see description of Fig. \ref{fig_CAZJ0211+1051_neut_MJD56579.909}. The grey dash-dotted vertical lines in plot (i) show spatial associations with a second (IC190317A with $W_\mathrm{T}=0.035$) and a third neutrino (IC220918A with $W_\mathrm{T}=0.089$).}
    \label{fig_CAZJ0509+0541_neut_MJD58018.871}
\end{figure*}

\begin{figure*}
    \centering
    \includegraphics[width=18cm]{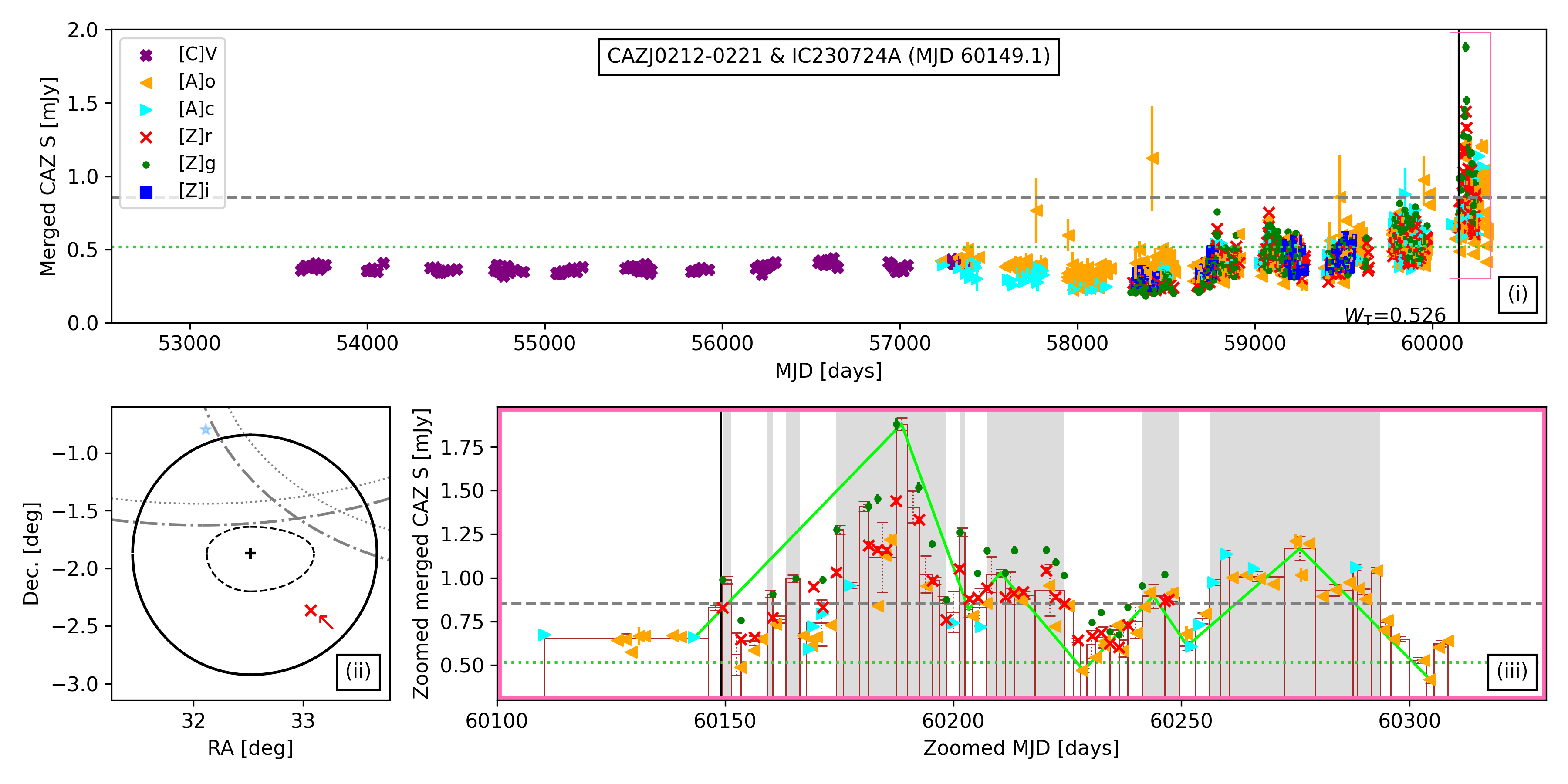}
    \caption{Light curve and sky map of the blazar CAZJ0212$-$0221 which is spatiotemporally associated with the neutrino IC230724A ($W_\mathrm{T}=0.526$). For plot details, see description of Fig. \ref{fig_CAZJ0211+1051_neut_MJD56579.909}. We note that, spatially, CAZJ0212$-$0221 is 0.32$^\circ$ outside of the published error region of IC230724A.}
    \label{fig_CAZJ0212-0221_neut_MJD60149.076}
\end{figure*}

\subsection{Individual spatiotemporal blazar-neutrino associations} \label{sec_results_individual_assoc}
In this subsection, we individually investigate a few interesting spatiotemporal blazar-neutrino associations. We begin with all four associations arising from the BB95 metric with neutrinos that have $W_\mathrm{T}>0.5$. This $W_\mathrm{T}$ threshold selects the most reliably reconstructed neutrino events and is similar to a common quality threshold used in the literature; namely, $\mathcal{S} \ge 0.5$ and $\Omega < 10~\mathrm{deg}^2$. In the updated IceCat1+ sample there are 79 (22\% of 356) neutrino events that meet $W_\mathrm{T}>0.5$, of which 64 (81\%) are classified as ``gold'' events by the IceCube collaboration. Their minimum, median, and maximum $W_\mathrm{T}$ (and $\mathcal{S}$) are 0.502, 0.663, and 0.997, respectively. Likewise, the minimum, median, and maximum areas ($\Omega$) of the enlarged error regions are 3.3, 5.2, and 12.3~$\mathrm{deg}^2$, respectively.

Two of the four BB95 associations with $W_\mathrm{T}>0.5$ (namely, the IBL CAZJ0211+1051 with IC131014A, and the IBL CAZJ0509+0541 with IC170922A) also fulfill the BB95 as the peak of a BBHOP temporal metric. These two associations are the most prominent in our study. In Appendix \ref{appendix_additional_LCs}, we also explore three additional associations that arise from this metric regardless of their low $W_\mathrm{T}$. The complete list of spatial and spatiotemporal associations is given in an electronic table.

The light curve and sky map of the IBL CAZJ0211+1051 associated with IC131014A ($W_\mathrm{T}$=$0.665$) are found in Fig. \ref{fig_CAZJ0211+1051_neut_MJD56579.909}. Notably, this neutrino arrived at the peak of the brightest radio flare ever observed from CAZJ0211+1051 (see, \citetalias{Hovatta2021}, \citetalias{Kouch2024_CGRaBSvIC}). Nevertheless, in the \textit{Fermi} $\gamma$-ray band, the source only showed modest activity at the time of the neutrino arrival (e.g., \citealt{Righi2019}), which does not necessarily disfavor a neutrino connection (e.g., \citealt{Franckowiak2020, Kun2023_gamma_suppression, Garrappa2024_IC_v_Fermi}). Regarding the spatial association, we note that the location is firmly within the enlarged error region of IC131014A but narrowly misses its published error region by 0.17$^\circ$. 

Dedicated studies on CAZJ0211+1051 are needed to confirm if it really is a neutrino emitter. Recent studies have suggested a link between neutrino emission and changes in the multiwavelength polarization degree and angle of blazars (directly, e.g., \citealt{Paraschos2025_neutrino_v_MWL_polarization}; and indirectly, e.g., \citealt{Novikova2023, Blinov2025_PKS1502_repeating_pattern}). Additionally, polarization measurements can reliably trace shocked regions in the jet (e.g., \citealt{Liodakis2022_IXPE_Nature, Kouch2024_IXPE_PKS2155}) that could accelerate protons (a key ingredient of unraveling the blazar-neutrino mystery; e.g., \citealt{Stathopoulos2025_MagRec_neutrino_emission_pc_vs_subpc_scales}). Interestingly, CAZJ0211+1051 is one of the prime targets of multiwavelength photo-polarimetric studies aimed at distinguishing between leptonic and hadronic emissions in blazar jets (e.g., \citealt{Peirson2022_prime_IXPE_targets, Kouch2025_IXPE_0954, Agudo2025_IXPE_BLLac}) and is known to exhibit extreme, rapid changes in its polarization properties (e.g., \citealt{Liodakis2024_J0211_fast_pol_variations}). Thus, by combining such photo-polarimetric studies with dedicated polarization monitoring, future studies could confirm the suspected neutrino-emitting nature of CAZJ0211+1051.

The next spatiotemporal association we focus on is that of CAZJ0207+0950 (4FGL~J0207.9+0953; a blazar candidate) with IC131014A, which is, as discussed above, also spatiotemporally associated with the neighboring blazar CAZJ0211+1051. As seen in Fig. \ref{fig_CAZJ0207+0950_neut_MJD56579.909}, the neutrino arrived $\sim$20~d after the peak of a major flare. Notably, while this meets both the BB95 and BBHOP metrics separately, it does not meet the BB95 at the peak of a BBHOP metrics since the neutrino did not arrive exactly at the peak of the BBHOP flare. Due to the statistical nature of the neutrino data, it is not possible to definitively determine from which (if any) of the two sources this neutrino emerged. Nevertheless, due to CAZJ0211+1051 reaching a historically high radio flux at the time IC131014A arrived, fulfilling the BB95 at the peak of a BBHOP metric in the optical band, and being closer to the most likely arrival direction of IC131014A ($d_{\mathrm{BN},\phi}$ of 0.59$^\circ$ as opposed to 1.04$^\circ$), it could be favored over CAZJ0207+0950 as the emitter of IC131014A.

Subsequently, we mention the well-known association of CAZJ0509+0541 with IC170922A ($W_\mathrm{T}$=$0.631$), whose light curve and sky map are given in Fig. \ref{fig_CAZJ0509+0541_neut_MJD58018.871}. As this association has been studied extensively (e.g., \citealt{Keivani2018_TXS0506, Gao2019_TXS_AM3_modeling, Cerruti2019_TXS_electron_SSC_as_seed, Padovani2019_TXS0506_masqBLL, Rodrigues2019_TXS_AM3_modeling, Reimer2019_Xray_photon_field_in_TXS, Ros2020_TXS0506_via_VLBI, Yang2025_neut_from_disk_of_TXS0506, Fiorillo2025_TXS0506_neut_not_from_corona}), we do not discuss it further. We do however note that CAZJ0509+0541 has two more neutrino associations with IC190317A ($W_\mathrm{T}$=$0.035$) and IC220918A ($W_\mathrm{T}$=$0.089$), which respectively precede and succeed its all-time largest radio flare at MJD$\sim$59000. In the optical band, the arrival time of IC220918A does not show any sign of increased activity, while that of IC190317A coincides with a period of increased activity (as noted in \citetalias{Kouch2024_CGRaBSvIC}) which narrowly misses the BB95 threshold of this study.

Next, we discuss the association of the IBL CAZ J0212$-$0221 with IC230724A ($W_\mathrm{T}$=$0.526$), whose light curve and sky map are given in Fig. \ref{fig_CAZJ0212-0221_neut_MJD60149.076}. While the neutrino arrival time coincides with a BB95 which is locally small in amplitude, it precedes the all-time largest optical flare of the source by $\sim$40~d. As such, this is another example where both the BB95 and BBHOP metrics are met separately, but the BB95 at the peak of a BBHOP metric is not. Spatially, the blazar falls well within the enlarged error region but 0.32$^\circ$ outside of the published one.

\subsection{Comparison to \citetalias{Kouch2024_CGRaBSvIC}} \label{sec_results_K24_comparison}
The spatiotemporal results of this work are not directly comparable to those of \citetalias{Kouch2024_CGRaBSvIC} due to differences in how the temporal associations are defined. First, here we focus on rapid changes in the optical flux density, whereas in \citetalias{Kouch2024_CGRaBSvIC} we focused on the long-term, seasonal changes, allowing us to compare the optical behavior to the generally longer term radio behavior. Second, here the spatiotemporal tests are only applied to confidently variable sources, while in  \citetalias{Kouch2024_CGRaBSvIC} we did not distinguish between variable or non-variable sources. Although the effect of variability in practice should be comparable in both studies, because non-variable sources had minimal effect on the tests of \citetalias{Kouch2024_CGRaBSvIC} due to their lack of major flares. Third, the more rapid temporal metrics of this study are more strongly affected by the presence of seasonal gaps (see Sect. \ref{sec_results_observational_gaps}) than the longer-term metrics of \citetalias{Kouch2024_CGRaBSvIC}. As a result, there are several spatiotemporal associations in \citetalias{Kouch2024_CGRaBSvIC} which are identified only as spatial associations in this study.

For example, the most notable \citetalias{Kouch2024_CGRaBSvIC} association missing here is of the LBL CAZJ0502+1338 (4FGL~J0502.5+1340) with IC151114A ($W_\mathrm{T}$=$0.957$). Notably, this blazar has a second neutrino association (IC190712A; $W_\mathrm{T}$=$0.041$) and was associated with a neutrino hotspot (\citealt{Buson2023}). As seen in Fig.~7 of \citetalias{Kouch2024_CGRaBSvIC}, IC151114A arrived at a historically high radio state and a relatively high optical state. However, around this time, the optical brightness is not high enough to be a BB95 as it only reaches the 93th percentile, and there is not enough cadence to identify a potential BBHOP flare. Likewise, the second neutrino (IC190712A) arrived close to an optical flare but in a seasonal gap. Therefore, even though CAZJ0502+1338 is not identified to be spatiotemporally neutrino-associated in this study, it is still a promising neutrino-emitting candidate.

Despite such individual differences, one can still broadly compare the correlation strengths of this study to those of \citetalias{Kouch2024_CGRaBSvIC}. In the most comparable scenario, here we have $p=0.1246$ (for all RFC$^\dagger$ via BB95) while we had $p \approx 0.04$ in \citetalias{Kouch2024_CGRaBSvIC} (for optical-only, counted TS in the enlarged scenario). The decrease in correlation strength is interesting because the radio-selected blazar sample of \citetalias{Kouch2024_CGRaBSvIC} (i.e., CGRaBS) contained only 1157 sources while the current radio-selected sample (i.e., RFC$^\dagger$) has 3225 sources (nearly three times as many sources and nearly two times as many optically variable sources). If a spatiotemporal correlation exists between neutrinos and the whole blazar population, increasing the size of the blazar sample should strengthen the correlation (e.g., \citealt{Liodakis2022_WH}). Thus, this decrease in the correlation strength is evidence against a general spatiotemporal connection between blazars and neutrinos, at least in the optical band. Crucially, this does not exclude a potential spatiotemporal connection in other bands than optical. The challenge is that in other bands it is not possible to check the effect of increasing the number of blazar light curves as done for optical here. For example, in \citetalias{Hovatta2021} and \citetalias{Kouch2024_CGRaBSvIC}, we already tested the largest possible number of long-term light curves currently available in the radio band. Nonetheless, the lack of a population-based blazar-neutrino correlation suggested by our optical analyses is in agreement with recent population studies (e.g., \citealt{Abbasi2025_mm_blz_vs_neut}) and generally inline with the estimate that only a small fraction IceCube neutrinos originate from blazars (e.g., \citealt{Oikonomou2022_frac_of_neutrinos_by_blazars, Yoshida2023_low_blz_contr_to_neut, Robinson2024_low_blz_contr_to_neut}; also see Sect. \ref{sec_results_frac_of_neutrinos_coming_from_blazars}).

\begin{figure*}
    \centering
    \includegraphics[width=18cm]{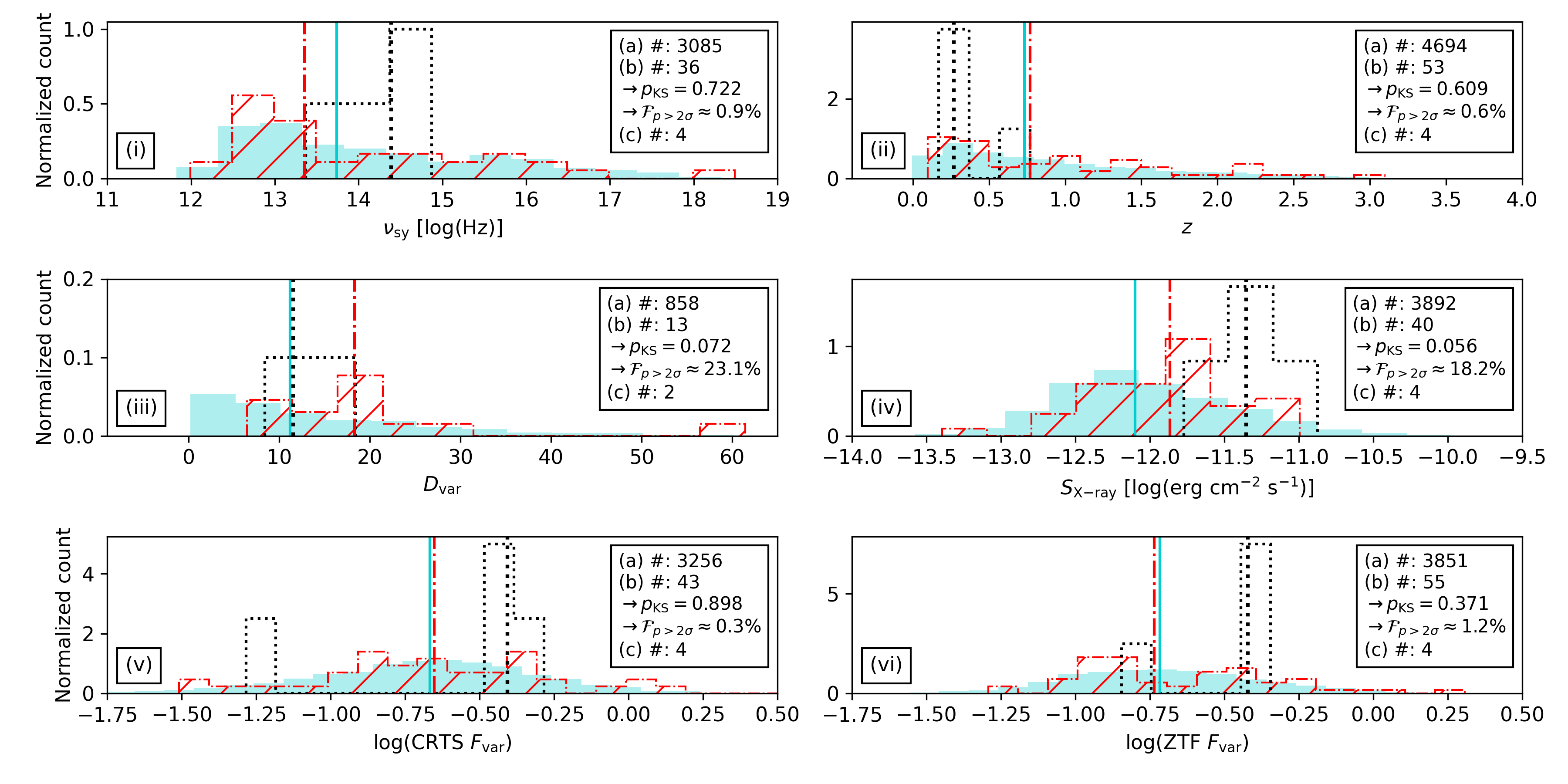}
    \caption{Distribution of $\nu_\mathrm{sy}$ (subplot i), $z$ (subplot ii), $D_\mathrm{var}$ (subplot iii), median $S_\text{X-ray}$ (subplot iv), CRTS $F_\mathrm{var}$ (subplot v), and ZTF $F_\mathrm{var}$ (subplot vi) of: (a) all blazars of the RFC$^\dagger$ and 4LAC samples (with blue, solid lines); (b) blazars spatially associated with neutrinos having $W_\mathrm{T}>0.5$ (with red, dash-dotted lines); and (c) blazars spatiotemporally associated with neutrinos via BB95 having $W_\mathrm{T}>0.5$ (with black, dotted lines). ``$p_\mathrm{KS}$'' gives the $p$-value of the KS test, and ``$\mathcal{F}_{p>2\sigma}$'' the fraction of resampled KS $p$-values reaching the 2$\sigma$ threshold. ``\#'' denotes the size of each sample.}
    \label{fig_CAZ_basic_dist}
\end{figure*}

\subsection{Physical properties of the neutrino-associated blazars} \label{sec_results_candidate_blazar_properties}
Here, we investigate if the physical properties of the neutrino-associated blazars are different from the rest of the blazar population. As discussed in Sect. \ref{sec_results_observational_gaps}, we may be missing $\sim$22\% of potential spatiotemporal associations due to seasonal gaps. Therefore, in this section we consider all spatially neutrino-associated blazars.

We selected neutrino-associated blazars on the basis of an evident spatial association with high-weight neutrinos ($W_\mathrm{T}>0.5$; see Sect. \ref{sec_results_individual_assoc} for justification) regardless of their variability. While many of these are expected to be false candidates, at least we ensured that any spatially neutrino-associated blazar that could also potentially be temporally associated is included in the comparison. This approach has the added benefit of allowing for some meaningful statistics due to a larger number of associations, which is not possible when only focusing on the few high-weight spatiotemporal associations. We find that a total of 61 blazars from either the RFC$^\dagger$ or the 4LAC samples are spatially associated with these 79 high-weight events. In terms of classification, 4 (7\% of 61) are AGN, 16 (26\%) unclassified blazars, 16 (26\%) BLLs, and 25 (41\%) FSRQs.

In Fig. \ref{fig_CAZ_basic_dist}, we plot the distribution of four physical properties (synchrotron peak frequency, redshift, radio variability Doppler factor, and median X-ray flux density) and two light curve properties (CRTS and ZTF $F_\mathrm{var}$) of all 5880 blazars of the RFC$^\dagger$ and 4LAC samples (using blue, solid lines; denoted as ``a'') along with spatially neutrino-associated blazars which have $W_\mathrm{T}>0.5$ (using red, dash-dotted lines; denoted as ``b''). To check if these distributions are significantly different, we use the Kolmogorov-Smirnov (KS) and Anderson-Darling (AD) tests. Although, for brevity, we only report the results of the KS test, which leads to more conservative correlation strengths in all scenarios. Since the sample sizes between the (a) and (b) distributions are vastly different (two orders of magnitude), we also apply the KS and AD tests to a randomly selected subsample of the larger (a) which is equal in size to the smaller (b). The resampling process is repeated $10^4$ times. We then calculate the fraction of the resampled KS and AD $p$-values that cross the 2$\sigma$ threshold for each of the six physical quantities. By chance, only 4.6\% of the resampled $p$-values should cross the 2$\sigma$ threshold: a larger fraction would suggest that (a) and (b) are likely different distributions. For comparison, we also plot the respective distributions of the spatiotemporally neutrino-associated blazars via BB95 with $W_\mathrm{T}>0.5$, which were discussed in detail in Sect. \ref{sec_results_individual_assoc}. These are labeled as ``c'' and plotted using black, dotted lines. Unfortunately, the sample sizes of (c) are too small to allow for reliable KS or AD testing (even via resampling).

Figure \ref{fig_CAZ_basic_dist} (i) shows that spatially neutrino-associated blazars have similar average synchrotron peak frequencies with respect to other blazars in the RFC$^\dagger$ and 4LAC samples. Meanwhile, the four high-weight, spatiotemporally neutrino-associated blazars (including CAZJ0211+1051 and CAZJ0509+0541) clearly show an excess around the ISP regime $14 < \log(\nu_\mathrm{sy}) < 15$. Although the most likely neutrino-emitting blazars are IBLs, we do not find any evidence from the distribution of the spatially neutrino-associated blazars that the IBL subclass is more efficient at emitting neutrinos than other blazar subclasses.

As made evident from Fig. \ref{fig_CAZ_basic_dist} (ii), the neutrino-associated blazars seem to have a similar redshift distribution as the RFC$^\dagger$ and 4LAC blazars. This is generally in line with the results of \cite{Azzollini2025_properties_of_PeVatron_blz} who studied the physical properties of 52 blazars which were spatially correlated with IceCube neutrino hotspots. This suggests that blazars at any redshift epoch could be neutrino emitters.

Figure \ref{fig_CAZ_basic_dist} (iii) hints that spatially neutrino-associated blazars have a slightly higher than average radio variability Doppler factor, $D_\mathrm{var}$, than the rest of the RFC$^\dagger$ and 4LAC blazars. Although the significance of the simple KS test is less than 2$\sigma$, that of the simple AD test is $>$2$\sigma$ and the resampled KS test supports this implication (i.e., $\mathcal{F}_{p>2\sigma} = 23.1\% > 4.6\%$) as does the resampled AD test. This is in agreement with the results of \cite{Plavin2025_high_beaming_of_neutrino_assoc_blazars} who found that spatially neutrino-associated blazars have a higher than average Doppler factor with a significance of 2.4$\sigma$. To compare the results more directly, we repeat this test by selecting only gold events with $\Omega < 10~\mathrm{deg}^2$, exactly as done by \cite{Plavin2025_high_beaming_of_neutrino_assoc_blazars}. In this case, the KS test gives 2.2$\sigma$ and $\mathcal{F}_{p>2\sigma}$ reaches 28\%, reaffirming the agreement. Remarkably, this agreement persists even though we have a total of 858 Doppler factors while \cite{Plavin2025_high_beaming_of_neutrino_assoc_blazars} only had 205. Moreover, we have 13 neutrino-associated blazars with a Doppler factor estimate while they had 8. This apparent positive correlation between the Doppler factor and neutrino emission supports the idea that neutrinos are produced in the jet, such that a more highly beamed jet results in a higher neutrino flux (e.g., \citealt{Kovalev2025_PKS1424_eye_of_sauran}). Nevertheless, the two high-weight, spatiotemporally neutrino-associated blazars with Doppler factor estimates are not highly beamed ($D_\mathrm{var}$ of 8.4 and 14.7).

Figure \ref{fig_CAZ_basic_dist} (iv) suggests that the spatially neutrino-associated blazars are on average slightly brighter in X-rays ($0.3 < E < 10$~keV) than the rest of the RFC$^\dagger$ and 4LAC blazars. Although the KS significance is just below 2$\sigma$, that of the AD test is $>$2$\sigma$ and the resampled KS test supports this (i.e., $\mathcal{F}_{p>2\sigma} = 18.2\% > 4.6\%$) as does the resampled AD test. Additionally, the four high-weight, spatiotemporally neutrino-associated blazars clearly have an above-average X-ray flux density. This generally agrees with the results of \cite{Plavin2024} who found a hint of a spatial correlation between hard-X-ray-detected ($\sim$10~keV) blazars and IceCube neutrinos at least at the 2$\sigma$ level. These results would suggest that X-ray-bright blazars are more efficient at producing IceCube neutrinos, which is inline with the theory of neutrino production. This process is most efficient at producing $\sim$PeV neutrinos when the seed photons have $\sim$keV (X-ray) energies in the proton frame (e.g., \citealt{Roulet2021_proton_photon_energies}) from a dense ambient medium (e.g., \citealt{Reimer2019_Xray_photon_field_in_TXS}), the synchrotron emission of the protons themselves (e.g., \citealt{Mastichiadis2021_Xray_flare_in_pp}), or the synchrotron or inverse-Compton emission of cospatial electrons (e.g., \citealt{Cerruti2019_TXS_electron_SSC_as_seed}). Nevertheless, we caution that this result is not obtained when we run more rigorous correlation tests on the whole X-ray-detected and X-ray-complete samples of blazars in our companion paper Paggi et al. (in prep.).

Figure \ref{fig_CAZ_basic_dist} (v) and (vi) show that the optical variability of the neutrino-emitting blazar candidates is similar to the rest of the RFC$^\dagger$ and 4LAC blazars, although the four high-weight spatiotemporally neutrino-associated blazars have either above average CRTS or ZTF $F_\mathrm{var}$. This is due to a selection bias, since we only identify blazar flares in confidently variable light curves, which have critically large CRTS or ZTF $F_\mathrm{var}$ by construction (for more details, see \citetalias{Kouch2025_CAZ_catalog}).

\section{Summary and conclusions} \label{sec_conclusions}
We tested the spatiotemporal correlation of 356 high-energy neutrino events of the updated IceCat1+ sample with 3225 and 3814 blazars of the radio-selected RFC$^\dagger$ and $\gamma$-ray-selected 4LAC samples, respectively, altogether constituting 5880 unique blazars. We used the methodology applied in our previous studies, \citetalias{Hovatta2021} and \citetalias{Kouch2024_CGRaBSvIC}, and optimized in \citetalias{Kouch2025_WH2}. We utilized optical all-sky survey light curves, collected in \citetalias{Kouch2025_CAZ_catalog}, for as many blazars as possible. In short, we calculated the weighted sum of the number of neutrinos whose arrival time is coincident with a flare (defined in \citetalias{Kouch2025_CAZ_catalog}) in the light curve of a spatially associated blazar. The correlation strength was then obtained by comparing this weighted sum to those arising from randomized simulations. Lastly, we also performed a complementary analysis using only spatially neutrino-associated blazars. We summarize our conclusions below:

\begin{itemize}
    \item Despite a substantial increase in data, the number of confident spatiotemporal associations still remains relatively low, resulting in generally weak correlation strengths. However, the number of spatiotemporal associations could be underestimated by $\sim$22\% due to the prevalence of seasonal gaps in the CAZ light curves.
    \item The spatiotemporal association of the IBLs CAZ J0211+1051 and CAZJ0509+0541 with the two reliably reconstructed neutrino events IC131014A and IC170922A, respectively, dominate our correlation strengths. As a result, in only one of the 18 tests, the post-trial spatiotemporal correlation strength exceeds 2$\sigma$ (i.e., 2.61$\sigma$ between neutrinos and BB95 at the peak of a BBHOP flare in RFC$^\dagger$ BLLs).
    \item The strongest spatiotemporal correlations are obtained at the time of the peak of the most prominent optical flares.
    \item We estimated an upper limit of $\lesssim$8\% for the fraction of the updated IceCat1+ cosmic neutrinos emitted during major optical flares from blazars, which is in agreement with previous estimates.
    \item Spatially neutrino-associated blazars have a higher than average Doppler factor, hinting that the jet is the origin of those neutrinos that come from blazars.
    \item Spatially neutrino-associated blazars have a slightly higher than average X-ray brightness, in line with the need for X-ray seed photons to produce neutrinos in the detectable range of IceCube. However, more rigorous tests done on X-ray-selected blazars in our companion paper (Paggi et al., in prep.) do not support this.
    \item Spatially neutrino-associated blazars seem to have similar synchrotron peak frequency, redshift, and optical variability distributions to those seen for the general blazar population.
\end{itemize}

This study presents a rigorous test of the phenomenological hypothesis that major optical flares of radio- and $\gamma$-ray-bright blazars trace periods of increased neutrino emission. The results suggest that the contribution of such major optical flares is very minimal at best. Nevertheless, we emphasize that due to the complexity of the blazar variability patterns, we cannot extend this conclusion to other frequencies than optical.

In the future, we aim to test the spatiotemporal correlation of the CAZ blazars against the upcoming IceCat-2 neutrinos (e.g., \citealt{Zegarelli2025_IceCat2}). IceCat-2 is expected to contain more events and, crucially, provide substantially improved spatial error regions as compared to IceCat-1. Notably, our spatiotemporal methodology, in combination with multiwavelength spectro-polarimetric observations, can be applied to data from the next-generation neutrino and electromagnetic facilities to elucidate the blazar-neutrino mystery (e.g., by testing predictions made for time delays between neutrino emission and electromagnetic flares; \citealt{Podlesnyi2025}). This is especially true for the southern sky owing to such upcoming facilities as KM3NeT (e.g., \citealt{KM3NeT2016_next_gen}), Cherenkov Telescope Array Observatory (CTAO; \citealt{CTA2019_sci_book}), \textit{Vera C. Rubin} Observatory's Legacy Survey of Space and Time (LSST; e.g., \citealt{Ivezic2019_LSST_sci_cases}), Square Kilometre Array Observatory (SKAO; e.g., \citealt{Agudo2015_SKA_AGN_jets}), and the African Milimetre Telescope (AMT; \citealt{Backes2016_AMT}).

\section*{Data availability}
Two tables, ``\texttt{neutrino.dat}'' and ``\texttt{assoc.dat}'', are available at the CDS via \href{https://cdsarc.cds.unistra.fr/viz-bin/cat/J/A+A/708/A383}{https://cdsarc.cds.unistra.fr/viz-bin/cat/J/A+A/708/A383}. The table ``\texttt{neutrino.dat}'' contains the full list of the updated IceCat1+ neutrinos. The table ``\texttt{assoc.dat}'' contains the full list of spatial and spatiotemporal blazar-neutrino associations. We note that Table \ref{table_assoc} is a subset of ``\texttt{assoc.dat}'' in its entirety.

\begin{acknowledgements}
We thank the anonymous referee for constructive comments.
PK was supported by Academy of Finland projects 346071 and 345899.
EL was supported by Academy of Finland projects 317636, 320045, and 346071.
TH acknowledges support from the Academy of Finland projects 317383, 320085, 345899, and 362571 and from the European Union ERC-2024-COG - PARTICLES - 101169986. 
KK acknowledges support from the European Research Council (ERC) under the European Union's Horizon 2020 research and innovation programme (grant agreement No. 101002352).
IL was funded by the European Union ERC-2022-STG - BOOTES - 101076343. Views and opinions expressed are however those of the author(s) only and do not necessarily reflect those of the European Union or the European Research Council Executive Agency. Neither the European Union nor the granting authority can be held responsible for them.
Based on observations obtained with the Samuel Oschin Telescope 48-inch and the 60-inch Telescope at the Palomar Observatory as part of the \textit{Zwicky} Transient Facility project. ZTF is supported by the National Science Foundation under Grant No. AST-2034437 and a collaboration including Caltech, IPAC, the Weizmann Institute for Science, the Oskar Klein Center at Stockholm University, the University of Maryland, Deutsches Elektronen-Synchrotron and Humboldt University, the TANGO Consortium of Taiwan, the University of Wisconsin at Milwaukee, Trinity College Dublin, Lawrence Livermore National Laboratories, and IN2P3, France. Operations are conducted by COO, IPAC, and UW. The ZTF forced-photometry service was funded under the Heising-Simons Foundation grant \#12540303 (PI: M.J.Graham).
This work used data from the Asteroid Terrestrial-impact Last Alert System (ATLAS) project. The ATLAS project is primarily funded to search for near earth asteroids through NASA grants NN12AR55G, 80NSSC18K0284, and 80NSSC18K1575; byproducts of the NEO search include images and catalogs from the survey area. This work was partially funded by Kepler/K2 grant J1944/80NSSC19K0112 and HST GO-15889, and STFC grants ST/T000198/1 and ST/S006109/1. The ATLAS science products have been made possible through the contributions of the University of Hawaii Institute for Astronomy, the Queen's University Belfast, the Space Telescope Science Institute, the South African Astronomical Observatory, and The Millennium Institute of Astrophysics (MAS), Chile.
This work has made use of data from the Joan Or\'{o} Telescope (TJO) of the Montsec Observatory (OdM), which is owned by the Catalan Government and operated by the Institute for Space Studies of Catalonia (IEEC).
This work makes use of Matplotlib (\citealt{Hunter2007_Matplotlib}), NumPy (\citealt{Harris2020_NumPy}), SciPy (\citealt{Virtanen2020_SciPy}), and Astropy (\citealt{Astropy2022_v5}).
\end{acknowledgements}

%
%

\bibliographystyle{aa} 
\bibliography{ref.bib} 

\begin{thebibliography}{137}
\expandafter\ifx\csname natexlab\endcsname\relax\def\natexlab#1{#1}\fi

\bibitem[{{Aartsen} {et~al.}(2013){Aartsen}, {Abbasi}, {Abdou}, {Ackermann},
  {Adams}, {Aguilar}, {Ahlers}, {Altmann}, {Auffenberg}, {Bai}, {Baker},
  {Barwick}, {Baum}, {Bay}, {Beatty}, {Bechet}, {Becker Tjus}, {Becker},
  {Bell}, {Benabderrahmane}, {BenZvi}, {Berdermann}, {Berghaus}, {Berley},
  {Bernardini}, {Bernhard}, {Bertrand}, {Besson}, {Binder}, {Bindig}, {Bissok},
  {Blaufuss}, {Blumenthal}, {Boersma}, {Bohaichuk}, {Bohm}, {Bose},
  {B{\"o}ser}, {Botner}, {Brayeur}, {Bretz}, {Brown}, {Bruijn}, {Brunner},
  {Carson}, {Casey}, {Casier}, {Chirkin}, {Christov}, {Christy}, {Clark},
  {Clevermann}, {Coenders}, {Cohen}, {Cowen}, {Cruz Silva}, {Danninger},
  {Daughhetee}, {Davis}, {De Clercq}, {De Ridder}, {Desiati}, {de With},
  {DeYoung}, {D{\'\i}az-V{\'e}lez}, {Dunkman}, {Eagan}, {Eberhardt}, {Eisch},
  {Ellsworth}, {Euler}, {Evenson}, {Fadiran}, {Fazely}, {Fedynitch},
  {Feintzeig}, {Feusels}, {Filimonov}, {Finley}, {Fischer-Wasels}, {Flis},
  {Franckowiak}, {Franke}, {Frantzen}, {Fuchs}, {Gaisser}, {Gallagher},
  {Gerhardt}, {Gladstone}, {Gl{\"u}senkamp}, {Goldschmidt}, {Golup},
  {Gonzalez}, {Goodman}, {G{\'o}ra}, {Grant}, {Gro{\ss}}, {Gurtner}, {Ha}, {Haj
  Ismail}, {Hallen}, {Hallgren}, {Halzen}, {Hanson}, {Heereman}, {Heinen},
  {Helbing}, {Hellauer}, {Hickford}, {Hill}, {Hoffman}, {Hoffmann}, {Homeier},
  {Hoshina}, {Huelsnitz}, {Hulth}, {Hultqvist}, {Hussain}, {Ishihara},
  {Jacobi}, {Jacobsen}, {Jagielski}, {Japaridze}, {Jero}, {Jlelati},
  {Kaminsky}, {Kappes}, {Karg}, {Karle}, {Kelley}, {Kiryluk}, {Kislat},
  {Kl{\"a}s}, {Klein}, {K{\"o}hne}, {Kohnen}, {Kolanoski}, {K{\"o}pke},
  {Kopper}, {Kopper}, {Koskinen}, {Kowalski}, {Krasberg}, {Krings}, {Kroll},
  {Kunnen}, {Kurahashi}, {Kuwabara}, {Labare}, {Landsman}, {Larson},
  {Lesiak-Bzdak}, {Leuermann}, {Leute}, {L{\"u}nemann}, {Madsen}, {Maruyama},
  {Mase}, {Matis}, {McNally}, {Meagher}, {Merck}, {M{\'e}sz{\'a}ros}, {Meures},
  {Miarecki}, {Middell}, {Milke}, {Miller}, {Mohrmann}, {Montaruli}, {Morse},
  {Nahnhauer}, {Naumann}, {Niederhausen}, {Nowicki}, {Nygren}, {Obertacke},
  {Odrowski}, {Olivas}, {Olivo}, {O'Murchadha}, {Paul}, {Pepper}, {P{\'e}rez de
  los Heros}, {Pfendner}, {Pieloth}, {Pinat}, {Pirk}, {Posselt}, {Price},
  {Przybylski}, {R{\"a}del}, {Rameez}, {Rawlins}, {Redl}, {Reimann}, {Resconi},
  {Rhode}, {Ribordy}, {Richman}, {Riedel}, {Rodrigues}, {Rott}, {Ruhe},
  {Ruzybayev}, {Ryckbosch}, {Saba}, {Salameh}, {Sander}, {Santander}, {Sarkar},
  {Schatto}, {Scheel}, {Scheriau}, {Schmidt}, {Schmitz}, {Schoenen},
  {Sch{\"o}neberg}, {Sch{\"o}nwald}, {Schukraft}, {Schulte}, {Schulz},
  {Seckel}, {Sestayo}, {Seunarine}, {Sheremata}, {Smith}, {Soiron}, {Soldin},
  {Spiczak}, {Spiering}, {Stamatikos}, {Stanev}, {Stasik}, {Stezelberger},
  {Stokstad}, {St{\"o}{\ss}l}, {Strahler}, {Str{\"o}m}, {Sullivan}, {Taavola},
  {Taboada}, {Tamburro}, {Ter-Antonyan}, {Te{\v{s}}i{\'c}}, {Tilav}, {Toale},
  {Toscano}, {Usner}, {van der Drift}, {van Eijndhoven}, {Van Overloop}, {van
  Santen}, {Vehring}, {Voge}, {Vraeghe}, {Walck}, {Waldenmaier}, {Wallraff},
  {Wasserman}, {Weaver}, {Wellons}, {Wendt}, {Westerhoff}, {Whitehorn},
  {Wiebe}, {Wiebusch}, {Williams}, {Wissing}, {Wolf}, {Wood}, {Woschnagg},
  {Xu}, {Xu}, {Xu}, {Yanez}, {Yodh}, {Yoshida}, {Zarzhitsky}, {Ziemann},
  {Zierke}, {Zilles}, \& {Zoll}}]{aartsen2013_sys_err}
{Aartsen}, M.~G., {Abbasi}, R., {Abdou}, Y., {et~al.} 2013, \prl, 111, 021103

\bibitem[{{Aartsen} {et~al.}(2017{\natexlab{a}}){Aartsen}, {Abraham},
  {Ackermann}, {Adams}, {Aguilar}, {Ahlers}, {Ahrens}, {Altmann}, {Andeen},
  {Anderson}, {Ansseau}, {Anton}, {Archinger}, {Arguelles}, {Arlen},
  {Auffenberg}, {Axani}, {Bai}, {Barwick}, {Baum}, {Bay}, {Beatty}, {Becker
  Tjus}, {Becker}, {BenZvi}, {Berghaus}, {Berley}, {Bernardini}, {Bernhard},
  {Besson}, {Binder}, {Bindig}, {Bissok}, {Blaufuss}, {Blot}, {Boersma},
  {Bohm}, {B{\"o}rner}, {Bos}, {Bose}, {B{\"o}ser}, {Botner}, {Braun},
  {Brayeur}, {Bretz}, {Burgman}, {Casey}, {Casier}, {Cheung}, {Chirkin},
  {Christov}, {Clark}, {Classen}, {Coenders}, {Collin}, {Conrad}, {Cowen},
  {Cruz Silva}, {Daughhetee}, {Davis}, {Day}, {de Andr{\'e}}, {De Clercq}, {del
  Pino Rosendo}, {Dembinski}, {De Ridder}, {Desiati}, {de Vries}, {de
  Wasseige}, {de With}, {DeYoung}, {D{\'\i}az-V{\'e}lez}, {di Lorenzo},
  {Dujmovic}, {Dumm}, {Dunkman}, {Eberhardt}, {Ehrhardt}, {Eichmann}, {Euler},
  {Evenson}, {Fahey}, {Fazely}, {Feintzeig}, {Felde}, {Filimonov}, {Finley},
  {Flis}, {F{\"o}sig}, {Franckowiak}, {Fuchs}, {Gaisser}, {Gaior}, {Gallagher},
  {Gerhardt}, {Ghorbani}, {Giang}, {Gladstone}, {Glagla}, {Gl{\"u}senkamp},
  {Goldschmidt}, {Golup}, {Gonzalez}, {G{\'o}ra}, {Grant}, {Griffith}, {Haack},
  {Haj Ismail}, {Hallgren}, {Halzen}, {Hansen}, {Hansmann}, {Hansmann},
  {Hanson}, {Hebecker}, {Heereman}, {Helbing}, {Hellauer}, {Hickford},
  {Hignight}, {Hill}, {Hoffman}, {Hoffmann}, {Holzapfel}, {Homeier}, {Hoshina},
  {Huang}, {Huber}, {Huelsnitz}, {Hultqvist}, {In}, {Ishihara}, {Jacobi},
  {Japaridze}, {Jeong}, {Jero}, {Jones}, {Jurkovic}, {Kappes}, {Karg}, {Karle},
  {Katz}, {Kauer}, {Keivani}, {Kelley}, {Kemp}, {Kheirandish}, {Kim},
  {Kintscher}, {Kiryluk}, {Kittler}, {Klein}, {Kohnen}, {Koirala}, {Kolanoski},
  {Konietz}, {K{\"o}pke}, {Kopper}, {Kopper}, {Koskinen}, {Kowalski}, {Krings},
  {Kroll}, {Kr{\"u}ckl}, {Kr{\"u}ger}, {Kunnen}, {Kunwar}, {Kurahashi},
  {Kuwabara}, {Labare}, {Lanfranchi}, {Larson}, {Lennarz}, {Lesiak-Bzdak},
  {Leuermann}, {Leuner}, {Lu}, {L{\"u}nemann}, {Madsen}, {Maggi}, {Mahn},
  {Mancina}, {Mandelartz}, {Maruyama}, {Mase}, {Maunu}, {McNally}, {Meagher},
  {Medici}, {Meier}, {Meli}, {Menne}, {Merino}, {Meures}, {Miarecki},
  {Middell}, {Mohrmann}, {Montaruli}, {Moulai}, \&
  {Nahnhauer}}]{Aartsen2017_IceCube_stacking_blz_subpop}
{Aartsen}, M.~G., {Abraham}, K., {Ackermann}, M., {et~al.} 2017{\natexlab{a}},
  \apj, 835, 45

\bibitem[{{Aartsen} {et~al.}(2017{\natexlab{b}}){Aartsen}, {Abraham},
  {Ackermann}, {Adams}, {Aguilar}, {Ahlers}, {Ahrens}, {Altmann}, {Andeen},
  {Anderson}, {Ansseau}, {Anton}, {Archinger}, {Arguelles}, {Arlen},
  {Auffenberg}, {Axani}, {Bai}, {Barwick}, {Baum}, {Bay}, {Beatty}, {Becker
  Tjus}, {Becker}, {BenZvi}, {Berghaus}, {Berley}, {Bernardini}, {Bernhard},
  {Besson}, {Binder}, {Bindig}, {Bissok}, {Blaufuss}, {Blot}, {Boersma},
  {Bohm}, {B{\"o}rner}, {Bos}, {Bose}, {B{\"o}ser}, {Botner}, {Braun},
  {Brayeur}, {Bretz}, {Burgman}, {Casey}, {Casier}, {Cheung}, {Chirkin},
  {Christov}, {Clark}, {Classen}, {Coenders}, {Collin}, {Conrad}, {Cowen},
  {Cruz Silva}, {Daughhetee}, {Davis}, {Day}, {de Andr{\'e}}, {De Clercq}, {del
  Pino Rosendo}, {Dembinski}, {De Ridder}, {Desiati}, {de Vries}, {de
  Wasseige}, {de With}, {DeYoung}, {D{\'\i}az-V{\'e}lez}, {di Lorenzo},
  {Dujmovic}, {Dumm}, {Dunkman}, {Eberhardt}, {Ehrhardt}, {Eichmann}, {Euler},
  {Evenson}, {Fahey}, {Fazely}, {Feintzeig}, {Felde}, {Filimonov}, {Finley},
  {Flis}, {F{\"o}sig}, {Franckowiak}, {Fuchs}, {Gaisser}, {Gaior}, {Gallagher},
  {Gerhardt}, {Ghorbani}, {Giang}, {Gladstone}, {Glagla}, {Gl{\"u}senkamp},
  {Goldschmidt}, {Golup}, {Gonzalez}, {G{\'o}ra}, {Grant}, {Griffith}, {Haack},
  {Haj Ismail}, {Hallgren}, {Halzen}, {Hansen}, {Hansmann}, {Hansmann},
  {Hanson}, {Hebecker}, {Heereman}, {Helbing}, {Hellauer}, {Hickford},
  {Hignight}, {Hill}, {Hoffman}, {Hoffmann}, {Holzapfel}, {Homeier}, {Hoshina},
  {Huang}, {Huber}, {Huelsnitz}, {Hultqvist}, {In}, {Ishihara}, {Jacobi},
  {Japaridze}, {Jeong}, {Jero}, {Jones}, {Jurkovic}, {Kappes}, {Karg}, {Karle},
  {Katz}, {Kauer}, {Keivani}, {Kelley}, {Kemp}, {Kheirandish}, {Kim},
  {Kintscher}, {Kiryluk}, {Kittler}, {Klein}, {Kohnen}, {Koirala}, {Kolanoski},
  {Konietz}, {K{\"o}pke}, {Kopper}, {Kopper}, {Koskinen}, {Kowalski}, {Krings},
  {Kroll}, {Kr{\"u}ckl}, {Kr{\"u}ger}, {Kunnen}, {Kunwar}, {Kurahashi},
  {Kuwabara}, {Labare}, {Lanfranchi}, {Larson}, {Lennarz}, {Lesiak-Bzdak},
  {Leuermann}, {Leuner}, {Lu}, {L{\"u}nemann}, {Madsen}, {Maggi}, {Mahn},
  {Mancina}, {Mandelartz}, {Maruyama}, {Mase}, {Maunu}, {McNally}, {Meagher},
  {Medici}, {Meier}, {Meli}, {Menne}, {Merino}, {Meures}, {Miarecki},
  {Middell}, {Mohrmann}, {Montaruli}, {Moulai}, \&
  {Nahnhauer}}]{Aartsen2017_blz_maximal_contr}
{Aartsen}, M.~G., {Abraham}, K., {Ackermann}, M., {et~al.} 2017{\natexlab{b}},
  \apj, 835, 45

\bibitem[{{Aartsen} {et~al.}(2020){Aartsen}, {Ackermann}, {Adams}, {Aguilar},
  {Ahlers}, {Ahrens}, {Alispach}, {Andeen}, {Anderson}, {Ansseau}, {Anton},
  {Arg{\"u}elles}, {Auffenberg}, {Axani}, {Backes}, {Bagherpour}, {Bai},
  {Balagopal}, {Barbano}, {Barwick}, {Bastian}, {Baum}, {Baur}, {Bay},
  {Beatty}, {Becker}, {Becker Tjus}, {BenZvi}, {Berley}, {Bernardini},
  {Besson}, {Binder}, {Bindig}, {Blaufuss}, {Blot}, {Bohm}, {B{\"o}rner},
  {B{\"o}ser}, {Botner}, {B{\"o}ttcher}, {Bourbeau}, {Bourbeau}, {Bradascio},
  {Braun}, {Bron}, {Brostean-Kaiser}, {Burgman}, {Buscher}, {Busse}, {Carver},
  {Chen}, {Cheung}, {Chirkin}, {Choi}, {Clark}, {Classen}, {Coleman}, {Collin},
  {Conrad}, {Coppin}, {Correa}, {Cowen}, {Cross}, {Dave}, {De Clercq},
  {DeLaunay}, {Dembinski}, {Deoskar}, {De Ridder}, {Desiati}, {de Vries}, {de
  Wasseige}, {de With}, {DeYoung}, {Diaz}, {D{\'\i}az-V{\'e}lez}, {Dujmovic},
  {Dunkman}, {Dvorak}, {Eberhardt}, {Ehrhardt}, {Eller}, {Engel}, {Evenson},
  {Fahey}, {Fazely}, {Felde}, {Filimonov}, {Finley}, {Fox}, {Franckowiak},
  {Friedman}, {Fritz}, {Gaisser}, {Gallagher}, {Ganster}, {Garrappa},
  {Gerhardt}, {Ghorbani}, {Glauch}, {Gl{\"u}senkamp}, {Goldschmidt},
  {Gonzalez}, {Grant}, {Griffith}, {Griswold}, {G{\"u}nder}, {G{\"u}nd{\"u}z},
  {Haack}, {Hallgren}, {Halliday}, {Halve}, {Halzen}, {Hanson}, {Haungs},
  {Hebecker}, {Heereman}, {Heix}, {Helbing}, {Hellauer}, {Henningsen},
  {Hickford}, {Hignight}, {Hill}, {Hoffman}, {Hoffmann}, {Hoinka},
  {Hokanson-Fasig}, {Hoshina}, {Huang}, {Huber}, {Huber}, {Hultqvist},
  {H{\"u}nnefeld}, {Hussain}, {In}, {Iovine}, {Ishihara}, {Japaridze}, {Jeong},
  {Jero}, {Jones}, {Jonske}, {Joppe}, {Kang}, {Kang}, {Kappes}, {Kappesser},
  {Karg}, {Karl}, {Karle}, {Katz}, {Kauer}, {Kelley}, {Kheirandish}, {Kim},
  {Kintscher}, {Kiryluk}, {Kittler}, {Klein}, {Koirala}, {Kolanoski},
  {K{\"o}pke}, {Kopper}, {Kopper}, {Koskinen}, {Kowalski}, {Krings},
  {Kr{\"u}ckl}, {Kulacz}, {Kurahashi}, {Kyriacou}, {Labare}, {Lanfranchi},
  {Larson}, {Lauber}, {Lazar}, {Leonard}, {Leszczy{\'n}ska}, {Leuermann},
  {Liu}, {Lohfink}, {Lozano Mariscal}, {Lu}, {Lucarelli}, {L{\"u}nemann},
  {Luszczak}, {Lyu}, {Ma}, {Madsen}, {Maggi}, {Mahn}, {Makino}, {Mallik},
  {Mallot}, {Mancina}, {Mari{\c{s}}}, {Maruyama}, {Mase}, {Matis}, {Maunu},
  {McNally}, {Meagher}, {Medici}, {Medina}, {Meier}, {Meighen-Berger}, {Menne},
  {Merino}, {Meures}, {Micallef}, {Mockler}, {Moment{\'e}}, {Montaruli},
  {Moore}, {Morse}, {Moulai}, {Muth}, {Nagai}, {Naumann}, {Neer},
  {Niederhausen}, {Nisa}, {Nowicki}, {Nygren}, {Obertacke Pollmann}, {Oehler},
  {Olivas}, {O'Murchadha}, {O'Sullivan}, {Palczewski}, {Pandya}, {Pankova},
  {Park}, {Peiffer}, {P{\'e}rez de los Heros}, {Philippen}, {Pieloth}, {Pinat},
  {Pizzuto}, {Plum}, {Porcelli}, {Price}, {Przybylski}, {Raab}, {Raissi},
  {Rameez}, {Rauch}, {Rawlins}, {Rea}, {Reimann}, {Relethford}, {Renschler},
  {Renzi}, {Resconi}, {Rhode}, {Richman}, {Robertson}, {Rongen}, {Rott},
  {Ruhe}, {Ryckbosch}, {Rysewyk}, {Safa}, {Sanchez Herrera}, {Sandrock},
  {Sandroos}, {Santander}, {Sarkar}, {Sarkar}, {Satalecka}, {Schaufel},
  {Schieler}, {Schlunder}, {Schmidt}, {Schneider}, {Schneider}, {Schr{\"o}der},
  {Schumacher}, {Sclafani}, {Seckel}, {Seunarine}, {Shefali}, {Silva},
  {Snihur}, {Soedingrekso}, {Soldin}, {Song}, {Spiczak}, {Spiering},
  {Stachurska}, {Stamatikos}, {Stanev}, {Stein}, {Steinm{\"u}ller}, {Stettner},
  {Steuer}, {Stezelberger}, {Stokstad}, {St{\"o}{\ss}l}, {Strotjohann},
  {St{\"u}rwald}, {Stuttard}, {Sullivan}, {Taboada}, {Tenholt}, {Ter-Antonyan},
  {Terliuk}, {Tilav}, {Tollefson}, {Tomankova}, {T{\"o}nnis}, {Toscano},
  {Tosi}, {Trettin}, {Tselengidou}, {Tung}, {Turcati}, {Turcotte}, {Turley},
  {Ty}, {Unger}, {Unland Elorrieta}, {Usner}, {Vandenbroucke}, {Van Driessche},
  {van Eijk}, {van Eijndhoven}, {Vanheule}, {van Santen}, {Vraeghe}, {Walck},
  {Wallace}, {Wallraff}, {Wandkowsky}, {Watson}, {Weaver}, {Weindl}, {Weiss},
  {Weldert}, {Wendt}, {Werthebach}, {Whelan}, {Whitehorn}, {Wiebe}, {Wiebusch},
  {Wille}, {Williams}, {Wills}, {Wolf}, {Wood}, {Wood}, {Woschnagg}, {Wrede},
  {Xu}, {Xu}, {Xu}, {Yanez}, {Yodh}, {Yoshida}, {Yuan}, \&
  {Z{\"o}cklein}}]{Aartsen2020}
{Aartsen}, M.~G., {Ackermann}, M., {Adams}, J., {et~al.} 2020, \prl, 124,
  051103

\bibitem[{{Abbasi} {et~al.}(2023{\natexlab{a}}){Abbasi}, {Ackermann}, {Adams},
  {Agarwalla}, {Aguilar}, {Ahlers}, {Alameddine}, {Amin}, {Andeen}, {Anton},
  {Arg{\"u}elles}, {Ashida}, {Athanasiadou}, {Axani}, {Bai}, {Balagopal},
  {Baricevic}, {Barwick}, {Basu}, {Bay}, {Beatty}, {Becker}, {Becker Tjus},
  {Beise}, {Bellenghi}, {Benning}, {BenZvi}, {Berley}, {Bernardini}, {Besson},
  {Binder}, {Blaufuss}, {Blot}, {Bontempo}, {Book}, {Meneguolo}, {B{\"o}ser},
  {Botner}, {B{\"o}ttcher}, {Bourbeau}, {Braun}, {Brinson}, {Brostean-Kaiser},
  {Burley}, {Busse}, {Butterfield}, {Campana}, {Carloni}, {Carnie-Bronca},
  {Chattopadhyay}, {Chau}, {Chen}, {Chen}, {Chirkin}, {Choi}, {Clark},
  {Classen}, {Coleman}, {Collin}, {Connolly}, {Conrad}, {Coppin}, {Correa},
  {Countryman}, {Cowen}, {Dave}, {De Clercq}, {DeLaunay}, {Delgado},
  {Dembinski}, {Deng}, {Deoskar}, {Desai}, {Desiati}, {de Vries}, {de
  Wasseige}, {DeYoung}, {Diaz}, {D{\'\i}az-V{\'e}lez}, {Dittmer}, {Domi},
  {Dujmovic}, {DuVernois}, {Ehrhardt}, {Eller}, {El Mentawi}, {Engel},
  {Erpenbeck}, {Evans}, {Evenson}, {Fan}, {Fang}, {Farrag}, {Fazely},
  {Fedynitch}, {Feigl}, {Fiedlschuster}, {Finley}, {Fischer}, {Fox},
  {Franckowiak}, {Friedman}, {Fritz}, {F{\"u}rst}, {Gaisser}, {Gallagher},
  {Ganster}, {Garcia}, {Gerhardt}, {Ghadimi}, {Glaser}, {Glauch},
  {Gl{\"u}senkamp}, {Goehlke}, {Gonzalez}, {Goswami}, {Grant}, {Gray}, {Gries},
  {Griffin}, {Griswold}, {G{\"u}nther}, {Gutjahr}, {Haack}, {Hallgren},
  {Halliday}, {Halve}, {Halzen}, {Hamdaoui}, {Minh}, {Hanson}, {Hardin},
  {Harnisch}, {Hatch}, {Haungs}, {Helbing}, {Hellrung}, {Henningsen},
  {Heuermann}, {Heyer}, {Hickford}, {Hidvegi}, {Hill}, {Hill}, {Hoffman},
  {Hori}, {Hoshina}, {Hou}, {Huber}, {Hultqvist}, {H{\"u}nnefeld}, {Hussain},
  {Hymon}, {In}, {Ishihara}, {Jacquart}, {Janik}, {Jansson}, {Japaridze},
  {Jayakumar}, {Jeong}, {Jin}, {Jones}, {Kang}, {Kang}, {Kang}, {Kappes},
  {Kappesser}, {Kardum}, {Karg}, {Karl}, {Karle}, {Katz}, {Kauer}, {Kelley},
  {Zathul}, {Kheirandish}, {Kiryluk}, {Klein}, {Kochocki}, {Koirala},
  {Kolanoski}, {Kontrimas}, {K{\"o}pke}, {Kopper}, {Koskinen}, {Koundal},
  {Kovacevich}, {Kowalski}, {Kozynets}, {Kruiswijk}, {Krupczak}, {Kumar},
  {Kun}, {Kurahashi}, {Lad}, {Lagunas Gualda}, {Lamoureux}, {Larson},
  {Latseva}, {Lauber}, {Lazar}, {Lee}, {Leonard DeHolton}, {Leszczy{\'n}ska},
  {Lincetto}, {Liu}, {Liubarska}, {Lohfink}, {Love}, {Mariscal}, {Lu},
  {Lucarelli}, {Ludwig}, {Luszczak}, {Lyu}, {Madsen}, {Mahn}, {Makino},
  {Manao}, {Mancina}, {Sainte}, {Mari{\c{s}}}, {Marka}, {Marka}, {Marsee},
  {Martinez-Soler}, {Maruyama}, {Mayhew}, {McElroy}, {McNally}, {Mead},
  {Meagher}, {Mechbal}, {Medina}, {Meier}, {Merckx}, {Merten}, {Micallef},
  {Montaruli}, {Moore}, {Morii}, {Morse}, {Moulai}, {Mukherjee}, {Naab},
  {Nagai}, {Nakos}, {Naumann}, {Necker}, {Neumann}, {Niederhausen}, {Nisa},
  {Noell}, {Nowicki}, {Obertacke Pollmann}, {O'Dell}, {Oehler}, {Oeyen},
  {Olivas}, {Orsoe}, {Osborn}, {O'Sullivan}, {Pandya}, {Park}, {Parker},
  {Paudel}, {Paul}, {P{\'e}rez de los Heros}, {Peterson}, {Philippen},
  {Pieper}, {Pizzuto}, {Plum}, {Pont{\'e}n}, {Popovych}, {Prado Rodriguez},
  {Pries}, {Procter-Murphy}, {Przybylski}, {Rack-Helleis}, {Rawlins}, {Rechav},
  {Rehman}, {Reichherzer}, {Renzi}, {Resconi}, {Reusch}, {Rhode}, {Richman},
  {Riedel}, {Rifaie}, {Roberts}, {Robertson}, {Rodan}, {Roellinghoff},
  {Rongen}, {Rott}, {Ruhe}, {Ruohan}, {Ryckbosch}, {Safa}, {Saffer},
  {Salazar-Gallegos}, {Sampathkumar}, {Sanchez Herrera}, {Sandrock},
  {Santander}, {Sarkar}, {Sarkar}, {Savelberg}, {Savina}, {Schaufel},
  {Schieler}, {Schindler}, {Schlickmann}, {Schl{\"u}ter}, {Schl{\"u}ter},
  {Schmidt}, {Schneider}, {Schr{\"o}der}, {Schumacher}, {Schwefer}, {Sclafani},
  {Seckel}, {Seikh}, {Seunarine}, {Shah}, {Sharma}, {Shefali}, {Shimizu},
  {Silva}, {Skrzypek}, {Smithers}, {Snihur}, {Soedingrekso}, {S{\o}gaard},
  {Soldin}, {Soldin}, {Sommani}, {Spannfellner}, {Spiczak}, {Spiering},
  {Stamatikos}, {Stanev}, {Stezelberger}, {St{\"u}rwald}, {Stuttard},
  {Sullivan}, {Taboada}, {Ter-Antonyan}, {Thiesmeyer}, {Thompson}, {Thwaites},
  {Tilav}, {Tollefson}, {T{\"o}nnis}, {Toscano}, {Tosi}, {Trettin}, {Tung},
  {Turcotte}, {Twagirayezu}, {Ty}, {Unland Elorrieta}, {Upadhyay}, {Upshaw},
  {Valtonen-Mattila}, {Vandenbroucke}, {van Eijndhoven}, {Vannerom}, {van
  Santen}, {Vara}, {Veitch-Michaelis}, {Venugopal}, {Vereecken}, {Verpoest},
  {Veske}, {Walck}, {Watson}, {Weaver}, {Weigel}, {Weindl}, {Weldert}, {Wendt},
  {Werthebach}, {Weyrauch}, {Whitehorn}, {Wiebusch}, {Willey}, {Williams},
  {Wolf}, {Wolf}, {Wrede}, {Xu}, {Yanez}, {Yildizci}, {Yoshida}, {Young}, {Yu},
  {Yu}, {Yuan}, {Zhang}, {Zhelnin}, \& {IceCube
  Collaboration}}]{abbasi2023_plavin_test}
{Abbasi}, R., {Ackermann}, M., {Adams}, J., {et~al.} 2023{\natexlab{a}}, \apj,
  954, 75

\bibitem[{{Abbasi} {et~al.}(2023{\natexlab{b}}){Abbasi}, {Ackermann}, {Adams},
  {Agarwalla}, {Aguilar}, {Ahlers}, {Alameddine}, {Amin}, {Andeen}, {Anton},
  {Arg{\"u}elles}, {Ashida}, {Athanasiadou}, {Axani}, {Bai}, {Balagopal},
  {Baricevic}, {Barwick}, {Basu}, {Bay}, {Beatty}, {Becker}, {Becker Tjus},
  {Beise}, {Bellenghi}, {BenZvi}, {Berley}, {Bernardini}, {Besson}, {Binder},
  {Bindig}, {Blaufuss}, {Blot}, {Bontempo}, {Book}, {Boscolo Meneguolo},
  {B{\"o}ser}, {Botner}, {B{\"o}ttcher}, {Bourbeau}, {Braun}, {Brinson},
  {Brostean-Kaiser}, {Burley}, {Busse}, {Butterfield}, {Campana}, {Carloni},
  {Carnie-Bronca}, {Chattopadhyay}, {Chau}, {Chen}, {Chen}, {Chirkin}, {Choi},
  {Clark}, {Classen}, {Coleman}, {Collin}, {Connolly}, {Conrad}, {Coppin},
  {Correa}, {Countryman}, {Cowen}, {Dave}, {De Clercq}, {DeLaunay}, {Delgado},
  {Dembinski}, {Deng}, {Deoskar}, {Desai}, {Desiati}, {de Vries}, {de
  Wasseige}, {DeYoung}, {Diaz}, {D{\'\i}az-V{\'e}lez}, {Dittmer}, {Domi},
  {Dujmovic}, {DuVernois}, {Ehrhardt}, {Eller}, {Engel}, {Erpenbeck}, {Evans},
  {Evenson}, {Fan}, {Fang}, {Farrag}, {Fazely}, {Fedynitch}, {Feigl},
  {Fiedlschuster}, {Finley}, {Fischer}, {Fox}, {Franckowiak}, {Friedman},
  {Fritz}, {F{\"u}rst}, {Gaisser}, {Gallagher}, {Ganster}, {Garcia},
  {Gerhardt}, {Ghadimi}, {Glaser}, {Glauch}, {Gl{\"u}senkamp}, {Goehlke},
  {Gonzalez}, {Goswami}, {Grant}, {Gray}, {Griffin}, {Griswold}, {G{\"u}nther},
  {Gutjahr}, {Haack}, {Hallgren}, {Halliday}, {Halve}, {Halzen}, {Hamdaoui},
  {Ha Minh}, {Hanson}, {Hardin}, {Harnisch}, {Hatch}, {Haungs}, {Helbing},
  {Hellrung}, {Henningsen}, {Heuermann}, {Heyer}, {Hickford}, {Hidvegi},
  {Hill}, {Hill}, {Hoffman}, {Hoshina}, {Hou}, {Huber}, {Hultqvist},
  {H{\"u}nnefeld}, {Hussain}, {Hymon}, {In}, {Ishihara}, {Jacquart}, {Janik},
  {Jansson}, {Japaridze}, {Jayakumar}, {Jeong}, {Jin}, {Jones}, {Kang}, {Kang},
  {Kang}, {Kappes}, {Kappesser}, {Kardum}, {Karg}, {Karl}, {Karle}, {Katz},
  {Kauer}, {Kelley}, {Zathul}, {Kheirandish}, {Kiryluk}, {Klein}, {Kochocki},
  {Koirala}, {Kolanoski}, {Kontrimas}, {K{\"o}pke}, {Kopper}, {Koskinen},
  {Koundal}, {Kovacevich}, {Kowalski}, {Kozynets}, {Kruiswijk}, {Krupczak},
  {Kumar}, {Kun}, {Kurahashi}, {Lad}, {Lagunas Gualda}, {Lamoureux}, {Larson},
  {Lauber}, {Lazar}, {Lee}, {Leonard DeHolton}, {Leszczy{\'n}ska}, {Lincetto},
  {Liu}, {Liubarska}, {Lohfink}, {Love}, {Lozano Mariscal}, {Lu}, {Lucarelli},
  {Ludwig}, {Luszczak}, {Lyu}, {Madsen}, {Mahn}, {Makino}, {Manao}, {Mancina},
  {Marie Sainte}, {Mari{\c{s}}}, {Marka}, {Marka}, {Marsee}, {Martinez-Soler},
  {Maruyama}, {Mayhew}, {McElroy}, {McNally}, {Mead}, {Meagher}, {Mechbal},
  {Medina}, {Meier}, {Merckx}, {Merten}, {Micallef}, {Montaruli}, {Moore},
  {Morii}, {Morse}, {Moulai}, {Mukherjee}, {Naab}, {Nagai}, {Nakos}, {Naumann},
  {Necker}, {Neumann}, {Niederhausen}, {Nisa}, {Noell}, {Nowicki}, {Obertacke
  Pollmann}, {O'Dell}, {Oehler}, {Oeyen}, {Olivas}, {Orsoe}, {Osborn},
  {O'Sullivan}, {Pandya}, {Park}, {Parker}, {Paudel}, {Paul}, {P{\'e}rez de los
  Heros}, {Peterson}, {Philippen}, {Pieper}, {Pizzuto}, {Plum}, {Pont{\'e}n},
  {Popovych}, {Prado Rodriguez}, {Pries}, {Procter-Murphy}, {Przybylski},
  {Rack-Helleis}, {Rawlins}, {Rechav}, {Rehman}, {Reichherzer}, {Renzi},
  {Resconi}, {Reusch}, {Rhode}, {Richman}, {Riedel}, {Roberts}, {Robertson},
  {Rodan}, {Roellinghoff}, {Rongen}, {Rott}, {Ruhe}, {Ruohan}, {Ryckbosch},
  {Safa}, {Saffer}, {Salazar-Gallegos}, {Sampathkumar}, {Sanchez Herrera},
  {Sandrock}, {Santander}, {Sarkar}, {Sarkar}, {Savelberg}, {Savina},
  {Schaufel}, {Schieler}, {Schindler}, {Schl{\"u}ter}, {Schl{\"u}ter},
  {Schmidt}, {Schneider}, {Schr{\"o}der}, {Schumacher}, {Schwefer}, {Sclafani},
  {Seckel}, {Seunarine}, {Shah}, {Sharma}, {Shefali}, {Shimizu}, {Silva},
  {Skrzypek}, {Smithers}, {Snihur}, {Soedingrekso}, {S{\o}gaard}, {Soldin},
  {Sommani}, {Spannfellner}, {Spiczak}, {Spiering}, {Stamatikos}, {Stanev},
  {Stezelberger}, {St{\"u}rwald}, {Stuttard}, {Sullivan}, {Taboada},
  {Ter-Antonyan}, {Thiesmeyer}, {Thompson}, {Thwaites}, {Tilav}, {Tollefson},
  {T{\"o}nnis}, {Toscano}, {Tosi}, {Trettin}, {Tung}, {Turcotte},
  {Twagirayezu}, {Ty}, {Unland Elorrieta}, {Upadhyay}, {Upshaw},
  {Valtonen-Mattila}, {Vandenbroucke}, {van Eijndhoven}, {Vannerom}, {van
  Santen}, {Vara}, {Veitch-Michaelis}, {Venugopal}, {Verpoest}, {Veske},
  {Walck}, {Watson}, {Weaver}, {Weigel}, {Weindl}, {Weldert}, {Wendt},
  {Werthebach}, {Weyrauch}, {Whitehorn}, {Wiebusch}, {Willey}, {Williams},
  {Wolf}, {Wolf}, {Wrede}, {Xu}, {Yanez}, {Yildizci}, {Yoshida}, {Yu}, {Yu},
  {Yuan}, {Zhang}, \& {Zhelnin}}]{Abbasi2023_IceCat1}
{Abbasi}, R., {Ackermann}, M., {Adams}, J., {et~al.} 2023{\natexlab{b}}, \apjs,
  269, 25

\bibitem[{{Abbasi} {et~al.}(2024{\natexlab{a}}){Abbasi}, {Ackermann}, {Adams},
  {Agarwalla}, {Aguilar}, {Ahlers}, {Alameddine}, {Amin}, {Andeen}, {Anton},
  {Arg{\"u}elles}, {Ashida}, {Athanasiadou}, {Axani}, {Bai}, {Balagopal V.},
  {Baricevic}, {Barwick}, {Basu}, {Bay}, {Beatty}, {Becker Tjus}, {Beise},
  {Bellenghi}, {Benning}, {BenZvi}, {Berley}, {Bernardini}, {Besson},
  {Blaufuss}, {Blot}, {Bontempo}, {Book}, {Boscolo Meneguolo}, {B{\"o}ser},
  {Botner}, {B{\"o}ttcher}, {Bourbeau}, {Braun}, {Brinson}, {Brostean-Kaiser},
  {Burley}, {Busse}, {Butterfield}, {Campana}, {Carloni}, {Carnie-Bronca},
  {Chattopadhyay}, {Chau}, {Chen}, {Chen}, {Chirkin}, {Choi}, {Clark},
  {Coenders}, {Coleman}, {Collin}, {Connolly}, {Conrad}, {Coppin}, {Correa},
  {Cowen}, {Dave}, {De Clercq}, {DeLaunay}, {Delgado}, {Deng}, {Deoskar},
  {Desai}, {Desiati}, {de Vries}, {de Wasseige}, {DeYoung}, {Diaz},
  {D{\'\i}az-V{\'e}lez}, {Dittmer}, {Domi}, {Dujmovic}, {DuVernois},
  {Ehrhardt}, {Eimer}, {Eller}, {Ellinger}, {El Mentawi}, {Els{\"a}sser},
  {Engel}, {Erpenbeck}, {Evans}, {Evenson}, {Fan}, {Fang}, {Farrag}, {Fazely},
  {Fedynitch}, {Feigl}, {Fiedlschuster}, {Finley}, {Fischer}, {Fox},
  {Franckowiak}, {Fritz}, {F{\"u}rst}, {Gallagher}, {Ganster}, {Garcia},
  {Gerhardt}, {Ghadimi}, {Glaser}, {Glauch}, {Gl{\"u}senkamp}, {Goehlke},
  {Gonzalez}, {Goswami}, {Grant}, {Gray}, {Gries}, {Griffin}, {Griswold},
  {Groth}, {G{\"u}nther}, {Gutjahr}, {Haack}, {Hallgren}, {Halliday}, {Halve},
  {Halzen}, {Hamdaoui}, {Ha Minh}, {Hanson}, {Hardin}, {Harnisch}, {Hatch},
  {Haungs}, {Helbing}, {Hellrung}, {Henningsen}, {Heuermann}, {Heyer},
  {Hickford}, {Hidvegi}, {Hill}, {Hill}, {Hoffman}, {Hori}, {Hoshina}, {Hou},
  {Huber}, {Hultqvist}, {H{\"u}nnefeld}, {Hussain}, {Hymon}, {In}, {Ishihara},
  {Jacquart}, {Janik}, {Jansson}, {Japaridze}, {Jeong}, {Jin}, {Jones}, {Kamp},
  {Kang}, {Kang}, {Kang}, {Kappes}, {Kappesser}, {Kardum}, {Karg}, {Karl},
  {Karle}, {Katil}, {Katz}, {Kauer}, {Kelley}, {Khatee Zathul}, {Kheirandish},
  {Kiryluk}, {Klein}, {Kochocki}, {Koirala}, {Kolanoski}, {Kontrimas},
  {K{\"o}pke}, {Kopper}, {Koskinen}, {Koundal}, {Kovacevich}, {Kowalski},
  {Kozynets}, {Krishnamoorthi}, {Kruiswijk}, {Krupczak}, {Kumar}, {Kun},
  {Kurahashi}, {Lad}, {Lagunas Gualda}, {Lamoureux}, {Larson}, \&
  {Latseva}}]{Abbasi2024_no_corr_between_events_and_diffuse}
{Abbasi}, R., {Ackermann}, M., {Adams}, J., {et~al.} 2024{\natexlab{a}}, \apj,
  964, 40

\bibitem[{{Abbasi} {et~al.}(2024{\natexlab{b}}){Abbasi}, {Ackermann}, {Adams},
  {Agarwalla}, {Aguilar}, {Ahlers}, {Alameddine}, {Amin}, {Andeen},
  {Arg{\"u}elles}, {Ashida}, {Athanasiadou}, {Ausborm}, {Axani}, {Bai},
  {Balagopal}, {Baricevic}, {Barwick}, {Bash}, {Basu}, {Bay}, {Beatty}, {Becker
  Tjus}, {Beise}, {Bellenghi}, {Benning}, {BenZvi}, {Berley}, {Bernardini},
  {Besson}, {Blaufuss}, {Bloom}, {Blot}, {Bontempo}, {Book Motzkin}, {Boscolo
  Meneguolo}, {B{\"o}ser}, {Botner}, {B{\"o}ttcher}, {Braun}, {Brinson},
  {Brostean-Kaiser}, {Brusa}, {Burley}, {Butterfield}, {Campana}, {Caracas},
  {Carloni}, {Carpio}, {Chattopadhyay}, {Chau}, {Chen}, {Chirkin}, {Choi},
  {Clark}, {Coleman}, {Collin}, {Connolly}, {Conrad}, {Corley}, {Cowen},
  {Dave}, {De Clercq}, {DeLaunay}, {Delgado}, {Deng}, {Desai}, {Desiati}, {de
  Vries}, {de Wasseige}, {DeYoung}, {Diaz}, {D{\'\i}az-V{\'e}lez}, {Dierichs},
  {Dittmer}, {Domi}, {Draper}, {Dujmovic}, {Durnford}, {Dutta}, {DuVernois},
  {Ehrhardt}, {Eidenschink}, {Eimer}, {Eller}, {Ellinger}, {El Mentawi},
  {Els{\"a}sser}, {Engel}, {Erpenbeck}, {Evans}, {Evenson}, {Fan}, {Fang},
  {Farrag}, {Fazely}, {Fedynitch}, {Feigl}, {Fiedlschuster}, {Finley},
  {Fischer}, {Fox}, {Franckowiak}, {Fukami}, {F{\"u}rst}, {Gallagher},
  {Ganster}, {Garcia}, {Garcia}, {Garg}, {Genton}, {Gerhardt}, {Ghadimi},
  {Girard-Carillo}, {Glaser}, {Gl{\"u}senkamp}, {Gonzalez}, {Goswami},
  {Granados}, {Grant}, {Gray}, {Gries}, {Griffin}, {Griswold}, {Groth},
  {Guevel}, {G{\"u}nther}, {Gutjahr}, {Ha}, {Haack}, {Hallgren}, {Halve},
  {Halzen}, {Hamdaoui}, {Minh}, {Handt}, {Hanson}, {Hardin}, {Harnisch},
  {Hatch}, {Haungs}, {H{\"a}u{\ss}ler}, {Helbing}, {Hellrung},
  {Hermannsgabner}, {Heuermann}, {Heyer}, {Hickford}, {Hidvegi}, {Hill},
  {Hill}, {Hoffman}, {Hori}, {Hoshina}, {Hostert}, {Hou}, {Huber}, {Hultqvist},
  {H{\"u}nnefeld}, {Hussain}, {Hymon}, {Ishihara}, {Iwakiri}, {Jacquart},
  {Jain}, {Janik}, {Jansson}, {Japaridze}, {Jeong}, {Jin}, {Jones}, {Kamp},
  {Kang}, {Kang}, {Kang}, {Kappes}, {Kappesser}, {Kardum}, {Karg}, {Karl},
  {Karle}, {Katil}, {Katz}, {Kauer}, {Kelley}, {Khanal}, {Khatee Zathul},
  {Kheirandish}, {Kiryluk}, {Klein}, {Kochocki}, {Koirala}, {Kolanoski},
  {Kontrimas}, {K{\"o}pke}, {Kopper}, {Koskinen}, {Koundal}, {Kovacevich}, \&
  {Kowalski}}]{Abbasi2024_stacking_w_MOJAVE}
{Abbasi}, R., {Ackermann}, M., {Adams}, J., {et~al.} 2024{\natexlab{b}}, \apj,
  973, 97

\bibitem[{{Abbasi} {et~al.}(2025){Abbasi}, {Ackermann}, {Adams}, {Agarwalla},
  {Aguilar}, {Ahlers}, {Alameddine}, {Amin}, {Andeen}, {Arg{\"u}elles},
  {Ashida}, {Athanasiadou}, {Axani}, {Babu}, {Bai}, {Baines-Holmes}, {Balagopal
  V.}, {Barwick}, {Bash}, {Basu}, {Bay}, {Beatty}, {Becker Tjus}, {Behrens},
  {Beise}, {Bellenghi}, {Benkel}, {BenZvi}, {Berley}, {Bernardini}, {Besson},
  {Blaufuss}, {Bloom}, {Blot}, {Bodo}, {Bontempo}, {Book Motzkin}, {Boscolo
  Meneguolo}, {B{\"o}ser}, {Botner}, {B{\"o}ttcher}, {Braun}, {Brinson},
  {Brisson-Tsavoussis}, {Burley}, {Butterfield}, {Campana}, {Carloni},
  {Carpio}, {Chattopadhyay}, {Chau}, {Chen}, {Chirkin}, {Choi}, {Clark},
  {Coleman}, {Coleman}, {Collin}, {Connolly}, {Conrad}, {Corley}, {Cowen}, {De
  Clercq}, {DeLaunay}, {Delgado}, {Delmeulle}, {Deng}, {Desiati}, {de Vries},
  {de Wasseige}, {DeYoung}, {D{\'\i}az-V{\'e}lez}, {DiKerby}, {Dittmer},
  {Domi}, {Draper}, {Dueser}, {Durnford}, {Dutta}, {DuVernois}, {Ehrhardt},
  {Eidenschink}, {Eimer}, {Eller}, {Ellinger}, {Els{\"a}sser}, {Engel},
  {Erpenbeck}, {Esmail}, {Eulig}, {Evans}, {Evenson}, {Fan}, {Fang}, {Farrag},
  {Fazely}, {Fedynitch}, {Feigl}, {Finley}, {Fischer}, {Fox}, {Franckowiak},
  {Fukami}, {F{\"u}rst}, {Gallagher}, {Ganster}, {Garcia}, {Garcia}, {Garg},
  {Genton}, {Gerhardt}, {Ghadimi}, {Glaser}, {Gl{\"u}senkamp}, {Gonzalez},
  {Goswami}, {Granados}, {Grant}, {Gray}, {Griffin}, {Griswold}, {Groth},
  {Guevel}, {G{\"u}nther}, {Gutjahr}, {Ha}, {Haack}, {Hallgren}, {Halve},
  {Halzen}, {Hamacher}, {Minh}, {Handt}, {Hanson}, {Hardin}, {Harnisch},
  {Hatch}, {Haungs}, {H{\"a}u{\ss}ler}, {Helbing}, {Hellrung}, {Hennig},
  {Heuermann}, {Hewett}, {Heyer}, {Hickford}, {Hidvegi}, {Hill}, {Hill},
  {Hmaid}, {Hoffman}, {Hooper}, {Hori}, {Hoshina}, {Hostert}, {Hou}, {Huber},
  {Hultqvist}, {Hymon}, {Ishihara}, {Iwakiri}, {Jacquart}, {Jain}, {Janik},
  {Jeong}, {Jin}, {Kamp}, {Kang}, {Kang}, {Kappes}, {Kardum}, {Karg}, {Karl},
  {Karle}, {Katil}, {Kauer}, {Kelley}, {Khanal}, {Khatee Zathul},
  {Kheirandish}, {Kimku}, {Kiryluk}, {Klein}, {Klein}, {Kobayashi}, {Kochocki},
  {Koirala}, {Kolanoski}, {Kontrimas}, {K{\"o}pke}, {Kopper}, {Koskinen},
  {Koundal}, {Kowalski}, {Kozynets}, {Krieger}, {Krishnamoorthi}, {Krishnan},
  {Kruiswijk}, \& {Krupczak}}]{Abbasi2025_mm_blz_vs_neut}
{Abbasi}, R., {Ackermann}, M., {Adams}, J., {et~al.} 2025, arXiv e-prints,
  arXiv:2507.03989

\bibitem[{{Abbasi} {et~al.}(2022){Abbasi}, {Ackermann}, {Adams}, {Aguilar},
  {Ahlers}, {Ahrens}, {Alameddine}, {Alispach}, {Alves}, {Amin}, {Andeen},
  {Anderson}, {Anton}, {Arg{\"u}elles}, {Ashida}, {Axani}, {Bai}, {Balagopal
  V.}, {Barbano}, {Barwick}, {Bastian}, {Basu}, {Baur}, {Bay}, {Beatty},
  {Becker}, {Tjus}, {Bellenghi}, {BenZvi}, {Berley}, {Bernardini}, {Besson},
  {Binder}, {Bindig}, {Blaufuss}, {Blot}, {Boddenberg}, {Bontempo}, {Borowka},
  {B{\"o}ser}, {Botner}, {B{\"o}ttcher}, {Bourbeau}, {Bradascio}, {Braun},
  {Brinson}, {Bron}, {Brostean-Kaiser}, {Browne}, {Burgman}, {Burley}, {Busse},
  {Campana}, {Carnie-Bronca}, {Chen}, {Chen}, {Chirkin}, {Choi}, {Clark},
  {Clark}, {Classen}, {Coleman}, {Collin}, {Conrad}, {Coppin}, {Correa},
  {Cowen}, {Cross}, {Dappen}, {Dave}, {De Clercq}, {DeLaunay}, {L{\'o}pez},
  {Dembinski}, {Deoskar}, {Desai}, {Desiati}, {de Vries}, {de Wasseige}, {de
  With}, {DeYoung}, {Diaz}, {D{\'\i}az-V{\'e}lez}, {Dittmer}, {Dujmovic},
  {Dunkman}, {DuVernois}, {Dvorak}, {Ehrhardt}, {Eller}, {Engel}, {Erpenbeck},
  {Evans}, {Evenson}, {Fan}, {Fazely}, {Feigl}, {Fiedlschuster}, {Fienberg},
  {Filimonov}, {Finley}, {Fischer}, {Fox}, {Franckowiak}, {Friedman}, {Fritz},
  {F{\"u}rst}, {Gaisser}, {Gallagher}, {Ganster}, {Garcia}, {Garrappa},
  {Gerhardt}, {Ghadimi}, {Glaser}, {Glauch}, {Gl{\"u}senkamp}, {Gonzalez},
  {Goswami}, {Grant}, {Gr{\'e}goire}, {Griswold}, {G{\"u}nther}, {Gutjahr},
  {Haack}, {Hallgren}, {Halliday}, {Halve}, {Halzen}, {Minh}, {Hanson},
  {Hardin}, {Harnisch}, {Haungs}, {Hebecker}, {Helbing}, {Henningsen},
  {Hettinger}, {Hickford}, {Hignight}, {Hill}, {Hill}, {Hoffman}, {Hoffmann},
  {Hokanson-Fasig}, {Hoshina}, {Huang}, {Huber}, {Huber}, {Hultqvist},
  {H{\"u}nnefeld}, {Hussain}, {Hymon}, {In}, {Iovine}, {Ishihara}, {Jansson},
  {Japaridze}, {Jeong}, {Jin}, {Jones}, {Kang}, {Kang}, {Kang}, {Kappes},
  {Kappesser}, {Kardum}, {Karg}, {Karl}, {Karle}, {Katz}, {Kauer},
  {Kellermann}, {Kelley}, {Kheirandish}, {Kin}, {Kintscher}, {Kiryluk},
  {Klein}, {Koirala}, {Kolanoski}, {Kontrimas}, {K{\"o}pke}, {Kopper},
  {Kopper}, {Koskinen}, {Koundal}, {Kovacevich}, {Kowalski}, {Kozynets}, {Kun},
  {Kurahashi}, {Lad}, {Gualda}, {Lanfranchi}, {Larson}, {Lauber}, {Lazar},
  {Lee}, {Leonard}, {Leszczy{\'n}ska}, {Li}, {Lincetto}, {Liu}, {Liubarska},
  {Lohfink}, {Mariscal}, {Lu}, {Lucarelli}, {Ludwig}, {Luszczak}, {Lyu}, {Ma},
  {Madsen}, {Mahn}, {Makino}, {Mancina}, {Mari{\c{s}}}, {Martinez-Soler},
  {Maruyama}, {Mase}, {McElroy}, {McNally}, {Mead}, {Meagher}, {Mechbal},
  {Medina}, {Meier}, {Meighen-Berger}, {Micallef}, {Mockler}, {Montaruli},
  {Moore}, {Morse}, {Moulai}, {Naab}, {Nagai}, {Naumann}, {Necker}, {Nguyễn},
  {Niederhausen}, {Nisa}, {Nowicki}, {Pollmann}, {Oehler}, {Oeyen}, {Olivas},
  {O'Sullivan}, {Pandya}, {Pankova}, {Park}, {Parker}, {Paudel}, {Paul}, {de
  los Heros}, {Peters}, {Peterson}, {Philippen}, {Pieper}, {Pittermann},
  {Pizzuto}, {Plum}, {Popovych}, {Porcelli}, {Rodriguez}, {Price}, {Pries},
  {Przybylski}, {Raab}, {Raissi}, {Rameez}, {Rawlins}, {Rea}, {Rehman},
  {Reichherzer}, {Reimann}, {Renzi}, {Resconi}, {Reusch}, {Rhode}, {Richman},
  {Riedel}, {Roberts}, {Robertson}, {Roellinghoff}, {Rongen}, {Rott}, {Ruhe},
  {Ryckbosch}, {Cantu}, {Safa}, {Saffer}, {Herrera}, {Sandrock}, {Sandroos},
  {Santander}, {Sarkar}, {Sarkar}, {Satalecka}, {Schaufel}, {Schieler},
  {Schindler}, {Schmidt}, {Schneider}, {Schneider}, {Schr{\"o}der},
  {Schumacher}, {Schwefer}, {Sclafani}, {Seckel}, {Seunarine}, {Sharma},
  {Shefali}, {Silva}, {Skrzypek}, {Smithers}, {Snihur}, {Soedingrekso},
  {Soldin}, {Spannfellner}, {Spiczak}, {Spiering}, {Stachurska}, {Stamatikos},
  {Stanev}, {Stein}, {Stettner}, {Steuer}, {Stezelberger}, {St{\"u}rwald},
  {Stuttard}, {Sullivan}, {Taboada}, {Ter-Antonyan}, {Tilav}, {Tischbein},
  {Tollefson}, {T{\"o}nnis}, {Toscano}, {Tosi}, {Trettin}, {Tselengidou},
  {Tung}, {Turcati}, {Turcotte}, {Turley}, {Twagirayezu}, {Ty}, {Elorrieta},
  {Valtonen-Mattila}, {Vandenbroucke}, {van Eijndhoven}, {Vannerom}, {van
  Santen}, {Verpoest}, {Walck}, {Watson}, {Weaver}, {Weigel}, {Weindl},
  {Weiss}, {Weldert}, {Wendt}, {Werthebach}, {Weyrauch}, {Whitehorn},
  {Wiebusch}, {Williams}, {Wolf}, {Woschnagg}, {Wrede}, {Wulff}, {Xu}, {Yanez},
  {Yoshida}, {Yu}, {Yuan}, {Zhang}, {Zhelnin}, \& {IceCube
  Collaboration}}]{Abbasi2022_extra_neut}
{Abbasi}, R., {Ackermann}, M., {Adams}, J., {et~al.} 2022, \apj, 928, 50

\bibitem[{{Abbasi} {et~al.}(2021{\natexlab{a}}){Abbasi}, {Ackermann}, {Adams},
  {Aguilar}, {Ahlers}, {Ahrens}, {Alispach}, {Alves}, {Amin}, {An}, {Andeen},
  {Anderson}, {Anton}, {Arg{\"u}elles}, {Ashida}, {Axani}, {Bai}, {Balagopal},
  {Barbano}, {Barwick}, {Bastian}, {Basu}, {Baur}, {Bay}, {Beatty}, {Becker},
  {Becker Tjus}, {Bellenghi}, {BenZvi}, {Berley}, {Bernardini}, {Besson},
  {Binder}, {Bindig}, {Blaufuss}, {Blot}, {Boddenberg}, {Bontempo}, {Borowka},
  {B{\"o}ser}, {Botner}, {B{\"o}ttcher}, {Bourbeau}, {Bradascio}, {Braun},
  {Bron}, {Brostean-Kaiser}, {Browne}, {Burgman}, {Burley}, {Busse}, {Campana},
  {Carnie-Bronca}, {Chen}, {Chirkin}, {Choi}, {Clark}, {Clark}, {Classen},
  {Coleman}, {Collin}, {Conrad}, {Coppin}, {Correa}, {Cowen}, {Cross},
  {Dappen}, {Dave}, {De Clercq}, {DeLaunay}, {Dembinski}, {Deoskar}, {De
  Ridder}, {Desai}, {Desiati}, {de Vries}, {de Wasseige}, {de With}, {DeYoung},
  {Dharani}, {Diaz}, {D{\'\i}az-V{\'e}lez}, {Dittmer}, {Dujmovic}, {Dunkman},
  {DuVernois}, {Dvorak}, {Ehrhardt}, {Eller}, {Engel}, {Erpenbeck}, {Evans},
  {Evenson}, {Fan}, {Fazely}, {Fiedlschuster}, {Fienberg}, {Filimonov},
  {Finley}, {Fischer}, {Fox}, {Franckowiak}, {Friedman}, {Fritz}, {F{\"u}rst},
  {Gaisser}, {Gallagher}, {Ganster}, {Garcia}, {Garrappa}, {Gerhardt},
  {Ghadimi}, {Glaser}, {Glauch}, {Gl{\"u}senkamp}, {Goldschmidt}, {Gonzalez},
  {Goswami}, {Grant}, {Gr{\'e}goire}, {Griswold}, {G{\"u}nd{\"u}z},
  {G{\"u}nther}, {Haack}, {Hallgren}, {Halliday}, {Halve}, {Halzen}, {Ha Minh},
  {Hanson}, {Hardin}, {Harnisch}, {Haungs}, {Hauser}, {Hebecker}, {Helbing},
  {Henningsen}, {Hettinger}, {Hickford}, {Hignight}, {Hill}, {Hill}, {Hoffman},
  {Hoffmann}, {Hoinka}, {Hokanson-Fasig}, {Hoshina}, {Huang}, {Huber}, {Huber},
  {Hultqvist}, {H{\"u}nnefeld}, {Hussain}, {In}, {Iovine}, {Ishihara},
  {Jansson}, {Japaridze}, {Jeong}, {Jones}, {Kang}, {Kang}, {Kang}, {Kappes},
  {Kappesser}, {Karg}, {Karl}, {Karle}, {Katz}, {Kauer}, {Kellermann},
  {Kelley}, {Kheirandish}, {Kin}, {Kintscher}, {Kiryluk}, {Klein}, {Koirala},
  {Kolanoski}, {Kontrimas}, {K{\"o}pke}, {Kopper}, {Kopper}, {Koskinen},
  {Koundal}, {Kovacevich}, {Kowalski}, {Kozynets}, {Kun}, {Kurahashi}, {Lad},
  {Lagunas Gualda}, {Lanfranchi}, {Larson}, {Lauber}, {Lazar}, {Lee},
  {Leonard}, {Leszczy{\'n}ska}, {Li}, {Lincetto}, {Liu}, {Liubarska},
  {Lohfink}, {Mariscal}, {Lu}, {Lucarelli}, {Ludwig}, {Luszczak}, {Lyu}, {Ma},
  {Madsen}, {Mahn}, {Makino}, {Mancina}, {Mari{\c{s}}}, {Maruyama}, {Mase},
  {McElroy}, {McNally}, {Mead}, {Meagher}, {Medina}, {Meier}, {Meighen-Berger},
  {Micallef}, {Mockler}, {Montaruli}, {Moore}, {Morse}, {Moulai}, {Naab},
  {Nagai}, {Naumann}, {Necker}, {Nguy{\^e}n}, {Niederhausen}, {Nisa},
  {Nowicki}, {Nygren}, {Obertacke Pollmann}, {Oehler}, {Oeyen}, {Olivas},
  {O'Sullivan}, {Pandya}, {Pankova}, {Park}, {Parker}, {Paudel}, {Paul},
  {P{\'e}rez de los Heros}, {Peters}, {Peterson}, {Philippen}, {Pieloth},
  {Pieper}, {Pittermann}, {Pizzuto}, {Plum}, {Popovych}, {Porcelli}, {Prado
  Rodriguez}, {Price}, {Pries}, {Przybylski}, {Raab}, {Raissi}, {Rameez},
  {Rawlins}, {Rea}, {Rehman}, {Reichherzer}, {Reimann}, {Renzi}, {Resconi},
  {Reusch}, {Rhode}, {Richman}, {Riedel}, {Roberts}, {Robertson},
  {Roellinghoff}, {Rongen}, {Rott}, {Ruhe}, {Ryckbosch}, {Rysewyk Cantu},
  {Safa}, {Saffer}, {Herrera}, {Sandrock}, {Sandroos}, {Santander}, {Sarkar},
  {Sarkar}, {Satalecka}, {Scharf}, {Schaufel}, {Schieler}, {Schindler},
  {Schlunder}, {Schmidt}, {Schneider}, {Schneider}, {Schr{\"o}der},
  {Schumacher}, {Schwefer}, {Sclafani}, {Seckel}, {Seunarine}, {Sharma},
  {Shefali}, {Silva}, {Skrzypek}, {Smithers}, {Snihur}, {Soedingrekso},
  {Soldin}, {Spannfellner}, {Spiczak}, {Spiering}, {Stachurska}, {Stamatikos},
  {Stanev}, {Stein}, {Stettner}, {Steuer}, {Stezelberger}, {St{\"u}rwald},
  {Stuttard}, {Sullivan}, {Taboada}, {Tenholt}, {Ter-Antonyan}, {Tilav},
  {Tischbein}, {Tollefson}, {Tomankova}, {T{\"o}nnis}, {Toscano}, {Tosi},
  {Trettin}, {Tselengidou}, {Tung}, {Turcati}, {Turcotte}, {Turley},
  {Twagirayezu}, {Ty}, {Unland Elorrieta}, {Valtonen-Mattila}, {Vandenbroucke},
  {van Eijndhoven}, {Vannerom}, {van Santen}, {Verpoest}, {Vraeghe}, {Walck},
  {Watson}, {Weaver}, {Weigel}, {Weindl}, {Weiss}, {Weldert}, {Wendt},
  {Werthebach}, {Weyrauch}, {Whitehorn}, {Wiebusch}, {Williams}, {Wolf},
  {Woschnagg}, {Wrede}, {Wulff}, {Xu}, {Xu}, {Yanez}, {Yoshida}, {Yu}, {Yuan},
  {Zhang}, \& {IceCube Collaboration}}]{IC2021_PKS1424_GB6J1542}
{Abbasi}, R., {Ackermann}, M., {Adams}, J., {et~al.} 2021{\natexlab{a}}, \apjl,
  920, L45

\bibitem[{{Abbasi} {et~al.}(2021{\natexlab{b}}){Abbasi}, {Ackermann}, {Adams},
  {Aguilar}, {Ahlers}, {Ahrens}, {Alispach}, {Alves}, {Amin}, {Andeen},
  {Anderson}, {Ansseau}, {Anton}, {Arg{\"u}elles}, {Axani}, {Bai}, {Balagopal
  V.}, {Barbano}, {Barwick}, {Bastian}, {Basu}, {Baum}, {Baur}, {Bay},
  {Beatty}, {Becker}, {Becker Tjus}, {Bellenghi}, {BenZvi}, {Berley},
  {Bernardini}, {Besson}, {Binder}, {Bindig}, {Blaufuss}, {Blot}, {B{\"o}ser},
  {Botner}, {B{\"o}ttcher}, {Bourbeau}, {Bourbeau}, {Bradascio}, {Braun},
  {Bron}, {Brostean-Kaiser}, {Burgman}, {Busse}, {Campana}, {Chen}, {Chirkin},
  {Choi}, {Clark}, {Clark}, {Classen}, {Coleman}, {Collin}, {Conrad}, {Coppin},
  {Correa}, {Cowen}, {Cross}, {Dave}, {De Clercq}, {DeLaunay}, {Dembinski},
  {Deoskar}, {De Ridder}, {Desai}, {Desiati}, {de Vries}, {de Wasseige}, {de
  With}, {DeYoung}, {Dharani}, {Diaz}, {D{\'\i}az-V{\'e}lez}, {Dujmovic},
  {Dunkman}, {DuVernois}, {Dvorak}, {Ehrhardt}, {Eller}, {Engel}, {Evans},
  {Evenson}, {Fahey}, {Fazely}, {Fiedlschuster}, {Fienberg}, {Filimonov},
  {Finley}, {Fischer}, {Fox}, {Franckowiak}, {Friedman}, {Fritz}, {F{\"u}rst},
  {Gaisser}, {Gallagher}, {Ganster}, {Garrappa}, {Gerhardt}, {Ghadimi},
  {Glauch}, {Gl{\"u}senkamp}, {Goldschmidt}, {Gonzalez}, {Goswami}, {Grant},
  {Gr{\'e}goire}, {Griffith}, {Griswold}, {G{\"u}nd{\"u}z}, {Haack},
  {Hallgren}, {Halliday}, {Halve}, {Halzen}, {Ha Minh}, {Hanson}, {Hardin},
  {Haungs}, {Hauser}, {Hebecker}, {Helbing}, {Henningsen}, {Hickford},
  {Hignight}, {Hill}, {Hill}, {Hoffman}, {Hoffmann}, {Hoinka},
  {Hokanson-Fasig}, {Hoshina}, {Huang}, {Huber}, {Huber}, {Hultqvist},
  {H{\"u}nnefeld}, {Hussain}, {In}, {Iovine}, {Ishihara}, {Jansson},
  {Japaridze}, {Jeong}, {Jones}, {Joppe}, {Kang}, {Kang}, {Kang}, {Kappes},
  {Kappesser}, {Karg}, {Karl}, {Karle}, {Katori}, {Katz}, {Kauer},
  {Kellermann}, {Kelley}, {Kheirandish}, {Kim}, {Kin}, {Kintscher}, {Kiryluk},
  {Klein}, {Koirala}, {Kolanoski}, {K{\"o}pke}, {Kopper}, {Kopper}, {Koskinen},
  {Koundal}, {Kovacevich}, {Kowalski}, {Krings}, {Kr{\"u}ckl}, {Kulacz},
  {Kurahashi}, {Kyriacou}, {Lagunas Gualda}, {Lanfranchi}, {Larson}, {Lauber},
  {Lazar}, {Leonard}, {Leszczy{\'n}ska}, {Li}, {Liu}, {Lohfink}, {Lozano
  Mariscal}, {Lu}, {Lucarelli}, {Ludwig}, {Luszczak}, {Lyu}, {Ma}, {Madsen},
  {Mahn}, {Makino}, {Mallik}, {Mancina}, {Mandalia}, {Mari{\c{s}}}, {Maruyama},
  {Mase}, {McNally}, {Meagher}, {Medina}, {Meier}, {Meighen-Berger}, {Merz},
  {Micallef}, {Mockler}, {Moment{\'e}}, {Montaruli}, {Moore}, {Morse},
  {Moulai}, {Naab}, {Nagai}, {Naumann}, {Necker}, {Neer},
  {Nguy{\'a}{\guillemotright} n}, {Niederhausen}, {Nisa}, {Nowicki}, {Nygren},
  {Obertacke Pollmann}, {Oehler}, {Olivas}, {O'Sullivan}, {Pandya}, {Pankova},
  {Park}, {Parker}, {Paudel}, {Peiffer}, {P{\'e}rez de los Heros}, {Philippen},
  {Pieloth}, {Pieper}, {Pizzuto}, {Plum}, {Popovych}, {Porcelli}, {Prado
  Rodriguez}, {Price}, {Przybylski}, {Raab}, {Raissi}, {Rameez}, {Rawlins},
  {Rea}, {Rehman}, {Reimann}, {Renschler}, {Renzi}, {Resconi}, {Reusch},
  {Rhode}, {Richman}, {Riedel}, {Robertson}, {Roellinghoff}, {Rongen}, {Rott},
  {Ruhe}, {Ryckbosch}, {Rysewyk Cantu}, {Safa}, {Sanchez Herrera}, {Sandrock},
  {Sandroos}, {Santander}, {Sarkar}, {Sarkar}, {Satalecka}, {Scharf},
  {Schaufel}, {Schieler}, {Schlunder}, {Schmidt}, {Schneider}, {Schneider},
  {Schr{\"o}der}, {Schumacher}, {Sclafani}, {Seckel}, {Seunarine}, {Shefali},
  {Silva}, {Smithers}, {Snihur}, {Soedingrekso}, {Soldin}, {Spiczak},
  {Spiering}, {Stachurska}, {Stamatikos}, {Stanev}, {Stein}, {Stettner},
  {Steuer}, {Stezelberger}, {Stokstad}, {Strotjohann}, {Stuttard}, {Sullivan},
  {Taboada}, {Tenholt}, {Ter-Antonyan}, {Tilav}, {Tischbein}, {Tollefson},
  {Tomankova}, {T{\"o}nnis}, {Toscano}, {Tosi}, {Trettin}, {Tselengidou},
  {Tung}, {Turcati}, {Turcotte}, {Turley}, {Twagirayezu}, {Ty}, {Unger},
  {Unland Elorrieta}, {Vandenbroucke}, {van Eijk}, {van Eijndhoven},
  {Vannerom}, {van Santen}, {Verpoest}, {Vraeghe}, {Walck}, {Wallace},
  {Wandkowsky}, {Watson}, {Weaver}, {Weindl}, {Weiss}, {Weldert}, {Wendt},
  {Werthebach}, {Weyrauch}, {Whelan}, {Whitehorn}, {Wiebe}, {Wiebusch},
  {Williams}, {Wolf}, {Wood}, {Woschnagg}, {Wrede}, {Wulff}, {Xu}, {Xu},
  {Yanez}, {Yoshida}, {Yuan}, {Zhang}, \& {IceCube
  Collaboration}}]{IC2021_neut_review}
{Abbasi}, R., {Ackermann}, M., {Adams}, J., {et~al.} 2021{\natexlab{b}}, \prd,
  104, 022002

\bibitem[{{Acharyya} {et~al.}(2023){Acharyya}, {Adams}, {Archer}, {Bangale},
  {Bartkoske}, {Batista}, {Benbow}, {Brill}, {Buckley}, {Christiansen},
  {Chromey}, {Errando}, {Falcone}, {Feng}, {Foote}, {Fortson}, {Furniss},
  {Gallagher}, {Hanlon}, {Hanna}, {Hervet}, {Hinrichs}, {Hoang}, {Holder},
  {Humensky}, {Jin}, {Kaaret}, {Kertzman}, {Kherlakian}, {Kieda}, {Kleiner},
  {Korzoun}, {Kumar}, {Lang}, {Lundy}, {Maier}, {McGrath}, {Millard}, {Millis},
  {Mooney}, {Moriarty}, {Mukherjee}, {O'Brien}, {Ong}, {Pohl}, {Pueschel},
  {Quinn}, {Ragan}, {Reynolds}, {Ribeiro}, {Roache}, {Sadeh}, {Sadun}, {Saha},
  {Santander}, {Sembroski}, {Shang}, {Splettstoesser}, {Talluri}, {Tucci},
  {Vassiliev}, {Weinstein}, {Williams}, {Wong}, {Woo}, {Aharonian},
  {Aschersleben}, {Backes}, {Martins}, {Batzofin}, {Becherini}, {Berge},
  {Bernl{\"o}hr}, {Bi}, {B{\"o}ttcher}, {Boisson}, {Bolmont}, {de Bony de
  Lavergne}, {Borowska}, {Bouyahiaoui}, {Bradascio}, {Breuhaus}, {Brose},
  {Brun}, {Bruno}, {Bulik}, {Burger-Scheidlin}, {Caroff}, {Casanova}, {Cecil},
  {Celic}, {Cerruti}, {Chand}, {Chandra}, {Chen}, {Chibueze}, {Chibueze},
  {Cotter}, {Dai}, {Mbarubucyeye}, {Djannati-Ata{\"\i}}, {Dmytriiev},
  {Doroshenko}, {Einecke}, {Ernenwein}, {de Clairfontaine}, {Filipovic},
  {Fontaine}, {F{\"u}{\ss}ling}, {Funk}, {Gabici}, {Ghafourizadeh}, {Giavitto},
  {Glawion}, {Glicenstein}, {Goswami}, {Grolleron}, {Haerer}, {Hinton},
  {Holch}, {Holler}, {Horns}, {Jamrozy}, {Jankowsky}, {Joshi}, {Jung-Richardt},
  {Kasai}, {Katarzy{\'n}ski}, {Khatoon}, {Kh{\'e}lifi}, {Klepser},
  {Klu{\'z}niak}, {Kosack}, {Kostunin}, {Lang}, {Le Stum}, {Lemi{\`e}re},
  {Lenain}, {Leuschner}, {Lohse}, {Luashvili}, {Lypova}, {Mackey}, {Malyshev},
  {Marandon}, {Marchegiani}, {Marcowith}, {Mart{\'\i}-Devesa}, {Marx},
  {Mitchell}, {Moderski}, {Mohrmann}, {Montanari}, {Moulin}, {Murach},
  {Nakashima}, {Niemiec}, {Noel}, {O'Brien}, {Olivera-Nieto}, {de Ona
  Wilhelmi}, {Ostrowski}, {Panny}, {Panter}, {Peron}, {Prokhorov},
  {P{\"u}hlhofer}, {Punch}, {Quirrenbach}, {Reichherzer}, {Reimer}, {Reimer},
  {Ren}, {Renaud}, {Rieger}, {Rudak}, {Ruiz-Velasco}, {Sahakian}, {Santangelo},
  {Sasaki}, {Sch{\"a}fer}, {Sch{\"u}ssler}, {Schutte}, {Schwanke}, {Shapopi},
  {Specovius}, {Spencer}, {Stawarz}, {Steenkamp}, {Steinmassl}, {Sushch},
  {Suzuki}, {Takahashi}, {Tanaka}, {Terrier}, {van Eldik}, {Vecchi}, {Veh},
  {Venter}, \& {Vink}}]{Acharyya2023_MWL_assoc_w_0735}
{Acharyya}, A., {Adams}, C.~B., {Archer}, A., {et~al.} 2023, \apj, 954, 70

\bibitem[{{Adri{\'a}n-Mart{\'\i}nez} {et~al.}(2016){Adri{\'a}n-Mart{\'\i}nez},
  {Ageron}, {Aharonian}, {Aiello}, {Albert}, {Ameli}, {Anassontzis}, {Andre},
  {Androulakis}, {Anghinolfi}, {Anton}, {Ardid}, {Avgitas}, {Barbarino},
  {Barbarito}, {Baret}, {Barrios-Mart{\'\i}}, {Belhorma}, {Belias}, {Berbee},
  {van den Berg}, {Bertin}, {Beurthey}, {van Beveren}, {Beverini}, {Biagi},
  {Biagioni}, {Billault}, {Bond{\`\i}}, {Bormuth}, {Bouhadef}, {Bourlis},
  {Bourret}, {Boutonnet}, {Bouwhuis}, {Bozza}, {Bruijn}, {Brunner}, {Buis},
  {Busto}, {Cacopardo}, {Caillat}, {Calamai}, {Calvo}, {Capone}, {Caramete},
  {Cecchini}, {Celli}, {Champion}, {Cherkaoui El Moursli}, {Cherubini},
  {Chiarusi}, {Circella}, {Classen}, {Cocimano}, {Coelho}, {Coleiro},
  {Colonges}, {Coniglione}, {Cordelli}, {Cosquer}, {Coyle}, {Creusot},
  {Cuttone}, {D'Amico}, {De Bonis}, {De Rosa}, {De Sio}, {Di Capua}, {Di
  Palma}, {D{\'\i}az Garc{\'\i}a}, {Distefano}, {Donzaud}, {Dornic},
  {Dorosti-Hasankiadeh}, {Drakopoulou}, {Drouhin}, {Drury}, {Durocher},
  {Eberl}, {Eichie}, {van Eijk}, {El Bojaddaini}, {El Khayati}, {Elsaesser},
  {Enzenh{\"o}fer}, {Fassi}, {Favali}, {Fermani}, {Ferrara}, {Filippidis},
  {Frascadore}, {Fusco}, {Gal}, {Galat{\`a}}, {Garufi}, {Gay}, {Gebyehu},
  {Giordano}, {Gizani}, {Gracia}, {Graf}, {Gr{\'e}goire}, {Grella}, {Habel},
  {Hallmann}, {van Haren}, {Harissopulos}, {Heid}, {Heijboer}, {Heine},
  {Henry}, {Hern{\'a}ndez-Rey}, {Hevinga}, {Hofest{\"a}dt}, {Hugon},
  {Illuminati}, {James}, {Jansweijer}, {Jongen}, {de Jong}, {Kadler},
  {Kalekin}, {Kappes}, {Katz}, {Keller}, {Kieft}, {Kie{\ss}ling}, {Koffeman},
  {Kooijman}, {Kouchner}, {Kulikovskiy}, {Lahmann}, {Lamare}, {Leisos},
  {Leonora}, {Clark}, {Liolios}, {Llorens Alvarez}, {Lo Presti}, {L{\"o}hner},
  {Lonardo}, {Lotze}, {Loucatos}, {Maccioni}, {Mannheim}, {Margiotta},
  {Marinelli}, {Mari{\c{s}}}, {Markou}, {Mart{\'\i}nez-Mora}, {Martini},
  {Mele}, {Melis}, {Michael}, {Migliozzi}, {Migneco}, {Mijakowski}, {Miraglia},
  {Mollo}, {Mongelli}, {Morganti}, {Moussa}, {Musico}, {Musumeci}, {Navas},
  {Nicolau}, {Olcina}, {Olivetto}, {Orlando}, {Papaikonomou}, {Papaleo},
  {P{\u{a}}v{\u{a}}la{\c{s}}}, {Peek}, {Pellegrino}, {Perrina}, {Pfutzner},
  {Piattelli}, {Pikounis}, {Poma}, {Popa}, {Pradier}, {Pratolongo},
  {P{\"u}hlhofer}, {Pulvirenti}, {Quinn}, {Racca}, {Raffaelli}, {Randazzo},
  {Rapidis}, {Razis}, {Real}, {Resvanis}, {Reubelt}, {Riccobene}, {Rossi},
  {Rovelli}, {Salda{\~n}a}, {Salvadori}, {Samtleben}, {S{\'a}nchez
  Garc{\'\i}a}, {S{\'a}nchez Losa}, {Sanguineti}, {Santangelo}, {Santonocito},
  {Sapienza}, {Schimmel}, {Schmelling}, {Sciacca}, {Sedita}, {Seitz}, {Sgura},
  {Simeone}, {Siotis}, {Sipala}, {Spisso}, {Spurio}, {Stavropoulos},
  {Steijger}, {Stellacci}, {Stransky}, {Taiuti}, {Tayalati}, {T{\'e}zier},
  {Theraube}, {Thompson}, {Timmer}, {T{\"o}nnis}, {Trasatti}, {Trovato},
  {Tsirigotis}, {Tzamarias}, {Tzamariudaki}, {Vallage}, {Van Elewyck},
  {Vermeulen}, {Vicini}, {Viola}, {Vivolo}, {Volkert}, {Voulgaris}, {Wiggers},
  {Wilms}, {de Wolf}, {Zachariadou}, {Zornoza}, \&
  {Z{\'u}{\~n}iga}}]{KM3NeT2016_next_gen}
{Adri{\'a}n-Mart{\'\i}nez}, S., {Ageron}, M., {Aharonian}, F., {et~al.} 2016,
  Journal of Physics G Nuclear Physics, 43, 084001

\bibitem[{{Agudo} {et~al.}(2015){Agudo}, {Boettcher}, {Falcke},
  {Georganopoulos}, {Ghisellini}, {Giovannini}, {Giroletti}, {Gurvits},
  {G{\'o}mez}, {Laing}, {Lister}, {Mart{\'\i}}, {Meyer}, {Mizuno},
  {O'Sullivan}, {Padovani}, {Paragi}, {Perucho}, {Schleicher}, {Stawarz},
  {Vlahakis}, \& {Wardle}}]{Agudo2015_SKA_AGN_jets}
{Agudo}, I., {Boettcher}, M., {Falcke}, H.~D.~E., {et~al.} 2015, in Advancing
  Astrophysics with the Square Kilometre Array (AASKA14), 93

\bibitem[{{Agudo} {et~al.}(2025){Agudo}, {Liodakis}, {Otero-Santos}, {Middei},
  {Marscher}, {Jorstad}, {Zhang}, {Li}, {Di Gesu}, {Romani}, {Kim}, {Fenu},
  {Marshall}, {Pacciani}, {Escudero Pedrosa}, {Aceituno},
  {Ag{\'\i}s-Gonz{\'a}lez}, {Bonnoli}, {Casanova}, {Morcuende}, {Piirola},
  {Sota}, {Kouch}, {Lindfors}, {McCall}, {Jermak}, {Steele}, {Borman},
  {Grishina}, {Hagen-Thorn}, {Kopatskaya}, {Larionova}, {Morozova},
  {Savchenko}, {Shishkina}, {Troitskiy}, {Troitskaya}, {Vasilyev}, {Zhovtan},
  {Myserlis}, {Gurwell}, {Keating}, {Rao}, {Kang}, {Lee}, {Kim}, {Cheong},
  {Jeong}, {Angelakis}, {Kraus}, {Blinov}, {Maharana}, {Bachev}, {Jormanainen},
  {Nilsson}, {Fallah Ramazani}, {Casadio}, {Fuentes}, {Traianou}, {Thum},
  {G{\'o}mez}, {Antonelli}, {Bachetti}, {Baldini}, {Baumgartner}, {Bellazzini},
  {Bianchi}, {Bongiorno}, {Bonino}, {Brez}, {Bucciantini}, {Capitanio},
  {Castellano}, {Cavazzuti}, {Chen}, {Ciprini}, {Costa}, {De Rosa}, {Del
  Monte}, {Di Lalla}, {Di Marco}, {Donnarumma}, {Doroshenko}, {Dov{\v{c}}iak},
  {Ehlert}, {Enoto}, {Evangelista}, {Fabiani}, {Ferrazzoli}, {Garc{\'\i}a},
  {Gunji}, {Hayashida}, {Heyl}, {Iwakiri}, {Kaaret}, {Karas}, {Kislat},
  {Kitaguchi}, {Kolodziejczak}, {Krawczynski}, {La Monaca}, {Latronico},
  {Maldera}, {Manfreda}, {Marin}, {Marinucci}, {Massaro}, {Matt}, {Mitsuishi},
  {Mizuno}, {Muleri}, {Negro}, {Ng}, {O'Dell}, {Omodei}, {Oppedisano},
  {Papitto}, {Pavlov}, {Peirson}, {Perri}, {Pesce-Rollins}, {Petrucci},
  {Pilia}, {Possenti}, {Poutanen}, {Puccetti}, {Ramsey}, {Rankin}, {Ratheesh},
  {Roberts}, {Sgr{\`o}}, {Slane}, {Soffitta}, {Spandre}, {Swartz}, {Tamagawa},
  {Tavecchio}, {Taverna}, {Tawara}, {Tennant}, {Thomas}, {Tombesi}, {Trois},
  {Tsygankov}, {Turolla}, {Vink}, {Weisskopf}, {Wu}, {Xie}, \&
  {Zane}}]{Agudo2025_IXPE_BLLac}
{Agudo}, I., {Liodakis}, I., {Otero-Santos}, J., {et~al.} 2025, \apjl, 985, L15

\bibitem[{{Aguilar} {et~al.}(2021){Aguilar}, {Allison}, {Beatty}, {Bernhoff},
  {Besson}, {Bingefors}, {Botner}, {Buitink}, {Carter}, {Clark}, {Connolly},
  {Dasgupta}, {de Kockere}, {de Vries}, {Deaconu}, {DuVernois}, {Feigl},
  {Garc{\'\i}a-Fern{\'a}ndez}, {Glaser}, {Hallgren}, {Hallmann}, {Hanson},
  {Hendricks}, {Hokanson-Fasig}, {Hornhuber}, {Hughes}, {Karle}, {Kelley},
  {Klein}, {Krebs}, {Lahmann}, {Magnuson}, {Meures}, {Meyers}, {Nelles},
  {Novikov}, {Oberla}, {Oeyen}, {Pandya}, {Plaisier}, {Pyras}, {Ryckbosch},
  {Scholten}, {Seckel}, {Smith}, {Southall}, {Torres}, {Toscano}, {Van Den
  Broeck}, {van Eijndhoven}, {Vieregg}, {Welling}, {Wissel}, {Young}, \&
  {Zink}}]{Aguilar2021_RNO_Greenland}
{Aguilar}, J.~A., {Allison}, P., {Beatty}, J.~J., {et~al.} 2021, Journal of
  Instrumentation, 16, P03025

\bibitem[{{Ajello} {et~al.}(2022){Ajello}, {Baldini}, {Ballet}, {Bastieri},
  {Becerra Gonzalez}, {Bellazzini}, {Berretta}, {Bissaldi}, {Bonino}, {Brill},
  {Bruel}, {Buson}, {Caputo}, {Caraveo}, {Cheung}, {Chiaro}, {Cibrario},
  {Ciprini}, {Crnogorcevic}, {Cutini}, {D'Ammando}, {De Gaetano}, {Di Lalla},
  {Di Venere}, {Dom{\'\i}nguez}, {Ramazani}, {Ferrara}, {Fiori}, {Fukazawa},
  {Funk}, {Fusco}, {Gammaldi}, {Gargano}, {Garrappa}, {Gasparrini},
  {Giglietto}, {Giordano}, {Giroletti}, {Green}, {Grenier}, {Guiriec}, {Horan},
  {Hou}, {Kayanoki}, {Kuss}, {Larsson}, {Latronico}, {Lewis}, {Li}, {Liodakis},
  {Longo}, {Loparco}, {Lott}, {Lovellette}, {Lubrano}, {Madejski}, {Maldera},
  {Manfreda}, {Mart{\'\i}-Devesa}, {Mazziotta}, {Mereu}, {Michelson},
  {Mirabal}, {Mitthumsiri}, {Mizuno}, {Monzani}, {Morselli}, {Moskalenko},
  {Negro}, {Ojha}, {Orienti}, {Orlando}, {Ormes}, {Pei}, {Pe{\~n}a-Herazo},
  {Persic}, {Pesce-Rollins}, {Petrosian}, {Pillera}, {Poon}, {Porter},
  {Principe}, {Rain{\`o}}, {Rando}, {Rani}, {Razzano}, {Razzaque}, {Reimer},
  {Reimer}, {Scotton}, {Serini}, {Sgr{\`o}}, {Siskind}, {Spandre}, {Spinelli},
  {Suson}, {Tajima}, {Torres}, {Valverde}, {Yassin}, \&
  {Zaharijas}}]{Ajello2022_4lac_dr3}
{Ajello}, M., {Baldini}, L., {Ballet}, J., {et~al.} 2022, \apjs, 263, 24

\bibitem[{{Albert} {et~al.}(2024){Albert}, {Alves}, {Andr{\'e}}, {Ardid},
  {Ardid}, {Aubert}, {Aublin}, {Baret}, {Basa}, {Becherini}, {Belhorma},
  {Bendahman}, {Benfenati}, {Bertin}, {Biagi}, {Bissinger}, {Boumaaza},
  {Bouta}, {Bouwhuis}, {Br{\^a}nza{\c{s}}}, {Bruijn}, {Brunner}, {Busto},
  {Caiffi}, {Calvo}, {Campion}, {Capone}, {Caramete}, {Carenini}, {Carr},
  {Carretero}, {Celli}, {Cerisy}, {Chabab}, {Cherkaoui El Moursli}, {Chiarusi},
  {Circella}, {Coelho}, {Coleiro}, {Coniglione}, {Coyle}, {Creusot}, {Cruz},
  {D{\'\i}az}, {de Martino}, {Distefano}, {di Palma}, {Domi}, {Donzaud},
  {Dornic}, {Drouhin}, {Eberl}, {van Eeden}, {van Eijk}, {El Hedri}, {El
  Khayati}, {Enzenh{\"o}fer}, {Fermani}, {Ferrara}, {Filippini}, {Fusco},
  {Gagliardini}, {Garc{\'\i}a}, {Gatius Oliver}, {Gay}, {Gei{\ss}elbrecht},
  {Glotin}, {Gozzini}, {Gracia Ruiz}, {Graf}, {Guidi}, {Haegel}, {Hallmann},
  {van Haren}, {Heijboer}, {Hello}, {Hern{\'a}ndez-Rey}, {H{\"o}{\ss}l},
  {Hofest{\"a}dt}, {Huang}, {Illuminati}, {James}, {Jisse-Jung}, {de Jong}, {de
  Jong}, {Kadler}, {Kalekin}, {Katz}, {Kouchner}, {Kovalev}, {Kovalev},
  {Kreykenbohm}, {Kulikovskiy}, {Lahmann}, {Lamoureux}, {Lazo}, {Lef{\`e}vre},
  {Leonora}, {Levi}, {Le Stum}, {Lopez-Coto}, {Loucatos}, {Maderer}, {Manczak},
  {Marcelin}, {Margiotta}, {Marinelli}, {Mart{\'\i}nez-Mora}, {Migliozzi},
  {Moussa}, {Muller}, {Navas}, {Nezri}, {Fearraigh}, {Oukacha}, {P{\u{a}}un},
  {P{\u{a}}v{\u{a}}la{\c{s}}}, {Pe{\~n}a-Mart{\'\i}nez}, {Perrin-Terrin},
  {Pestel}, {Piattelli}, {Plavin}, {Poir{\`e}}, {Popa}, {Pradier}, {Pushkarev},
  {Randazzo}, {Real}, {Reck}, {Riccobene}, {Romanov}, {S{\'a}nchez-Losa},
  {Saina}, {Salesa Greus}, {Samtleben}, {Sanguineti}, {Sapienza}, {Schnabel},
  {Schumann}, {Sch{\"u}ssler}, {Seneca}, {Spurio}, {Stolarczyk}, {Taiuti},
  {Tayalati}, {Tingay}, {Troitsky}, {Vallage}, {Vannoye}, {van Elewyck},
  {Viola}, {Vivolo}, {Wilms}, {Zavatarelli}, {Zegarelli}, {Zornoza},
  {Z{\'u}{\~n}iga}, {(ANTARES Collaboration)}, {Hovatta}, {Kiehlmann},
  {Liodakis}, {Pavlidou}, {Readhead}, \& {(Ovro
  Collaboration)}}]{Albert2024_ANTARES_v_blz_assoc}
{Albert}, A., {Alves}, S., {Andr{\'e}}, M., {et~al.} 2024, \apj, 964, 3

\bibitem[{{Albert} {et~al.}(2021){Albert}, {Andr{\'e}}, {Anghinolfi}, {Anton},
  {Ardid}, {Aubert}, {Aublin}, {Baret}, {Basa}, {Belhorma}, {Bertin}, {Biagi},
  {Bissinger}, {Boumaaza}, {Bouta}, {Bouwhuis}, {Br{\^a}nza{\c{s}}}, {Bruijn},
  {Brunner}, {Busto}, {Capone}, {Caramete}, {Carr}, {Carretero}, {Celli},
  {Chabab}, {Chau}, {Cherkaoui El Moursli}, {Chiarusi}, {Circella}, {Coleiro},
  {Colomer-Molla}, {Coniglione}, {Coyle}, {Creusot}, {D{\'\i}az}, {de
  Wasseige}, {Deschamps}, {Distefano}, {di Palma}, {Domi}, {Donzaud}, {Dornic},
  {Drouhin}, {Eberl}, {El Khayati}, {Enzenh{\"o}fer}, {Fermani}, {Ferrara},
  {Filippini}, {Fusco}, {Gatelet}, {Gay}, {Glotin}, {Gozzini}, {Grac{\'\i}a},
  {Graf}, {Guidi}, {Hallmann}, {van Haren}, {Heijboer}, {Hello},
  {Hern{\'a}ndez-Rey}, {H{\"o}{\ss}l}, {Hofest{\"a}dt}, {Huang}, {Illuminati},
  {James}, {Jisse-Jung}, {de Jong}, {de Jong}, {Jongen}, {Kadler}, {Kalekin},
  {Katz}, {Khan-Chowdhury}, {Kouchner}, {Kreykenbohm}, {Kulikovskiy},
  {Lahmann}, {Le Breton}, {Lef{\`e}vre}, {Leonora}, {Levi}, {Lincetto},
  {Lopez-Coto}, {Loucatos}, {Maderer}, {Manczak}, {Marcelin}, {Margiotta},
  {Marinelli}, {Mart{\'\i}nez-Mora}, {Mazzou}, {Melis}, {Migliozzi}, {Moser},
  {Moussa}, {Muller}, {Nauta}, {Navas}, {Nezri}, {Nu{\~n}ez-Casti{\~n}eyra},
  {O'Fearraigh}, {Organokov}, {P{\u{a}}v{\u{a}}la{\c{s}}}, {Pellegrino},
  {Perrin-Terrin}, {Piattelli}, {Pieterse}, {Poir{\`e}}, {Popa}, {Pradier},
  {Randazzo}, {Reck}, {Riccobene}, {Salesa Greus}, {Samtleben},
  {S{\'a}nchez-Losa}, {Sanguineti}, {Sapienza}, {Schnabel}, {Sch{\"u}ssler},
  {Spurio}, {Stolarczyk}, {Taiuti}, {Tayalati}, {Thakore}, {Tingay}, {Vallage},
  {van Elewyck}, {Versari}, {Viola}, {Vivolo}, {Wilms}, {Zegarelli}, {Zornoza},
  {Z{\'u}{\~n}iga}, {ANTARES Collaboration}, \& {Buson}}]{Albert2021_ANTARES}
{Albert}, A., {Andr{\'e}}, M., {Anghinolfi}, M., {et~al.} 2021, \apj, 911, 48

\bibitem[{{Astropy Collaboration} {et~al.}(2022){Astropy Collaboration},
  {Price-Whelan}, {Lim}, {Earl}, {Starkman}, {Bradley}, {Shupe}, {Patil},
  {Corrales}, {Brasseur}, {N{\"o}the}, {Donath}, {Tollerud}, {Morris},
  {Ginsburg}, {Vaher}, {Weaver}, {Tocknell}, {Jamieson}, {van Kerkwijk},
  {Robitaille}, {Merry}, {Bachetti}, {G{\"u}nther}, {Aldcroft},
  {Alvarado-Montes}, {Archibald}, {B{\'o}di}, {Bapat}, {Barentsen},
  {Baz{\'a}n}, {Biswas}, {Boquien}, {Burke}, {Cara}, {Cara}, {Conroy},
  {Conseil}, {Craig}, {Cross}, {Cruz}, {D'Eugenio}, {Dencheva}, {Devillepoix},
  {Dietrich}, {Eigenbrot}, {Erben}, {Ferreira}, {Foreman-Mackey}, {Fox},
  {Freij}, {Garg}, {Geda}, {Glattly}, {Gondhalekar}, {Gordon}, {Grant},
  {Greenfield}, {Groener}, {Guest}, {Gurovich}, {Handberg}, {Hart},
  {Hatfield-Dodds}, {Homeier}, {Hosseinzadeh}, {Jenness}, {Jones}, {Joseph},
  {Kalmbach}, {Karamehmetoglu}, {Ka{\l}uszy{\'n}ski}, {Kelley}, {Kern},
  {Kerzendorf}, {Koch}, {Kulumani}, {Lee}, {Ly}, {Ma}, {MacBride}, {Maljaars},
  {Muna}, {Murphy}, {Norman}, {O'Steen}, {Oman}, {Pacifici}, {Pascual},
  {Pascual-Granado}, {Patil}, {Perren}, {Pickering}, {Rastogi}, {Roulston},
  {Ryan}, {Rykoff}, {Sabater}, {Sakurikar}, {Salgado}, {Sanghi}, {Saunders},
  {Savchenko}, {Schwardt}, {Seifert-Eckert}, {Shih}, {Jain}, {Shukla}, {Sick},
  {Simpson}, {Singanamalla}, {Singer}, {Singhal}, {Sinha}, {Sip{\H{o}}cz},
  {Spitler}, {Stansby}, {Streicher}, {{\v{S}}umak}, {Swinbank}, {Taranu},
  {Tewary}, {Tremblay}, {de Val-Borro}, {Van Kooten}, {Vasovi{\'c}}, {Verma},
  {de Miranda Cardoso}, {Williams}, {Wilson}, {Winkel}, {Wood-Vasey}, {Xue},
  {Yoachim}, {Zhang}, {Zonca}, \& {Astropy Project
  Contributors}}]{Astropy2022_v5}
{Astropy Collaboration}, {Price-Whelan}, A.~M., {Lim}, P.~L., {et~al.} 2022,
  \apj, 935, 167

\bibitem[{{Atwood} {et~al.}(2009){Atwood}, {Abdo}, {Ackermann}, {Althouse},
  {Anderson}, {Axelsson}, {Baldini}, {Ballet}, {Band}, {Barbiellini},
  {Bartelt}, {Bastieri}, {Baughman}, {Bechtol}, {B{\'e}d{\'e}r{\`e}de},
  {Bellardi}, {Bellazzini}, {Berenji}, {Bignami}, {Bisello}, {Bissaldi},
  {Blandford}, {Bloom}, {Bogart}, {Bonamente}, {Bonnell}, {Borgland},
  {Bouvier}, {Bregeon}, {Brez}, {Brigida}, {Bruel}, {Burnett}, {Busetto},
  {Caliandro}, {Cameron}, {Caraveo}, {Carius}, {Carlson}, {Casandjian},
  {Cavazzuti}, {Ceccanti}, {Cecchi}, {Charles}, {Chekhtman}, {Cheung},
  {Chiang}, {Chipaux}, {Cillis}, {Ciprini}, {Claus}, {Cohen-Tanugi},
  {Condamoor}, {Conrad}, {Corbet}, {Corucci}, {Costamante}, {Cutini}, {Davis},
  {Decotigny}, {DeKlotz}, {Dermer}, {de Angelis}, {Digel}, {do Couto e Silva},
  {Drell}, {Dubois}, {Dumora}, {Edmonds}, {Fabiani}, {Farnier}, {Favuzzi},
  {Flath}, {Fleury}, {Focke}, {Funk}, {Fusco}, {Gargano}, {Gasparrini},
  {Gehrels}, {Gentit}, {Germani}, {Giebels}, {Giglietto}, {Giommi}, {Giordano},
  {Glanzman}, {Godfrey}, {Grenier}, {Grondin}, {Grove}, {Guillemot}, {Guiriec},
  {Haller}, {Harding}, {Hart}, {Hays}, {Healey}, {Hirayama}, {Hjalmarsdotter},
  {Horn}, {Hughes}, {J{\'o}hannesson}, {Johansson}, {Johnson}, {Johnson},
  {Johnson}, {Johnson}, {Kamae}, {Katagiri}, {Kataoka}, {Kavelaars}, {Kawai},
  {Kelly}, {Kerr}, {Klamra}, {Kn{\"o}dlseder}, {Kocian}, {Komin}, {Kuehn},
  {Kuss}, {Landriu}, {Latronico}, {Lee}, {Lee}, {Lemoine-Goumard}, {Lionetto},
  {Longo}, {Loparco}, {Lott}, {Lovellette}, {Lubrano}, {Madejski}, {Makeev},
  {Marangelli}, {Massai}, {Mazziotta}, {McEnery}, {Menon}, {Meurer},
  {Michelson}, {Minuti}, {Mirizzi}, {Mitthumsiri}, {Mizuno}, {Moiseev},
  {Monte}, {Monzani}, {Moretti}, {Morselli}, {Moskalenko}, {Murgia},
  {Nakamori}, {Nishino}, {Nolan}, {Norris}, {Nuss}, {Ohno}, {Ohsugi}, {Omodei},
  {Orlando}, {Ormes}, {Paccagnella}, {Paneque}, {Panetta}, {Parent}, {Pearce},
  {Pepe}, {Perazzo}, {Pesce-Rollins}, {Picozza}, {Pieri}, {Pinchera}, {Piron},
  {Porter}, {Poupard}, {Rain{\`o}}, {Rando}, {Rapposelli}, {Razzano}, {Reimer},
  {Reimer}, {Reposeur}, {Reyes}, {Ritz}, {Rochester}, {Rodriguez}, {Romani},
  {Roth}, {Russell}, {Ryde}, {Sabatini}, {Sadrozinski}, {Sanchez}, {Sander},
  {Sapozhnikov}, {Parkinson}, {Scargle}, {Schalk}, \&
  {Scolieri}}]{Atwood2009_FermiLAT}
{Atwood}, W.~B., {Abdo}, A.~A., {Ackermann}, M., {et~al.} 2009, \apj, 697, 1071

\bibitem[{{Azzollini} {et~al.}(2025){Azzollini}, {Buson}, {Coleiro}, {de
  Clairfontaine}, {Pfeiffer}, {Zaballa}, {Boughelilba}, \&
  {Lincetto}}]{Azzollini2025_properties_of_PeVatron_blz}
{Azzollini}, A., {Buson}, S., {Coleiro}, A., {et~al.} 2025, \aap, 700, A228

\bibitem[{{Backes} {et~al.}(2016){Backes}, {M{\"u}ller}, {Conway}, {Deane},
  {Evans}, {Falcke}, {Fraga-Encinas}, {Goddi}, {Klein Wolt}, {Krichbaum},
  {MacLeod}, {Ribeiro}, {Roelofs}, {Shen}, \& {van
  Langevelde}}]{Backes2016_AMT}
{Backes}, M., {M{\"u}ller}, C., {Conway}, J.~E., {et~al.} 2016, in The 4th
  Annual Conference on High Energy Astrophysics in Southern Africa (HEASA
  2016), 29

\bibitem[{{Baikal-GVD Collaboration} {et~al.}(2018){Baikal-GVD Collaboration},
  {Avrorin}, {Avrorin}, {Aynutdinov}, {Bannash}, {Belolaptikov}, {Brudanin},
  {Budnev}, {Doroshenko}, {Domogatsky}, {Dvornick{\'y}}, {Dyachok},
  {Dzhilkibaev}, {Fajt}, {Fialkovsky}, {Gafarov}, {Golubkov}, {Gres}, {Honz},
  {Kebkal}, {Kebkal}, {Khramov}, {Kolbin}, {Konischev}, {Korobchenko},
  {Koshechkin}, {Kozhin}, {Kulepov}, {Kuleshov}, {Milenin}, {Mirgazov},
  {Osipova}, {Panfilov}, {Pan'kov}, {Petukhov}, {Pliskovsky}, {Rozanov},
  {Rjabov}, {Rushay}, {Safronov}, {Simkovic}, {Shoibonov}, {Solovjev},
  {Sorokovikov}, {Shelepov}, {Suvorova}, {Shtekl}, {Tabolenko}, {Tarashansky},
  {Yakovlev}, {Zagorodnikov}, \& {Zurbanov}}]{Baikal2018_status}
{Baikal-GVD Collaboration}, {Avrorin}, A.~D., {Avrorin}, A.~V., {et~al.} 2018,
  arXiv e-prints, arXiv:1808.10353

\bibitem[{{Bartos} {et~al.}(2021){Bartos}, {Veske}, {Kowalski}, {M{\'a}rka}, \&
  {M{\'a}rka}}]{Bartos2021}
{Bartos}, I., {Veske}, D., {Kowalski}, M., {M{\'a}rka}, Z., \& {M{\'a}rka}, S.
  2021, \apj, 921, 45

\bibitem[{{Bellenghi} {et~al.}(2023){Bellenghi}, {Padovani}, {Resconi}, \&
  {Giommi}}]{Bellenghi2023}
{Bellenghi}, C., {Padovani}, P., {Resconi}, E., \& {Giommi}, P. 2023, \apjl,
  955, L32

\bibitem[{{Bellm} {et~al.}(2019){Bellm}, {Kulkarni}, {Graham}, {Dekany},
  {Smith}, {Riddle}, {Masci}, {Helou}, {Prince}, {Adams}, {Barbarino},
  {Barlow}, {Bauer}, {Beck}, {Belicki}, {Biswas}, {Blagorodnova}, {Bodewits},
  {Bolin}, {Brinnel}, {Brooke}, {Bue}, {Bulla}, {Burruss}, {Cenko}, {Chang},
  {Connolly}, {Coughlin}, {Cromer}, {Cunningham}, {De}, {Delacroix}, {Desai},
  {Duev}, {Eadie}, {Farnham}, {Feeney}, {Feindt}, {Flynn}, {Franckowiak},
  {Frederick}, {Fremling}, {Gal-Yam}, {Gezari}, {Giomi}, {Goldstein},
  {Golkhou}, {Goobar}, {Groom}, {Hacopians}, {Hale}, {Henning}, {Ho}, {Hover},
  {Howell}, {Hung}, {Huppenkothen}, {Imel}, {Ip}, {Ivezi{\'c}}, {Jackson},
  {Jones}, {Juric}, {Kasliwal}, {Kaspi}, {Kaye}, {Kelley}, {Kowalski},
  {Kramer}, {Kupfer}, {Landry}, {Laher}, {Lee}, {Lin}, {Lin}, {Lunnan},
  {Giomi}, {Mahabal}, {Mao}, {Miller}, {Monkewitz}, {Murphy}, {Ngeow},
  {Nordin}, {Nugent}, {Ofek}, {Patterson}, {Penprase}, {Porter}, {Rauch},
  {Rebbapragada}, {Reiley}, {Rigault}, {Rodriguez}, {van Roestel}, {Rusholme},
  {van Santen}, {Schulze}, {Shupe}, {Singer}, {Soumagnac}, {Stein}, {Surace},
  {Sollerman}, {Szkody}, {Taddia}, {Terek}, {Van Sistine}, {van Velzen},
  {Vestrand}, {Walters}, {Ward}, {Ye}, {Yu}, {Yan}, \& {Zolkower}}]{bellm2019}
{Bellm}, E.~C., {Kulkarni}, S.~R., {Graham}, M.~J., {et~al.} 2019, \pasp, 131,
  018002

\bibitem[{Benjamini \& Hochberg(1995)}]{Benjamini_Hochberg1995}
Benjamini, Y. \& Hochberg, Y. 1995, Journal of the Royal Statistical Society.
  Series B (Methodological), 57, 289

\bibitem[{{Blandford} {et~al.}(2019){Blandford}, {Meier}, \&
  {Readhead}}]{Blandford2019}
{Blandford}, R., {Meier}, D., \& {Readhead}, A. 2019, \araa, 57, 467

\bibitem[{{Blinov} \& {Novikova}(2025)}]{Blinov2025_PKS1502_repeating_pattern}
{Blinov}, D. \& {Novikova}, P. 2025, \aap, 694, L10

\bibitem[{{B{\"o}ttcher}(2019)}]{Bottcher2019}
{B{\"o}ttcher}, M. 2019, Galaxies, 7, 20

\bibitem[{{Buson} {et~al.}(2023){Buson}, {Tramacere}, {Oswald}, {Barbano},
  {Fichet de Clairfontaine}, {Pfeiffer}, {Azzollini}, {Baghmanyan}, \&
  {Ajello}}]{Buson2023}
{Buson}, S., {Tramacere}, A., {Oswald}, L., {et~al.} 2023, arXiv e-prints,
  arXiv:2305.11263

\bibitem[{{Buson} {et~al.}(2022){Buson}, {Tramacere}, {Pfeiffer}, {Oswald}, {de
  Menezes}, {Azzollini}, \& {Ajello}}]{Buson2022}
{Buson}, S., {Tramacere}, A., {Pfeiffer}, L., {et~al.} 2022, \apjl, 933, L43

\bibitem[{{Cerruti} {et~al.}(2019){Cerruti}, {Zech}, {Boisson}, {Emery},
  {Inoue}, \& {Lenain}}]{Cerruti2019_TXS_electron_SSC_as_seed}
{Cerruti}, M., {Zech}, A., {Boisson}, C., {et~al.} 2019, \mnras, 483, L12

\bibitem[{{Cherenkov Telescope Array Consortium} {et~al.}(2019){Cherenkov
  Telescope Array Consortium}, {Acharya}, {Agudo}, {Al Samarai}, {Alfaro},
  {Alfaro}, {Alispach}, {Alves Batista}, {Amans}, {Amato}, {Ambrosi},
  {Antolini}, {Antonelli}, {Aramo}, {Araya}, {Armstrong}, {Arqueros},
  {Arrabito}, {Asano}, {Ashley}, {Backes}, {Balazs}, {Balbo}, {Ballester},
  {Ballet}, {Bamba}, {Barkov}, {Barres de Almeida}, {Barrio}, {Bastieri},
  {Becherini}, {Belfiore}, {Benbow}, {Berge}, {Bernardini}, {Bernardini},
  {Bernardos}, {Bernl{\"o}hr}, {Bertucci}, {Biasuzzi}, {Bigongiari}, {Biland},
  {Bissaldi}, {Biteau}, {Blanch}, {Blazek}, {Boisson}, {Bolmont}, {Bonanno},
  {Bonardi}, {Bonavolont{\`a}}, {Bonnoli}, {Bosnjak}, {B{\"o}ttcher},
  {Braiding}, {Bregeon}, {Brill}, {Brown}, {Brun}, {Brunetti}, {Buanes},
  {Buckley}, {Bugaev}, {B{\"u}hler}, {Bulgarelli}, {Bulik}, {Burton},
  {Burtovoi}, {Busetto}, {Canestrari}, {Capalbi}, {Capitanio}, {Caproni},
  {Caraveo}, {C{\'a}rdenas}, {Carlile}, {Carosi}, {Carqu{\'\i}n}, {Carr},
  {Casanova}, {Cascone}, {Catalani}, {Catalano}, {Cauz}, {Cerruti}, {Chadwick},
  {Chaty}, {Chaves}, {Chen}, {Chen}, {Chernyakova}, {Chikawa}, {Christov},
  {Chudoba}, {Cie{\'s}lar}, {Coco}, {Colafrancesco}, {Colin}, {Conforti},
  {Connaughton}, {Conrad}, {Contreras}, {Cortina}, {Costa}, {Costantini},
  {Cotter}, {Covino}, {Crocker}, {Cuadra}, {Cuevas}, {Cumani}, {D'A{\`\i}},
  {D'Ammando}, {D'Avanzo}, {D'Urso}, {Daniel}, {Davids}, {Dawson}, {Dazzi}, {De
  Angelis}, {de C{\'a}ssia dos Anjos}, {De Cesare}, {De Franco}, {de Gouveia
  Dal Pino}, {de la Calle}, {de los Reyes Lopez}, {De Lotto}, {De Luca}, {De
  Lucia}, {de Naurois}, {de O{\~n}a Wilhelmi}, {De Palma}, {De Persio}, {de
  Souza}, {Deil}, {Del Santo}, {Delgado}, {della Volpe}, {Di Girolamo}, {Di
  Pierro}, {Di Venere}, {D{\'\i}az}, {Dib}, {Diebold}, {Djannati-Ata{\"\i}},
  {Dom{\'\i}nguez}, {Dominis Prester}, {Dorner}, {Doro}, {Drass}, {Dravins},
  {Dubus}, {Dwarkadas}, {Ebr}, {Eckner}, {Egberts}, {Einecke}, {Ekoume},
  {Els{\"a}sser}, {Ernenwein}, {Espinoza}, {Evoli}, {Fairbairn},
  {Falceta-Goncalves}, {Falcone}, {Farnier}, {Fasola}, {Fedorova}, {Fegan},
  {Fernandez-Alonso}, {Fern{\'a}ndez-Barral}, {Ferrand}, {Fesquet},
  {Filipovic}, {Fioretti}, {Fontaine}, {Fornasa}, {Fortson}, {Freixas
  Coromina}, {Fruck}, {Fujita}, {Fukazawa}, {Funk}, {F{\"u}{\ss}ling},
  {Gabici}, {Gadola}, {Gallant}, {Garcia}, {Garcia L{\'o}pez}, {Garczarczyk},
  {Gaskins}, {Gasparetto}, {Gaug}, {Gerard}, {Giavitto}, {Giglietto}, {Giommi},
  {Giordano}, {Giro}, \& {Giroletti}}]{CTA2019_sci_book}
{Cherenkov Telescope Array Consortium}, {Acharya}, B.~S., {Agudo}, I., {et~al.}
  2019, {Science with the Cherenkov Telescope Array}

\bibitem[{{Cohen} {et~al.}(2014){Cohen}, {Romani}, {Filippenko}, {Cenko},
  {Lott}, {Zheng}, \& {Li}}]{Cohen2014_KAIT_blazars}
{Cohen}, D.~P., {Romani}, R.~W., {Filippenko}, A.~V., {et~al.} 2014, \apj, 797,
  137

\bibitem[{Davison \& Hinkley(1997)}]{davison_hinkley_1997}
Davison, A.~C. \& Hinkley, D.~V. 1997, Bootstrap Methods and their Application,
  Cambridge Series in Statistical and Probabilistic Mathematics (Cambridge
  University Press)

\bibitem[{{de Jaeger} {et~al.}(2023){de Jaeger}, {Shappee}, {Kochanek},
  {Hinkle}, {Garrappa}, {Liodakis}, {Franckowiak}, {Stanek}, {Beacom}, \&
  {Prieto}}]{deJaeger2023_optical_gamma_flare_correlation}
{de Jaeger}, T., {Shappee}, B.~J., {Kochanek}, C.~S., {et~al.} 2023, \mnras,
  519, 6349

\bibitem[{{Drake} {et~al.}(2009){Drake}, {Djorgovski}, {Mahabal}, {Beshore},
  {Larson}, {Graham}, {Williams}, {Christensen}, {Catelan}, {Boattini},
  {Gibbs}, {Hill}, \& {Kowalski}}]{drake2009}
{Drake}, A.~J., {Djorgovski}, S.~G., {Mahabal}, A., {et~al.} 2009, \apj, 696,
  870

\bibitem[{{Eisenstein} \& {Hut}(1998)}]{Eisenstein1998_HOP}
{Eisenstein}, D.~J. \& {Hut}, P. 1998, \apj, 498, 137

\bibitem[{{Fang} {et~al.}(2017){Fang}, {Alvarez-Muniz}, {Alves Batista},
  {Bustamante}, {Carvalho}, {Charrier}, {Cognard}, {de Jong}, {de Vries},
  {Finley}, {Gou}, {Gu}, {Gu{\'e}pin}, {Hanson}, {Hu}, {Kotera}, {Le Coz},
  {Mao}, {Martineau-Huynh}, {Medina}, {Mostafa}, {Mottez}, {Murase}, {Niess},
  {Oikonomou}, {Schr{\"o}der}, {Tasse}, {Timmermans}, {Renault-Tinacci},
  {Tueros}, {Wu}, {Zarka}, {Zech}, {Yi}, {Zheng}, {Zilles}, {Cognard}, {de
  Vries}, {Gu{\'e}pin}, {Hanson}, {Mao}, {Mottez}, \& {Zheng}}]{Fang2017_GRAND}
{Fang}, K., {Alvarez-Muniz}, J., {Alves Batista}, R., {et~al.} 2017, in
  International Cosmic Ray Conference, Vol. 301, 35th International Cosmic Ray
  Conference (ICRC2017), 996

\bibitem[{{Filippenko} {et~al.}(2001){Filippenko}, {Li}, {Treffers}, \&
  {Modjaz}}]{filippenko2001}
{Filippenko}, A.~V., {Li}, W.~D., {Treffers}, R.~R., \& {Modjaz}, M. 2001, in
  Astronomical Society of the Pacific Conference Series, Vol. 246, IAU Colloq.
  183: Small Telescope Astronomy on Global Scales, ed. B.~{Paczynski}, W.-P.
  {Chen}, \& C.~{Lemme}, 121

\bibitem[{{Fiorillo} {et~al.}(2025){Fiorillo}, {Testagrossa}, {Petropoulou}, \&
  {Winter}}]{Fiorillo2025_TXS0506_neut_not_from_corona}
{Fiorillo}, D. F.~G., {Testagrossa}, F., {Petropoulou}, M., \& {Winter}, W.
  2025, \apj, 986, 104

\bibitem[{{Franckowiak} {et~al.}(2020){Franckowiak}, {Garrappa}, {Paliya},
  {Shappee}, {Stein}, {Strotjohann}, {Kowalski}, {Buson}, {Kiehlmann},
  {Max-Moerbeck}, \& {Angioni}}]{Franckowiak2020}
{Franckowiak}, A., {Garrappa}, S., {Paliya}, V., {et~al.} 2020, \apj, 893, 162

\bibitem[{{Gao} {et~al.}(2019){Gao}, {Fedynitch}, {Winter}, \&
  {Pohl}}]{Gao2019_TXS_AM3_modeling}
{Gao}, S., {Fedynitch}, A., {Winter}, W., \& {Pohl}, M. 2019, Nature Astronomy,
  3, 88

\bibitem[{{Garrappa} {et~al.}(2024){Garrappa}, {Buson}, {Sinapius},
  {Franckowiak}, {Liodakis}, {Bartolini}, {Giroletti}, {Nanci}, {Principe}, \&
  {Venters}}]{Garrappa2024_IC_v_Fermi}
{Garrappa}, S., {Buson}, S., {Sinapius}, J., {et~al.} 2024, \aap, 687, A59

\bibitem[{{Ghisellini} {et~al.}(2011){Ghisellini}, {Tavecchio}, {Foschini}, \&
  {Ghirlanda}}]{Ghisellini2011_transitional_BLL_FSRQ}
{Ghisellini}, G., {Tavecchio}, F., {Foschini}, L., \& {Ghirlanda}, G. 2011,
  \mnras, 414, 2674

\bibitem[{{Giommi} {et~al.}(2020){Giommi}, {Glauch}, {Padovani}, {Resconi},
  {Turcati}, \& {Chang}}]{Giommi2020}
{Giommi}, P., {Glauch}, T., {Padovani}, P., {et~al.} 2020, \mnras, 497, 865

\bibitem[{{Gordon} {et~al.}(2016){Gordon}, {Jacobs}, {Beasley}, {Peck},
  {Gaume}, {Charlot}, {Fey}, {Ma}, {Titov}, \& {Boboltz}}]{gordon2016_vcsii}
{Gordon}, D., {Jacobs}, C., {Beasley}, A., {et~al.} 2016, \aj, 151, 154

\bibitem[{{Harris} {et~al.}(2020){Harris}, {Millman}, {van der Walt},
  {Gommers}, {Virtanen}, {Cournapeau}, {Wieser}, {Taylor}, {Berg}, {Smith},
  {Kern}, {Picus}, {Hoyer}, {van Kerkwijk}, {Brett}, {Haldane}, {del R{\'\i}o},
  {Wiebe}, {Peterson}, {G{\'e}rard-Marchant}, {Sheppard}, {Reddy}, {Weckesser},
  {Abbasi}, {Gohlke}, \& {Oliphant}}]{Harris2020_NumPy}
{Harris}, C.~R., {Millman}, K.~J., {van der Walt}, S.~J., {et~al.} 2020, \nat,
  585, 357

\bibitem[{{Hooper} \&
  {Plant}(2023)}]{Hooper_Plant2023_leptonic_neut_production}
{Hooper}, D. \& {Plant}, K. 2023, \prl, 131, 231001

\bibitem[{{Hovatta} \& {Lindfors}(2019)}]{Hovatta_Lindfors2019}
{Hovatta}, T. \& {Lindfors}, E. 2019, \nar, 87, 101541

\bibitem[{{Hovatta} {et~al.}(2021){Hovatta}, {Lindfors}, {Kiehlmann},
  {Max-Moerbeck}, {Hodges}, {Liodakis}, {L{\"a}hteem{\"a}ki}, {Pearson},
  {Readhead}, {Reeves}, {Suutarinen}, {Tammi}, \& {Tornikoski}}]{Hovatta2021}
{Hovatta}, T., {Lindfors}, E., {Kiehlmann}, S., {et~al.} 2021, \aap, 650, A83

\bibitem[{{Huang} {et~al.}(2024){Huang}, {Cao}, {Chen}, {Liu}, {Wang}, {You},
  \& {Qi}}]{Huang2024_HUNT}
{Huang}, T.~Q., {Cao}, Z., {Chen}, M., {et~al.} 2024, in 38th International
  Cosmic Ray Conference, 1080

\bibitem[{{Huber}(2019)}]{Huber2019_blz_maximal_contr}
{Huber}, M. 2019, in International Cosmic Ray Conference, Vol.~36, 36th
  International Cosmic Ray Conference (ICRC2019), 916

\bibitem[{{Hunter}(2007)}]{Hunter2007_Matplotlib}
{Hunter}, J.~D. 2007, Computing in Science and Engineering, 9, 90

\bibitem[{{IceCube Collaboration} {et~al.}(2018){IceCube Collaboration},
  {Aartsen}, {Ackermann}, {Adams}, {Aguilar}, {Ahlers}, {Ahrens}, {Samarai},
  {Altmann}, {Andeen}, {Anderson}, {Ansseau}, {Anton}, {Arg{\"u}elles},
  {Arsioli}, {Auffenberg}, {Axani}, {Bagherpour}, {Bai}, {Barron}, {Barwick},
  {Baum}, {Bay}, {Beatty}, {Becker Tjus}, {Becker}, {BenZvi}, {Berley},
  {Bernardini}, {Besson}, {Binder}, {Bindig}, {Blaufuss}, {Blot}, {Bohm},
  {B{\"o}rner}, {Bos}, {B{\"o}ser}, {Botner}, {Bourbeau}, {Bourbeau},
  {Bradascio}, {Braun}, {Brenzke}, {Bretz}, {Bron}, {Brostean-Kaiser},
  {Burgman}, {Busse}, {Carver}, {Cheung}, {Chirkin}, {Christov}, {Clark},
  {Classen}, {Coenders}, {Collin}, {Conrad}, {Coppin}, {Correa}, {Cowen},
  {Cross}, {Dave}, {Day}, {de Andr{\'e}}, {De Clercq}, {DeLaunay}, {Dembinski},
  {DeRidder}, {Desiati}, {de Vries}, {de Wasseige}, {de With}, {DeYoung},
  {D{\'\i}az-V{\'e}lez}, {di Lorenzo}, {Dujmovic}, {Dumm}, {Dunkman}, {Dvorak},
  {Eberhardt}, {Ehrhardt}, {Eichmann}, {Eller}, {Evenson}, {Fahey}, {Fazely},
  {Felde}, {Filimonov}, {Finley}, {Flis}, {Franckowiak}, {Friedman}, {Fritz},
  {Gaisser}, {Gallagher}, {Gerhardt}, {Ghorbani}, {Giommi}, {Glauch},
  {Gl{\"u}senkamp}, {Goldschmidt}, {Gonzalez}, {Grant}, {Griffith}, {Haack},
  {Hallgren}, {Halzen}, {Hanson}, {Hebecker}, {Heereman}, {Helbing},
  {Hellauer}, {Hickford}, {Hignight}, {Hill}, {Hoffman}, {Hoffmann}, {Hoinka},
  {Hokanson-Fasig}, {Hoshina}, {Huang}, {Huber}, {Hultqvist}, {H{\"u}nnefeld},
  {Hussain}, {In}, {Iovine}, {Ishihara}, {Jacobi}, {Japaridze}, {Jeong},
  {Jero}, {Jones}, {Kalaczynski}, {Kang}, {Kappes}, {Kappesser}, {Karg},
  {Karle}, {Katz}, {Kauer}, {Keivani}, {Kelley}, {Kheirandish}, {Kim}, {Kim},
  {Kintscher}, {Kiryluk}, {Kittler}, {Klein}, {Koirala}, {Kolanoski},
  {K{\"o}pke}, {Kopper}, {Kopper}, {Koschinsky}, {Koskinen}, {Kowalski},
  {Krammer}, {Krings}, {Kroll}, {Kr{\"u}ckl}, {Kunwar}, {Kurahashi},
  {Kuwabara}, {Kyriacou}, {Labare}, {Lanfranchi}, {Larson}, {Lauber},
  {Leonard}, {Lesiak-Bzdak}, {Leuermann}, {Liu}, {Lozano Mariscal}, {Lu},
  {L{\"u}nemann}, {Luszczak}, {Madsen}, {Maggi}, {Mahn}, {Mancina}, {Maruyama},
  {Mase}, {Maunu}, {Meagher}, {Medici}, {Meier}, {Menne}, {Merino}, {Meures},
  {Miarecki}, {Micallef}, {Moment{\'e}}, {Montaruli}, {Moore}, {Morse},
  {Moulai}, {Nahnhauer}, {Nakarmi}, {Naumann}, {Neer}, {Niederhausen},
  {Nowicki}, {Nygren}, {Obertacke Pollmann}, {Olivas}, {O'Murchadha},
  {O'Sullivan}, {Padovani}, {Palczewski}, {Pandya}, {Pankova}, {Peiffer},
  {Pepper}, {P{\'e}rez de los Heros}, {Pieloth}, {Pinat}, {Plum}, {Price},
  {Przybylski}, {Raab}, {R{\"a}del}, {Rameez}, {Rawlins}, {Rea}, {Reimann},
  {Relethford}, {Relich}, {Resconi}, {Rhode}, {Richman}, {Robertson}, {Rongen},
  {Rott}, {Ruhe}, {Ryckbosch}, {Rysewyk}, {Safa}, {Sahakyan}, {S{\"a}lzer},
  {Sanchez Herrera}, {Sandrock}, {Sandroos}, {Santander}, {Sarkar}, {Sarkar},
  {Satalecka}, {Schlunder}, {Schmidt}, {Schneider}, {Schoenen},
  {Sch{\"o}neberg}, {Schumacher}, {Sclafani}, {Seckel}, {Seunarine},
  {Soedingrekso}, {Soldin}, {Song}, {Spiczak}, {Spiering}, {Stachurska},
  {Stamatikos}, {Stanev}, {Stasik}, {Stettner}, {Steuer}, {Stezelberger},
  {Stokstad}, {St{\"o}{\ss}l}, {Strotjohann}, {Stuttard}, {Sullivan},
  {Sutherland}, {Taboada}, {Tatar}, {Tenholt}, {Ter-Antonyan}, {Terliuk},
  {Tilav}, {Toale}, {Tobin}, {Toennis}, {Toscano}, {Tosi}, {Tselengidou},
  {Tung}, {Turcati}, {Turley}, {Ty}, {Unger}, {Usner}, {Vandenbroucke}, {Van
  Driessche}, {van Eijk}, {van Eijndhoven}, {Vanheule}, {van Santen}, {Vogel},
  {Vraeghe}, {Walck}, {Wallace}, {Wallraff}, {Wandler}, {Wandkowsky}, {Waza},
  {Weaver}, {Weiss}, {Wendt}, {Werthebach}, {Westerhoff}, {Whelan},
  {Whitehorn}, {Wiebe}, {Wiebusch}, {Wille}, {Williams}, {Wills}, {Wolf},
  {Wood}, {Wood}, {Woschnagg}, {Xu}, {Xu}, {Xu}, {Yanez}, {Yodh}, {Yoshida}, \&
  {Yuan}}]{IC2018_TXS0506}
{IceCube Collaboration}, {Aartsen}, M.~G., {Ackermann}, M., {et~al.} 2018,
  Science, 361, 147

\bibitem[{{IceCube Collaboration} {et~al.}(2022){IceCube Collaboration},
  {Abbasi}, {Ackermann}, {Adams}, {Aguilar}, {Ahlers}, {Ahrens}, {Alameddine},
  {Alispach}, {Alves}, {Amin}, {Andeen}, {Anderson}, {Anton}, {Arg{\"u}elles},
  {Ashida}, {Axani}, {Bai}, {Balagopal}, {Barbano}, {Barwick}, {Bastian},
  {Basu}, {Baur}, {Bay}, {Beatty}, {Becker}, {Becker Tjus}, {Bellenghi},
  {Benzvi}, {Berley}, {Bernardini}, {Besson}, {Binder}, {Bindig}, {Blaufuss},
  {Blot}, {Boddenberg}, {Bontempo}, {Borowka}, {B{\"o}ser}, {Botner},
  {B{\"o}ttcher}, {Bourbeau}, {Bradascio}, {Braun}, {Brinson}, {Bron},
  {Brostean-Kaiser}, {Browne}, {Burgman}, {Burley}, {Busse}, {Campana},
  {Carnie-Bronca}, {Chen}, {Chen}, {Chirkin}, {Choi}, {Clark}, {Clark},
  {Classen}, {Coleman}, {Collin}, {Conrad}, {Coppin}, {Correa}, {Cowen},
  {Cross}, {Dappen}, {Dave}, {de Clercq}, {Delaunay}, {Delgado L{\'o}pez},
  {Dembinski}, {Deoskar}, {Desai}, {Desiati}, {de Vries}, {de Wasseige}, {de
  With}, {Deyoung}, {Diaz}, {D{\'\i}az-V{\'e}lez}, {Dittmer}, {Dujmovic},
  {Dunkman}, {Duvernois}, {Dvorak}, {Ehrhardt}, {Eller}, {Engel}, {Erpenbeck},
  {Evans}, {Evenson}, {Fan}, {Fazely}, {Fedynitch}, {Feigl}, {Fiedlschuster},
  {Fienberg}, {Filimonov}, {Finley}, {Fischer}, {Fox}, {Franckowiak},
  {Friedman}, {Fritz}, {F{\"u}rst}, {Gaisser}, {Gallagher}, {Ganster},
  {Garcia}, {Garrappa}, {Gerhardt}, {Ghadimi}, {Glaser}, {Glauch},
  {Gl{\"u}senkamp}, {Goldschmidt}, {Gonzalez}, {Goswami}, {Grant},
  {Gr{\'e}goire}, {Griswold}, {G{\"u}nther}, {Gutjahr}, {Haack}, {Hallgren},
  {Halliday}, {Halve}, {Halzen}, {Hanson}, {Hardin}, {Harnisch}, {Haungs},
  {Hebecker}, {Helbing}, {Henningsen}, {Hettinger}, {Hickford}, {Hignight},
  {Hill}, {Hill}, {Hoffman}, {Hoffmann}, {Hokanson-Fasig}, {Hoshina}, {Huang},
  {Huber}, {Huber}, {Hultqvist}, {H{\"u}nnefeld}, {Hussain}, {Hymon}, {in},
  {Iovine}, {Ishihara}, {Jansson}, {Japaridze}, {Jeong}, {Jin}, {Jones},
  {Kang}, {Kang}, {Kang}, {Kappes}, {Kappesser}, {Kardum}, {Karg}, {Karl},
  {Karle}, {Katz}, {Kauer}, {Kellermann}, {Kelley}, {Kheirandish}, {Kin},
  {Kintscher}, {Kiryluk}, {Klein}, {Koirala}, {Kolanoski}, {Kontrimas},
  {K{\"o}pke}, {Kopper}, {Kopper}, {Koskinen}, {Koundal}, {Kovacevich},
  {Kowalski}, {Kozynets}, {Kun}, {Kurahashi}, {Lad}, {Lagunas Gualda},
  {Lanfranchi}, {Larson}, {Lauber}, {Lazar}, {Lee}, {Leonard},
  {Leszczy{\'n}ska}, {Li}, {Lincetto}, {Liu}, {Liubarska}, {Lohfink}, {Lozano
  Mariscal}, {Lu}, {Lucarelli}, {Ludwig}, {Luszczak}, {Lyu}, {Ma}, {Madsen},
  {Mahn}, {Makino}, {Mancina}, {Mari{\c{s}}}, {Martinez-Soler}, {Maruyama},
  {Mase}, {McElroy}, {McNally}, {Mead}, {Meagher}, {Mechbal}, {Medina},
  {Meier}, {Meighen-Berger}, {Micallef}, {Mockler}, {Montaruli}, {Moore},
  {Morse}, {Moulai}, {Naab}, {Nagai}, {Nahnhauer}, {Naumann}, {Necker},
  {Nguyen}, {Niederhausen}, {Nisa}, {Nowicki}, {Nygren}, {Obertack},
  {Pollmann}, {Oehler}, {Oeyen}, {Olivas}, {O'Sullivan}, {Pandya}, {Pankova},
  {Park}, {Parker}, {Paudel}, {Paul}, {P{\'e}rez de Los Heros}, {Peters},
  {Peterson}, {Philippen}, {Pieper}, {Pittermann}, {Pizzuto}, {Plum},
  {Popovych}, {Porcelli}, {Prado Rodriguez}, {Price}, {Pries}, {Przybylski},
  {Rack-Helleis}, {Raissi}, {Rameez}, {Rawlins}, {Rea}, {Rehman},
  {Reichherzer}, {Reimann}, {Renzi}, {Resconi}, {Reusch}, {Rhode}, {Richman},
  {Riedel}, {Roberts}, {Robertson}, {Roellinghoff}, {Rongen}, {Rott}, {Ruhe},
  {Ryckbosch}, {Rysewyk Cantu}, {Safa}, {Saffer}, {Sanchez Herrera},
  {Sandrock}, {Sandroos}, {Santander}, {Sarkar}, {Sarkar}, {Satalecka},
  {Schaufel}, {Schieler}, {Schindler}, {Schmidt}, {Schneider}, {Schneider},
  {Schr{\"o}der}, {Schumacher}, {Schwefer}, {Sclafani}, {Seckel}, {Seunarine},
  {Sharma}, {Shefali}, {Silva}, {Skrzypek}, {Smithers}, {Snihur},
  {Soedingrekso}, {Soldin}, {Spannfellner}, {Spiczak}, {Spiering},
  {Stachurska}, {Stamatikos}, {Stanev}, {Stein}, {Stettner}, {Steuer},
  {Stezelberger}, {Stokstad}, {St{\"u}rwald}, {Stuttard}, {Sullivan},
  {Taboada}, {Ter-Antonyan}, {Tilav}, {Tischbein}, {Tollefson}, {T{\"o}nnis},
  {Toscano}, {Tosi}, {Trettin}, {Tselengidou}, {Tung}, {Turcati}, {Turcotte},
  {Turley}, {Twagirayezu}, {Ty}, {Unland Elorrieta}, {Valtonen-Mattila},
  {Vandenbroucke}, {van Eijndhoven}, {Vannerom}, {van Santen}, {Verpoest},
  {Walck}, {Watson}, {Weaver}, {Weigel}, {Weindl}, {Weiss}, {Weldert}, {Wendt},
  {Werthebach}, {Weyrauch}, {Whitehorn}, {Wiebusch}, {Williams}, {Wolf},
  {Woschnagg}, {Wrede}, {Wulff}, {Xu}, {Yanez}, {Yoshida}, {Yu}, {Yuan},
  {Zhangan}, \& {Zhelnin}}]{IC2022_NGC1068}
{IceCube Collaboration}, {Abbasi}, R., {Ackermann}, M., {et~al.} 2022, Science,
  378, 538

\bibitem[{{IceCube Collaboration} {et~al.}(2021){IceCube Collaboration},
  {Abbasi}, {Ackermann}, {Adams}, {Aguilar}, {Ahlers}, {Ahrens}, {Alispach},
  {Amin}, {Andeen}, {Anderson}, {Ansseau}, {Anton}, {Arg{\"u}elles}, {Axani},
  {Bai}, {Balagopal V.}, {Barbano}, {Barwick}, {Bastian}, {Basu}, {Baum},
  {Baur}, {Bay}, {Beatty}, {Becker}, {Becker Tjus}, {Bellenghi}, {BenZvi},
  {Berley}, {Bernardini}, {Besson}, {Binder}, {Bindig}, {Blaufuss}, {Blot},
  {Bohm}, {B{\"o}ser}, {Botner}, {B{\"o}ttcher}, {Bourbeau}, {Bourbeau},
  {Bradascio}, {Braun}, {Bron}, {Brostean-Kaiser}, {Burgman}, {Buscher},
  {Busse}, {Campana}, {Carver}, {Chen}, {Cheung}, {Chirkin}, {Choi}, {Clark},
  {Clark}, {Classen}, {Coleman}, {Collin}, {Conrad}, {Coppin}, {Correa},
  {Cowen}, {Cross}, {Dave}, {De Clercq}, {DeLaunay}, {Dembinski}, {Deoskar},
  {De Ridder}, {Desai}, {Desiati}, {de Vries}, {de Wasseige}, {de With},
  {DeYoung}, {Dharani}, {Diaz}, {D{\'\i}az-V{\'e}lez}, {Dujmovic}, {Dunkman},
  {DuVernois}, {Dvorak}, {Ehrhardt}, {Eller}, {Engel}, {Evenson}, {Fahey},
  {Fazely}, {Felde}, {Fienberg}, {Filimonov}, {Finley}, {Fischer}, {Fox},
  {Franckowiak}, {Friedman}, {Fritz}, {Gaisser}, {Gallagher}, {Ganster},
  {Garrappa}, {Gerhardt}, {Ghadimi}, {Glauch}, {Gl{\"u}senkamp}, {Goldschmidt},
  {Gonzalez}, {Goswami}, {Grant}, {Gr{\'e}goire}, {Griffith}, {Griswold},
  {G{\"u}nd{\"u}z}, {Haack}, {Hallgren}, {Halliday}, {Halve}, {Halzen}, {Minh},
  {Hanson}, {Hardin}, {Haungs}, {Hauser}, {Hebecker}, {Heix}, {Helbing},
  {Hellauer}, {Henningsen}, {Hickford}, {Hignight}, {Hill}, {Hill}, {Hoffman},
  {Hoffmann}, {Hoinka}, {Hokanson-Fasig}, {Hoshina}, {Huang}, {Huber}, {Huber},
  {Hultqvist}, {H{\"u}nnefeld}, {Hussain}, {In}, {Iovine}, {Ishihara},
  {Jansson}, {Japaridze}, {Jeong}, {Jones}, {Jonske}, {Joppe}, {Kang}, {Kang},
  {Kang}, {Kappes}, {Kappesser}, {Karg}, {Karl}, {Karle}, {Katz}, {Kauer},
  {Kellermann}, {Kelley}, {Kheirandish}, {Kim}, {Kin}, {Kintscher}, {Kiryluk},
  {Kittler}, {Klein}, {Koirala}, {Kolanoski}, {K{\"o}pke}, {Kopper}, {Kopper},
  {Koskinen}, {Koundal}, {Kovacevich}, {Kowalski}, {Krings}, {Kr{\"u}ckl},
  {Kulacz}, {Kurahashi}, {Kyriacou}, {Lagunas Gualda}, {Lanfranchi}, {Larson},
  {Lauber}, {Lazar}, {Leonard}, {Leszczy{\'n}ska}, {Li}, {Liu}, {Lohfink},
  {Lozano Mariscal}, {Lu}, {Lucarelli}, {Ludwig}, {L{\"u}nemann}, {Luszczak},
  {Lyu}, {Ma}, {Madsen}, {Maggi}, {Mahn}, {Makino}, {Mallik}, {Mancina},
  {Mari{\c{s}}}, {Maruyama}, {Mase}, {Maunu}, {McNally}, {Meagher}, {Medina},
  {Meier}, {Meighen-Berger}, {Merz}, {Micallef}, {Mockler}, {Moment{\'e}},
  {Montaruli}, {Moore}, {Morse}, {Moulai}, {Muth}, {Naab}, {Nagai}, {Naumann},
  {Necker}, {Neer}, {Nguy{\^e}n}, {Niederhausen}, {Nisa}, {Nowicki}, {Nygren},
  {Obertacke Pollmann}, {Oehler}, {Olivas}, {O'Sullivan}, {Pandya}, {Pankova},
  {Park}, {Parker}, {Paudel}, {Peiffer}, {P{\'e}rez de los Heros}, {Philippen},
  {Pieloth}, {Pieper}, {Pizzuto}, {Plum}, {Popovych}, {Porcelli}, {Prado
  Rodriguez}, {Price}, {Przybylski}, {Raab}, {Raissi}, {Rameez}, {Rawlins},
  {Rea}, {Rehman}, {Reimann}, {Renschler}, {Renzi}, {Resconi}, {Reusch},
  {Rhode}, {Richman}, {Riedel}, {Robertson}, {Roellinghoff}, {Rongen}, {Rott},
  {Ruhe}, {Ryckbosch}, {Rysewyk Cantu}, {Safa}, {Sanchez Herrera}, {Sandrock},
  {Sandroos}, {Santander}, {Sarkar}, {Sarkar}, {Satalecka}, {Scharf},
  {Schaufel}, {Schieler}, {Schlunder}, {Schmidt}, {Schneider}, {Schneider},
  {Schr{\"o}der}, {Schumacher}, {Sclafani}, {Seckel}, {Seunarine}, {Shefali},
  {Silva}, {Smithers}, {Snihur}, {Soedingrekso}, {Soldin}, {Song}, {Spiczak},
  {Spiering}, {Stachurska}, {Stamatikos}, {Stanev}, {Stein}, {Stettner},
  {Steuer}, {Stezelberger}, {Stokstad}, {Strotjohann}, {St{\"u}rwald},
  {Stuttard}, {Sullivan}, {Taboada}, {Tenholt}, {Ter-Antonyan}, {Tilav},
  {Tollefson}, {Tomankova}, {T{\"o}nnis}, {Toscano}, {Tosi}, {Trettin},
  {Tselengidou}, {Tung}, {Turcati}, {Turcotte}, {Turley}, {Twagirayezu}, {Ty},
  {Unger}, {Unland Elorrieta}, {Vandenbroucke}, {van Eijk}, {van Eijndhoven},
  {Vannerom}, {van Santen}, {Verpoest}, {Vraeghe}, {Walck}, {Wallace},
  {Watson}, {Weaver}, {Weindl}, {Weiss}, {Weldert}, {Wendt}, {Werthebach},
  {Whelan}, {Whitehorn}, {Wiebe}, {Wiebusch}, {Williams}, {Wolf}, {Wood},
  {Woschnagg}, {Wrede}, {Wulff}, {Xu}, {Xu}, {Yanez}, {Yoshida}, {Yuan},
  {Zhang}, \& {Z{\"o}cklein}}]{IC2021_history_ref}
{IceCube Collaboration}, {Abbasi}, R., {Ackermann}, M., {et~al.} 2021, arXiv
  e-prints, arXiv:2101.09836

\bibitem[{{Ivezi{\'c}} {et~al.}(2019){Ivezi{\'c}}, {Kahn}, {Tyson}, {Abel},
  {Acosta}, {Allsman}, {Alonso}, {AlSayyad}, {Anderson}, {Andrew}, {Angel},
  {Angeli}, {Ansari}, {Antilogus}, {Araujo}, {Armstrong}, {Arndt}, {Astier},
  {Aubourg}, {Auza}, {Axelrod}, {Bard}, {Barr}, {Barrau}, {Bartlett}, {Bauer},
  {Bauman}, {Baumont}, {Bechtol}, {Bechtol}, {Becker}, {Becla}, {Beldica},
  {Bellavia}, {Bianco}, {Biswas}, {Blanc}, {Blazek}, {Blandford}, {Bloom},
  {Bogart}, {Bond}, {Booth}, {Borgland}, {Borne}, {Bosch}, {Boutigny},
  {Brackett}, {Bradshaw}, {Brandt}, {Brown}, {Bullock}, {Burchat}, {Burke},
  {Cagnoli}, {Calabrese}, {Callahan}, {Callen}, {Carlin}, {Carlson},
  {Chandrasekharan}, {Charles-Emerson}, {Chesley}, {Cheu}, {Chiang}, {Chiang},
  {Chirino}, {Chow}, {Ciardi}, {Claver}, {Cohen-Tanugi}, {Cockrum}, {Coles},
  {Connolly}, {Cook}, {Cooray}, {Covey}, {Cribbs}, {Cui}, {Cutri}, {Daly},
  {Daniel}, {Daruich}, {Daubard}, {Daues}, {Dawson}, {Delgado}, {Dellapenna},
  {de Peyster}, {de Val-Borro}, {Digel}, {Doherty}, {Dubois},
  {Dubois-Felsmann}, {Durech}, {Economou}, {Eifler}, {Eracleous}, {Emmons},
  {Fausti Neto}, {Ferguson}, {Figueroa}, {Fisher-Levine}, {Focke}, {Foss},
  {Frank}, {Freemon}, {Gangler}, {Gawiser}, {Geary}, {Gee}, {Geha}, {Gessner},
  {Gibson}, {Gilmore}, {Glanzman}, {Glick}, {Goldina}, {Goldstein}, {Goodenow},
  {Graham}, {Gressler}, {Gris}, {Guy}, {Guyonnet}, {Haller}, {Harris},
  {Hascall}, {Haupt}, {Hernandez}, {Herrmann}, {Hileman}, {Hoblitt}, {Hodgson},
  {Hogan}, {Howard}, {Huang}, {Huffer}, {Ingraham}, {Innes}, {Jacoby}, {Jain},
  {Jammes}, {Jee}, {Jenness}, {Jernigan}, {Jevremovi{\'c}}, {Johns}, {Johnson},
  {Johnson}, {Jones}, {Juramy-Gilles}, {Juri{\'c}}, {Kalirai}, {Kallivayalil},
  {Kalmbach}, {Kantor}, {Karst}, {Kasliwal}, {Kelly}, {Kessler}, {Kinnison},
  {Kirkby}, {Knox}, {Kotov}, {Krabbendam}, {Krughoff}, {Kub{\'a}nek},
  {Kuczewski}, {Kulkarni}, {Ku}, {Kurita}, {Lage}, {Lambert}, {Lange},
  {Langton}, {Le Guillou}, {Levine}, {Liang}, {Lim}, {Lintott}, {Long},
  {Lopez}, {Lotz}, {Lupton}, {Lust}, {MacArthur}, {Mahabal}, {Mandelbaum},
  {Markiewicz}, {Marsh}, {Marshall}, {Marshall}, {May}, {McKercher}, {McQueen},
  {Meyers}, {Migliore}, {Miller}, \& {Mills}}]{Ivezic2019_LSST_sci_cases}
{Ivezi{\'c}}, {\v{Z}}., {Kahn}, S.~M., {Tyson}, J.~A., {et~al.} 2019, \apj,
  873, 111

\bibitem[{{Kadler} {et~al.}(2016){Kadler}, {Krau{\ss}}, {Mannheim}, {Ojha},
  {M{\"u}ller}, {Schulz}, {Anton}, {Baumgartner}, {Beuchert}, {Buson},
  {Carpenter}, {Eberl}, {Edwards}, {Eisenacher Glawion}, {Els{\"a}sser},
  {Gehrels}, {Gr{\"a}fe}, {Gulyaev}, {Hase}, {Horiuchi}, {James}, {Kappes},
  {Kappes}, {Katz}, {Kreikenbohm}, {Kreter}, {Kreykenbohm}, {Langejahn},
  {Leiter}, {Litzinger}, {Longo}, {Lovell}, {McEnery}, {Natusch}, {Phillips},
  {Pl{\"o}tz}, {Quick}, {Ros}, {Stecker}, {Steinbring}, {Stevens}, {Thompson},
  {Tr{\"u}stedt}, {Tzioumis}, {Weston}, {Wilms}, \& {Zensus}}]{Kadler2016}
{Kadler}, M., {Krau{\ss}}, F., {Mannheim}, K., {et~al.} 2016, Nature Physics,
  12, 807

\bibitem[{{Keivani} {et~al.}(2018){Keivani}, {Murase}, {Petropoulou}, {Fox},
  {Cenko}, {Chaty}, {Coleiro}, {DeLaunay}, {Dimitrakoudis}, {Evans}, {Kennea},
  {Marshall}, {Mastichiadis}, {Osborne}, {Santander}, {Tohuvavohu}, \&
  {Turley}}]{Keivani2018_TXS0506}
{Keivani}, A., {Murase}, K., {Petropoulou}, M., {et~al.} 2018, \apj, 864, 84

\bibitem[{{KM3NeT Collaboration} {et~al.}(2025{\natexlab{a}}){KM3NeT
  Collaboration}, {MessMapp Group}, {Fermi-LAT Collaboration}, {Owens Valley
  Radio Observatory 40-m Telescope Group}, {SVOM Collaboration}, {Baldini},
  {Buchner}, {Erkenov}, {Globus}, {Merloni}, {Paggi}, {Popkov}, {Porquet},
  {Salvato}, {Sotnikova}, \& {Voitsik}}]{KM3NeT2025_230213A_source_search}
{KM3NeT Collaboration}, {MessMapp Group}, {Fermi-LAT Collaboration}, {et~al.}
  2025{\natexlab{a}}, arXiv e-prints, arXiv:2502.08484

\bibitem[{{KM3NeT Collaboration} {et~al.}(2025{\natexlab{b}}){KM3NeT
  Collaboration}, {Albert}, {Alhebsi}, {Alshamsi}, {Alves Garre}, {Ambrosone},
  {Ameli}, {Andre}, {Anghinolfi}, {Aphecetche}, {Ardid}, {Ardid},
  {Arg{\"u}elles}, {Atmani}, {Aublin}, {Badaracco}, {Bailly-Salins},
  {Barda{\v{c}}ov{\'a}}, {Baret}, {Bariego-Quintana}, {Becherini}, {Bendahman},
  {Benfenati Gualandi}, {Benhassi}, {Bennani}, {Benoit}, {Berbee}, {Bertin},
  {Biagi}, {Boettcher}, {Bonanno}, {Bouasla}, {Boumaaza}, {Bouta}, {Bouwhuis},
  {Bozza}, {Bozza}, {Br{\^a}nza{\c{s}}}, {Bretaudeau}, {Breuhaus}, {Bruijn},
  {Brunner}, {Bruno}, {Buis}, {Buompane}, {Buson}, {Busto}, {Caiffi}, {Calvo},
  {Capone}, {Carenini}, {Carretero}, {Cartraud}, {Castaldi}, {Cecchini},
  {Celli}, {Cerisy}, {Chabab}, {Chen}, {Cherubini}, {Chiarusi}, {Circella},
  {Cocimano}, {Coelho}, {Coleiro}, {Colonges}, {Condorelli}, {Coniglione},
  {Coyle}, {Creusot}, {Cuttone}, {D'Amico}, {Dallier}, {De Benedittis}, {De
  Martino}, {De Wasseige}, {Decoene}, {Del Rosso}, {Di Mauro}, {Di Palma},
  {Diaz}, {Diego-Tortosa}, {Distefano}, {Domi}, {Donzaud}, {Dornic},
  {Drakopoulou}, {Drouhin}, {Ducoin}, {Dvornick{\'y}}, {Eberl}, {Eckerov{\'a}},
  {Eddymaoui}, {van Eeden}, {Eff}, {van Eijk}, {El Bojaddaini}, {El Hedri},
  {Ellajosyula}, {Enzenh{\"o}fer}, {Ferrara}, {Filipovi{\'c}}, {Filippini},
  {Franciotti}, {Fusco}, {Gagliardini}, {Gal}, {Garc{\'\i}a M{\'e}ndez},
  {Garcia Soto}, {Gatius Oliver}, {Gei{\ss}elbrecht}, {Genton}, {Ghaddari},
  {Gialanella}, {Gibson}, {Giorgio}, {Goos}, {Goswami}, {Gozzini}, {Gracia},
  {Graf}, {Guidi}, {Guillon}, {Guti{\'e}rrez}, {Haack}, {van Haren},
  {Heijboer}, {Hennig}, {Henry}, {Hern{\'a}ndez-Rey}, {Idrissi Ibnsalih},
  {Ilioni}, {Illuminati}, {Joly}, {de Jong}, {de Jong}, {Jung},
  {Kalaczy{\'n}ski}, {Kalekin}, {Kamp}, {Katz}, {Kistauri}, {Kopper},
  {Kouchner}, {Kovalev}, {Kueviakoe}, {Kulikovskiy}, {Kvatadze}, {Labalme},
  {Lahmann}, {Lamoureux}, {Lancelin}, {Larosa}, {Lastoria}, {Lazar}, {Lazo},
  {Le Stum}, {Lehaut}, {Lemaitre}, {Leonora}, {Lessing}, {Levi}, {Lincetto},
  {Lindsey Clark}, {Longhitano}, {Lumb}, {Magnani}, {Majumdar}, {Malerba},
  {Mamedov}, {Manfreda}, {Marconi}, {Margiotta}, {Marinelli}, {Markou},
  {Martin}, {Marzaioli}, {Mastrodicasa}, {Mastroianni}, {Mauro}, {Miele},
  {Migliozzi}, {Migneco}, {Mitsou}, {Mollo}, {Mongelli}, {Morales-Gallegos},
  {Moussa}, {Mozun Mateo}, {Muller}, {Musone}, {Musumeci}, {Navas},
  {Nayerhoda}, {Nicolau}, {Nkosi}, {{\'O} Fearraigh}, {Oliviero}, {Orlando}, \&
  {Oukacha}}]{KM3NeT2025_100PeV_detection}
{KM3NeT Collaboration}, S., A., {Albert}, A., {Alhebsi}, A.~R., {et~al.}
  2025{\natexlab{b}}, \nat, 638, 376

\bibitem[{{Kouch} {et~al.}(2025{\natexlab{a}}){Kouch}, {Lindfors}, {Hovatta},
  {Liodakis}, {Koljonen}, {Nilsson}, {Jormanainen}, {Fallah Ramazani}, \&
  {Graham}}]{Kouch2025_WH2}
{Kouch}, P.~M., {Lindfors}, E., {Hovatta}, T., {et~al.} 2025{\natexlab{a}},
  \aap, 696, A73

\bibitem[{{Kouch} {et~al.}(2024{\natexlab{a}}){Kouch}, {Lindfors}, {Hovatta},
  {Liodakis}, {Koljonen}, {Nilsson}, {Kiehlmann}, {Max-Moerbeck}, {Readhead},
  {Reeves}, {Pearson}, {Jormanainen}, {Ramazani}, \&
  {Graham}}]{Kouch2024_CGRaBSvIC}
{Kouch}, P.~M., {Lindfors}, E., {Hovatta}, T., {et~al.} 2024{\natexlab{a}},
  \aap, 690, A111

\bibitem[{{Kouch} {et~al.}(2025{\natexlab{b}}){Kouch}, {Lindfors}, {Hovatta},
  {Liodakis}, {Koljonen}, {Paggi}, {Nilsson}, {Jormanainen}, {Fallah Ramazani},
  {Kankkunen}, {Wierda}, {Wagner}, \& {Graham}}]{Kouch2025_CAZ_catalog}
{Kouch}, P.~M., {Lindfors}, E., {Hovatta}, T., {et~al.} 2025{\natexlab{b}},
  arXiv e-prints, arXiv:2510.16584

\bibitem[{{Kouch} {et~al.}(2025{\natexlab{c}}){Kouch}, {Liodakis}, {Fenu},
  {Zhang}, {Boula}, {Middei}, {Di Gesu}, {Paraschos}, {Agudo}, {Jorstad},
  {Lindfors}, {Marscher}, {Krawczynski}, {Negro}, {Hu}, {Kim}, {Cavazzuti},
  {Errando}, {Blinov}, {Gourni}, {Kiehlmann}, {Kourtidis}, {Mandarakas},
  {Triantafyllou}, {Vervelaki}, {Borman}, {Kopatskaya}, {Larionova},
  {Morozova}, {Savchenko}, {Vasilyev}, {Troitskiy}, {Grishina}, {Shishkina},
  {Zhovtan}, {Aceituno}, {Bonnoli}, {Casanova}, {Escudero},
  {Ag{\'\i}s-Gonz{\'a}lez}, {Husillos}, {Otero-Santos}, {Piirola}, {Sota},
  {Myserlis}, {Gurwell}, {Keating}, {Rao}, {Angelakis}, {Kraus}, {Antonelli},
  {Bachetti}, {Baldini}, {Baumgartner}, {Bellazzini}, {Bianchi}, {Bongiorno},
  {Bonino}, {Brez}, {Bucciantini}, {Capitanio}, {Castellano}, {Chen},
  {Ciprini}, {Costa}, {De Rosa}, {Del Monte}, {Di Lalla}, {Di Marco},
  {Donnarumma}, {Doroshenko}, {Dov{\v{c}}iak}, {Ehlert}, {Enoto},
  {Evangelista}, {Fabiani}, {Ferrazzoli}, {Garcia}, {Gunji}, {Hayashida},
  {Heyl}, {Iwakiri}, {Kaaret}, {Karas}, {Kislat}, {Kitaguchi}, {Kolodziejczak},
  {La Monaca}, {Latronico}, {Maldera}, {Manfreda}, {Marin}, {Marinucci},
  {Marshall}, {Massaro}, {Matt}, {Mitsuishi}, {Mizuno}, {Muleri}, {Ng},
  {O'Dell}, {Omodei}, {Oppedisano}, {Papitto}, {Pavlov}, {Peirson}, {Perri},
  {Pesce-Rollins}, {Petrucci}, {Pilia}, {Possenti}, {Poutanen}, {Puccetti},
  {Ramsey}, {Rankin}, {Ratheesh}, {Roberts}, {Sgr{\`o}}, {Slane}, {Soffitta},
  {Spandre}, {Swartz}, {Tamagawa}, {Tavecchio}, {Taverna}, {Tawara}, {Tennant},
  {Thomas}, {Tombesi}, {Trois}, {Tsygankov}, {Turolla}, {Romani}, {Vink},
  {Weisskopf}, {Wu}, {Xie}, \& {Zane}}]{Kouch2025_IXPE_0954}
{Kouch}, P.~M., {Liodakis}, I., {Fenu}, F., {et~al.} 2025{\natexlab{c}}, \aap,
  695, A99

\bibitem[{{Kouch} {et~al.}(2024{\natexlab{b}}){Kouch}, {Liodakis}, {Middei},
  {Kim}, {Tavecchio}, {Marscher}, {Marshall}, {Ehlert}, {Di Gesu}, {Jorstad},
  {Agudo}, {Madejski}, {Romani}, {Errando}, {Lindfors}, {Nilsson}, {Toppari},
  {Potter}, {Imazawa}, {Sasada}, {Fukazawa}, {Kawabata}, {Uemura}, {Mizuno},
  {Nakaoka}, {Akitaya}, {McCall}, {Jermak}, {Steele}, {Myserlis}, {Gurwell},
  {Keating}, {Rao}, {Kang}, {Lee}, {Kim}, {Cheong}, {Jeong}, {Angelakis},
  {Kraus}, {Aceituno}, {Bonnoli}, {Casanova}, {Escudero},
  {Ag{\'\i}s-Gonz{\'a}lez}, {Husillos}, {Morcuende}, {Otero-Santos}, {Sota},
  {Bachev}, {Antonelli}, {Bachetti}, {Baldini}, {Baumgartner}, {Bellazzini},
  {Bianchi}, {Bongiorno}, {Bonino}, {Brez}, {Bucciantini}, {Capitanio},
  {Castellano}, {Cavazzuti}, {Chen}, {Ciprini}, {Costa}, {De Rosa}, {Del
  Monte}, {Di Lalla}, {Di Marco}, {Donnarumma}, {Doroshenko}, {Dov{\v{c}}iak},
  {Enoto}, {Evangelista}, {Fabiani}, {Ferrazzoli}, {Garcia}, {Gunji},
  {Hayashida}, {Heyl}, {Iwakiri}, {Kaaret}, {Karas}, {Kislat}, {Kitaguchi},
  {Kolodziejczak}, {Krawczynski}, {La Monaca}, {Latronico}, {Maldera},
  {Manfreda}, {Marin}, {Marinucci}, {Massaro}, {Matt}, {Mitsuishi}, {Muleri},
  {Negro}, {Ng}, {O'Dell}, {Omodei}, {Oppedisano}, {Papitto}, {Pavlov},
  {Peirson}, {Perri}, {Pesce-Rollins}, {Petrucci}, {Pilia}, {Possenti},
  {Poutanen}, {Puccetti}, {Ramsey}, {Rankin}, {Ratheesh}, {Roberts},
  {Sgr{\`o}}, {Slane}, {Soffitta}, {Spandre}, {Swartz}, {Tamagawa}, {Taverna},
  {Tawara}, {Tennant}, {Thomas}, {Tombesi}, {Trois}, {Tsygankov}, {Turolla},
  {Vink}, {Weisskopf}, {Wu}, {Xie}, \& {Zane}}]{Kouch2024_IXPE_PKS2155}
{Kouch}, P.~M., {Liodakis}, I., {Middei}, R., {et~al.} 2024{\natexlab{b}},
  \aap, 689, A119

\bibitem[{{Kovalev} {et~al.}(2007){Kovalev}, {Petrov}, {Fomalont}, \&
  {Gordon}}]{kovalev2007_vcs5}
{Kovalev}, Y.~Y., {Petrov}, L., {Fomalont}, E.~B., \& {Gordon}, D. 2007, \aj,
  133, 1236

\bibitem[{{Kovalev} {et~al.}(2025){Kovalev}, {Pushkarev}, {G{\'o}mez}, {Homan},
  {Lister}, {Livingston}, {Pashchenko}, {Plavin}, {Savolainen}, \&
  {Troitsky}}]{Kovalev2025_PKS1424_eye_of_sauran}
{Kovalev}, Y.~Y., {Pushkarev}, A.~B., {G{\'o}mez}, J.~L., {et~al.} 2025, \aap,
  700, L12

\bibitem[{{Krau{\ss}} {et~al.}(2014){Krau{\ss}}, {Kadler}, {Mannheim},
  {Schulz}, {Tr{\"u}stedt}, {Wilms}, {Ojha}, {Ros}, {Anton}, {Baumgartner},
  {Beuchert}, {Blanchard}, {B{\"u}rkel}, {Carpenter}, {Eberl}, {Edwards},
  {Eisenacher}, {Els{\"a}sser}, {Fehn}, {Fritsch}, {Gehrels}, {Gr{\"a}fe},
  {Gro{\ss}berger}, {Hase}, {Horiuchi}, {James}, {Kappes}, {Katz},
  {Kreikenbohm}, {Kreykenbohm}, {Langejahn}, {Leiter}, {Litzinger}, {Lovell},
  {M{\"u}ller}, {Phillips}, {Pl{\"o}tz}, {Quick}, {Steinbring}, {Stevens},
  {Thompson}, \& {Tzioumis}}]{Krauss2014}
{Krau{\ss}}, F., {Kadler}, M., {Mannheim}, K., {et~al.} 2014, \aap, 566, L7

\bibitem[{{Kreter} {et~al.}(2020){Kreter}, {Kadler}, {Krau{\ss}}, {Mannheim},
  {Buson}, {Ojha}, {Wilms}, \& {B{\"o}ttcher}}]{Kreter2020_fermi_flare_v_neut}
{Kreter}, M., {Kadler}, M., {Krau{\ss}}, F., {et~al.} 2020, \apj, 902, 133

\bibitem[{{Kun} {et~al.}(2022){Kun}, {Bartos}, {Becker Tjus}, {Biermann},
  {Franckowiak}, \& {Halzen}}]{Kun2022}
{Kun}, E., {Bartos}, I., {Becker Tjus}, J., {et~al.} 2022, \apj, 934, 180

\bibitem[{{Kun} {et~al.}(2023){Kun}, {Bartos}, {Becker Tjus}, {Biermann},
  {Franckowiak}, {Halzen}, \& {Mez{\H{o}}}}]{Kun2023_gamma_suppression}
{Kun}, E., {Bartos}, I., {Becker Tjus}, J., {et~al.} 2023, \aap, 679, A46

\bibitem[{{Kun} {et~al.}(2024){Kun}, {Bartos}, {Tjus}, {Biermann},
  {Franckowiak}, {Halzen}, {del Palacio}, \& {Woo}}]{Kun2024}
{Kun}, E., {Bartos}, I., {Tjus}, J.~B., {et~al.} 2024, \prd, 110, 123014

\bibitem[{{L{\"a}hteenm{\"a}ki} \&
  {Valtaoja}(2003)}]{Lahteenmaki2003_radio_gamma_correlation}
{L{\"a}hteenm{\"a}ki}, A. \& {Valtaoja}, E. 2003, \apj, 590, 95

\bibitem[{{Le{\'o}n-Tavares} {et~al.}(2011){Le{\'o}n-Tavares}, {Valtaoja},
  {Tornikoski}, {L{\"a}hteenm{\"a}ki}, \&
  {Nieppola}}]{LeonTavares2011_radio_gamma_correlation}
{Le{\'o}n-Tavares}, J., {Valtaoja}, E., {Tornikoski}, M.,
  {L{\"a}hteenm{\"a}ki}, A., \& {Nieppola}, E. 2011, \aap, 532, A146

\bibitem[{{Liodakis} {et~al.}(2022{\natexlab{a}}){Liodakis}, {Hovatta},
  {Pavlidou}, {Readhead}, {Blandford}, {Kiehlmann}, {Lindfors}, {Max-Moerbeck},
  {Pearson}, \& {Petropoulou}}]{Liodakis2022_WH}
{Liodakis}, I., {Hovatta}, T., {Pavlidou}, V., {et~al.} 2022{\natexlab{a}},
  \aap, 666, A36

\bibitem[{{Liodakis} {et~al.}(2024){Liodakis}, {Kiehlmann}, {Marscher},
  {Zhang}, {Blinov}, {Jorstad}, {Agudo}, {Ben{\'\i}tez}, {Berdyugin},
  {Bonnoli}, {Casadio}, {Chen}, {Chen}, {Ehlert}, {Escudero}, {Grishina},
  {Hiriart}, {Hsu}, {Imazawa}, {Jermak}, {Jose}, {Kaaret}, {Kopatskaya},
  {Lalchand}, {Larionova}, {Lindfors}, {L{\'o}pez}, {McCall}, {Morozova},
  {Palaiologou}, {Pandey}, {Poutanen}, {Rakshit}, {Reig}, {Sasada},
  {Savchenko}, {Shablovinskaya}, {Neha}, {Shrestha}, {Steele}, {Troitskiy},
  {Troitskaya}, {Uemura}, {Vasilyev}, {Weaver}, {Wiersema}, \&
  {Weisskopf}}]{Liodakis2024_J0211_fast_pol_variations}
{Liodakis}, I., {Kiehlmann}, S., {Marscher}, A.~P., {et~al.} 2024, \aap, 689,
  A200

\bibitem[{{Liodakis} {et~al.}(2022{\natexlab{b}}){Liodakis}, {Marscher},
  {Agudo}, {Berdyugin}, {Bernardos}, {Bonnoli}, {Borman}, {Casadio},
  {Casanova}, {Cavazzuti}, {Rodriguez Cavero}, {Di Gesu}, {Di Lalla},
  {Donnarumma}, {Ehlert}, {Errando}, {Escudero}, {Garc{\'\i}a-Comas},
  {Ag{\'\i}s-Gonz{\'a}lez}, {Husillos}, {Jormanainen}, {Jorstad}, {Kagitani},
  {Kopatskaya}, {Kravtsov}, {Krawczynski}, {Lindfors}, {Larionova}, {Madejski},
  {Marin}, {Marchini}, {Marshall}, {Morozova}, {Massaro}, {Masiero}, {Mawet},
  {Middei}, {Millar-Blanchaer}, {Myserlis}, {Negro}, {Nilsson}, {O'Dell},
  {Omodei}, {Pacciani}, {Paggi}, {Panopoulou}, {Peirson}, {Perri}, {Petrucci},
  {Poutanen}, {Puccetti}, {Romani}, {Sakanoi}, {Savchenko}, {Sota},
  {Tavecchio}, {Tinyanont}, {Vasilyev}, {Weaver}, {Zhovtan}, {Antonelli},
  {Bachetti}, {Baldini}, {Baumgartner}, {Bellazzini}, {Bianchi}, {Bongiorno},
  {Bonino}, {Brez}, {Bucciantini}, {Capitanio}, {Castellano}, {Ciprini},
  {Costa}, {De Rosa}, {Del Monte}, {Di Marco}, {Doroshenko}, {Dov{\v{c}}iak},
  {Enoto}, {Evangelista}, {Fabiani}, {Ferrazzoli}, {Garcia}, {Gunji},
  {Hayashida}, {Heyl}, {Iwakiri}, {Karas}, {Kitaguchi}, {Kolodziejczak}, {La
  Monaca}, {Latronico}, {Maldera}, {Manfreda}, {Marinucci}, {Matt},
  {Mitsuishi}, {Mizuno}, {Muleri}, {Ng}, {Oppedisano}, {Papitto}, {Pavlov},
  {Pesce-Rollins}, {Pilia}, {Possenti}, {Ramsey}, {Rankin}, {Ratheesh},
  {Sgr{\'o}}, {Slane}, {Soffitta}, {Spandre}, {Tamagawa}, {Taverna}, {Tawara},
  {Tennant}, {Thomas}, {Tombesi}, {Trois}, {Tsygankov}, {Turolla}, {Vink},
  {Weisskopf}, {Wu}, {Xie}, \& {Zane}}]{Liodakis2022_IXPE_Nature}
{Liodakis}, I., {Marscher}, A.~P., {Agudo}, I., {et~al.} 2022{\natexlab{b}},
  \nat, 611, 677

\bibitem[{{Liodakis} {et~al.}(2019){Liodakis}, {Romani}, {Filippenko},
  {Kocevski}, \& {Zheng}}]{Liodakis2019_optical_gamma_flare_correlation}
{Liodakis}, I., {Romani}, R.~W., {Filippenko}, A.~V., {Kocevski}, D., \&
  {Zheng}, W. 2019, \apj, 880, 32

\bibitem[{{Lu} {et~al.}(2025){Lu}, {Liang}, {Ouyang}, {Li}, \& {Wang}}]{Lu2024}
{Lu}, M.-X., {Liang}, Y.-F., {Ouyang}, X., {Li}, R.-L., \& {Wang}, X.-G. 2025,
  \prd, 112, 103013

\bibitem[{{Malecki}(2024)}]{Malecki2024_P_ONE}
{Malecki}, P. 2024, Universe, 10, 53

\bibitem[{{Mannheim}(1993)}]{Mannheim1993_hot_hadrons_in_jets}
{Mannheim}, K. 1993, \aap, 269, 67

\bibitem[{{Mannheim} \& {Biermann}(1989)}]{Mannheim1989}
{Mannheim}, K. \& {Biermann}, P.~L. 1989, \aap, 221, 211

\bibitem[{{Mastichiadis} \&
  {Petropoulou}(2021)}]{Mastichiadis2021_Xray_flare_in_pp}
{Mastichiadis}, A. \& {Petropoulou}, M. 2021, \apj, 906, 131

\bibitem[{{Meyer} {et~al.}(2019){Meyer}, {Scargle}, \&
  {Blandford}}]{Meyer2019_BBHOP}
{Meyer}, M., {Scargle}, J.~D., \& {Blandford}, R.~D. 2019, \apj, 877, 39

\bibitem[{{Moretti} \& {Caccianiga}(2025)}]{Moretti2025_FSRQ_corr_w_neut}
{Moretti}, A. \& {Caccianiga}, A. 2025, \aap, 704, A184

\bibitem[{{M{\"u}cke} \& {Protheroe}(2001)}]{Mucke2001}
{M{\"u}cke}, A. \& {Protheroe}, R.~J. 2001, Astroparticle Physics, 15, 121

\bibitem[{{Muxlow} {et~al.}(1996){Muxlow}, {Pedlar}, {Holloway}, {Gallimore},
  \& {Antonucci}}]{Muxlow1996_NGC1068_jet}
{Muxlow}, T.~W.~B., {Pedlar}, A., {Holloway}, A.~J., {Gallimore}, J.~F., \&
  {Antonucci}, R.~R.~J. 1996, \mnras, 278, 854

\bibitem[{{Nilsson} {et~al.}(2018){Nilsson}, {Lindfors}, {Takalo}, {Reinthal},
  {Berdyugin}, {Sillanp{\"a}{\"a}}, {Ciprini}, {Halkola}, {Hein{\"a}m{\"a}ki},
  {Hovatta}, {Kadenius}, {Nurmi}, {Ostorero}, {Pasanen}, {Rekola}, {Saarinen},
  {Sainio}, {Tuominen}, {Villforth}, {Vornanen}, \& {Zaprudin}}]{nilsson2018}
{Nilsson}, K., {Lindfors}, E., {Takalo}, L.~O., {et~al.} 2018, \aap, 620, A185

\bibitem[{{Novikova} {et~al.}(2023){Novikova}, {Shishkina}, \&
  {Blinov}}]{Novikova2023}
{Novikova}, P., {Shishkina}, E., \& {Blinov}, D. 2023, \mnras, 526, 347

\bibitem[{{Oikonomou}(2022)}]{Oikonomou2022_frac_of_neutrinos_by_blazars}
{Oikonomou}, F. 2022, in 37th International Cosmic Ray Conference, 30

\bibitem[{{Oikonomou} {et~al.}(2019){Oikonomou}, {Murase}, {Padovani},
  {Resconi}, \& {M{\'e}sz{\'a}ros}}]{Oikonomou2019_indepth_blz_neut_connection}
{Oikonomou}, F., {Murase}, K., {Padovani}, P., {Resconi}, E., \&
  {M{\'e}sz{\'a}ros}, P. 2019, \mnras, 489, 4347

\bibitem[{{Omeliukh} {et~al.}(2025){Omeliukh}, {Garrappa}, {Fallah Ramazani},
  {Franckowiak}, {Winter}, {Lindfors}, {Nilsson}, {Jormanainen}, {Wierda},
  {Filippenko}, {Zheng}, {Tornikoski}, {L{\"a}hteenm{\"a}ki}, {Kankkunen}, \&
  {Tammi}}]{Omeliukh2025_PKS0735_leptohadronic_modeling}
{Omeliukh}, A., {Garrappa}, S., {Fallah Ramazani}, V., {et~al.} 2025, \aap,
  695, A266

\bibitem[{{Padovani} {et~al.}(2022){Padovani}, {Boccardi}, {Falomo}, \&
  {Giommi}}]{Padovani2022_PKS1424_masqBLL}
{Padovani}, P., {Boccardi}, B., {Falomo}, R., \& {Giommi}, P. 2022, \mnras,
  511, 4697

\bibitem[{{Padovani} {et~al.}(2024{\natexlab{a}}){Padovani}, {Gilli},
  {Resconi}, {Bellenghi}, \& {Henningsen}}]{Padovani2024}
{Padovani}, P., {Gilli}, R., {Resconi}, E., {Bellenghi}, C., \& {Henningsen},
  F. 2024{\natexlab{a}}, \aap, 684, L21

\bibitem[{{Padovani} {et~al.}(2019){Padovani}, {Oikonomou}, {Petropoulou},
  {Giommi}, \& {Resconi}}]{Padovani2019_TXS0506_masqBLL}
{Padovani}, P., {Oikonomou}, F., {Petropoulou}, M., {Giommi}, P., \& {Resconi},
  E. 2019, \mnras, 484, L104

\bibitem[{{Padovani} {et~al.}(2024{\natexlab{b}}){Padovani}, {Resconi},
  {Ajello}, {Bellenghi}, {Bianchi}, {Blasi}, {Huang}, {Gabici}, {G{\'a}mez
  Rosas}, {Niederhausen}, {Peretti}, {Eichmann}, {Guetta}, {Lamastra}, \&
  {Shimizu}}]{Padovani2024_NGC1068}
{Padovani}, P., {Resconi}, E., {Ajello}, M., {et~al.} 2024{\natexlab{b}},
  Nature Astronomy, 8, 1077

\bibitem[{{Paraschos} {et~al.}(2025){Paraschos}, {Traianou}, {Debbrecht},
  {Liodakis}, \& {Ros}}]{Paraschos2025_neutrino_v_MWL_polarization}
{Paraschos}, G.~F., {Traianou}, E., {Debbrecht}, L.~C., {Liodakis}, I., \&
  {Ros}, E. 2025, \apj, 989, 208

\bibitem[{{Peirson} {et~al.}(2022){Peirson}, {Liodakis}, \&
  {Romani}}]{Peirson2022_prime_IXPE_targets}
{Peirson}, A.~L., {Liodakis}, I., \& {Romani}, R.~W. 2022, \apj, 931, 59

\bibitem[{{Petrov} {et~al.}(2019){Petrov}, {de Witt}, {Sadler}, {Phillips}, \&
  {Horiuchi}}]{petrov2019_lba}
{Petrov}, L., {de Witt}, A., {Sadler}, E.~M., {Phillips}, C., \& {Horiuchi}, S.
  2019, \mnras, 485, 88

\bibitem[{{Plavin} {et~al.}(2020){Plavin}, {Kovalev}, {Kovalev}, \&
  {Troitsky}}]{Plavin2020}
{Plavin}, A., {Kovalev}, Y.~Y., {Kovalev}, Y.~A., \& {Troitsky}, S. 2020, \apj,
  894, 101

\bibitem[{{Plavin} {et~al.}(2024){Plavin}, {Burenin}, {Kovalev}, {Lutovinov},
  {Starobinsky}, {Troitsky}, \& {Zakharov}}]{Plavin2024}
{Plavin}, A.~V., {Burenin}, R.~A., {Kovalev}, Y.~Y., {et~al.} 2024, \jcap,
  2024, 133

\bibitem[{{Plavin} {et~al.}(2021){Plavin}, {Kovalev}, {Kovalev}, \&
  {Troitsky}}]{Plavin2021}
{Plavin}, A.~V., {Kovalev}, Y.~Y., {Kovalev}, Y.~A., \& {Troitsky}, S.~V. 2021,
  \apj, 908, 157

\bibitem[{{Plavin} {et~al.}(2023){Plavin}, {Kovalev}, {Kovalev}, \&
  {Troitsky}}]{Plavin2023}
{Plavin}, A.~V., {Kovalev}, Y.~Y., {Kovalev}, Y.~A., \& {Troitsky}, S.~V. 2023,
  \mnras, 523, 1799

\bibitem[{{Plavin} {et~al.}(2025){Plavin}, {Kovalev}, \&
  {Troitsky}}]{Plavin2025_high_beaming_of_neutrino_assoc_blazars}
{Plavin}, A.~V., {Kovalev}, Y.~Y., \& {Troitsky}, S.~V. 2025, \apj, 991, 33

\bibitem[{{Podlesnyi} \& {Oikonomou}(2025)}]{Podlesnyi2025}
{Podlesnyi}, E. \& {Oikonomou}, F. 2025, \mnras, 544, 2897

\bibitem[{{Readhead} {et~al.}(1978){Readhead}, {Cohen}, {Pearson}, \&
  {Wilkinson}}]{readhead1978_flux_limited_sample_mostly_blz}
{Readhead}, A.~C.~S., {Cohen}, M.~H., {Pearson}, T.~J., \& {Wilkinson}, P.~N.
  1978, \nat, 276, 768

\bibitem[{{Reimer} {et~al.}(2019){Reimer}, {B{\"o}ttcher}, \&
  {Buson}}]{Reimer2019_Xray_photon_field_in_TXS}
{Reimer}, A., {B{\"o}ttcher}, M., \& {Buson}, S. 2019, \apj, 881, 46

\bibitem[{{Righi} {et~al.}(2019){Righi}, {Tavecchio}, \&
  {Pacciani}}]{Righi2019}
{Righi}, C., {Tavecchio}, F., \& {Pacciani}, L. 2019, \mnras, 484, 2067

\bibitem[{{Robinson} \&
  {B{\"o}ttcher}(2024)}]{Robinson2024_low_blz_contr_to_neut}
{Robinson}, J. \& {B{\"o}ttcher}, M. 2024, \apj, 977, 42

\bibitem[{{Rodrigues} {et~al.}(2019){Rodrigues}, {Gao}, {Fedynitch},
  {Palladino}, \& {Winter}}]{Rodrigues2019_TXS_AM3_modeling}
{Rodrigues}, X., {Gao}, S., {Fedynitch}, A., {Palladino}, A., \& {Winter}, W.
  2019, \apjl, 874, L29

\bibitem[{{Rodrigues} {et~al.}(2021){Rodrigues}, {Garrappa}, {Gao}, {Paliya},
  {Franckowiak}, \& {Winter}}]{Rodrigues2021_PKS1502_AM3_modeling}
{Rodrigues}, X., {Garrappa}, S., {Gao}, S., {et~al.} 2021, \apj, 912, 54

\bibitem[{{Rodrigues} {et~al.}(2024){Rodrigues}, {Karl}, {Padovani}, {Giommi},
  {Paiano}, {Falomo}, {Petropoulou}, \&
  {Oikonomou}}]{Rodrigues2024_spectra_of_neut_blz}
{Rodrigues}, X., {Karl}, M., {Padovani}, P., {et~al.} 2024, \aap, 689, A147

\bibitem[{{Rodrigues} {et~al.}(2025){Rodrigues}, {Rieger}, {Bohdan}, \&
  {Padovani}}]{Rodrigues2025_modeling_of_TXS}
{Rodrigues}, X., {Rieger}, F., {Bohdan}, A., \& {Padovani}, P. 2025, arXiv
  e-prints, arXiv:2508.18345

\bibitem[{{Ros} {et~al.}(2020){Ros}, {Kadler}, {Perucho}, {Boccardi}, {Cao},
  {Giroletti}, {Krau{\ss}}, \& {Ojha}}]{Ros2020_TXS0506_via_VLBI}
{Ros}, E., {Kadler}, M., {Perucho}, M., {et~al.} 2020, \aap, 633, L1

\bibitem[{{Roulet} \& {Vissani}(2021)}]{Roulet2021_proton_photon_energies}
{Roulet}, E. \& {Vissani}, F. 2021, \jcap, 2021, 050

\bibitem[{{Sahakyan} {et~al.}(2023){Sahakyan}, {Giommi}, {Padovani},
  {Petropoulou}, {B{\'e}gu{\'e}}, {Boccardi}, \&
  {Gasparyan}}]{Sahakyan2023_PKS0735_masq_BLL}
{Sahakyan}, N., {Giommi}, P., {Padovani}, P., {et~al.} 2023, \mnras, 519, 1396

\bibitem[{{Saurenhaus} {et~al.}(2026){Saurenhaus}, {Capel}, {Oikonomou}, \&
  {Buchner}}]{Saurenhaus2025_diffuse_neutr_from_Seyferts}
{Saurenhaus}, L., {Capel}, F., {Oikonomou}, F., \& {Buchner}, J. 2026, \prd,
  113, 023019

\bibitem[{{Scargle} {et~al.}(2013){Scargle}, {Norris}, {Jackson}, \&
  {Chiang}}]{scargle2013}
{Scargle}, J.~D., {Norris}, J.~P., {Jackson}, B., \& {Chiang}, J. 2013, \apj,
  764, 167

\bibitem[{{Smith} {et~al.}(2021){Smith}, {Hooper}, \& {Vieregg}}]{Smith2021}
{Smith}, D., {Hooper}, D., \& {Vieregg}, A. 2021, \jcap, 2021, 031

\bibitem[{{Sokolovsky} {et~al.}(2017){Sokolovsky}, {Gavras}, {Karampelas},
  {Antipin}, {Bellas-Velidis}, {Benni}, {Bonanos}, {Burdanov}, {Derlopa},
  {Hatzidimitriou}, {Khokhryakova}, {Kolesnikova}, {Korotkiy}, {Lapukhin},
  {Moretti}, {Popov}, {Pouliasis}, {Samus}, {Spetsieri}, {Veselkov}, {Volkov},
  {Yang}, \& {Zubareva}}]{sokolovsky2017}
{Sokolovsky}, K.~V., {Gavras}, P., {Karampelas}, A., {et~al.} 2017, \mnras,
  464, 274

\bibitem[{{Stathopoulos} \&
  {Petropoulou}(2026)}]{Stathopoulos2025_MagRec_neutrino_emission_pc_vs_subpc_scales}
{Stathopoulos}, S.~I. \& {Petropoulou}, M. 2026, \mnras, 545, staf2083

\bibitem[{{Stathopoulos} {et~al.}(2022){Stathopoulos}, {Petropoulou}, {Giommi},
  {Vasilopoulos}, {Padovani}, \&
  {Mastichiadis}}]{Stathopoulos2022_xray_flare_v_neut}
{Stathopoulos}, S.~I., {Petropoulou}, M., {Giommi}, P., {et~al.} 2022, \mnras,
  510, 4063

\bibitem[{{Suray} \& {Troitsky}(2024)}]{Suray2024}
{Suray}, A. \& {Troitsky}, S. 2024, \mnras, 527, L26

\bibitem[{{Tonry} {et~al.}(2018){Tonry}, {Denneau}, {Heinze}, {Stalder},
  {Smith}, {Smartt}, {Stubbs}, {Weiland}, \& {Rest}}]{tonry2018}
{Tonry}, J.~L., {Denneau}, L., {Heinze}, A.~N., {et~al.} 2018, \pasp, 130,
  064505

\bibitem[{{Troitsky}(2021)}]{Troitsky2021_neut_review}
{Troitsky}, S.~V. 2021, Physics Uspekhi, 64, 1261

\bibitem[{{Virtanen} {et~al.}(2020){Virtanen}, {Gommers}, {Oliphant},
  {Haberland}, {Reddy}, {Cournapeau}, {Burovski}, {Peterson}, {Weckesser},
  {Bright}, {van der Walt}, {Brett}, {Wilson}, {Millman}, {Mayorov}, {Nelson},
  {Jones}, {Kern}, {Larson}, {Carey}, {Polat}, {Feng}, {Moore}, {VanderPlas},
  {Laxalde}, {Perktold}, {Cimrman}, {Henriksen}, {Quintero}, {Harris},
  {Archibald}, {Ribeiro}, {Pedregosa}, {van Mulbregt}, \& {SciPy 1. 0
  Contributors}}]{Virtanen2020_SciPy}
{Virtanen}, P., {Gommers}, R., {Oliphant}, T.~E., {et~al.} 2020, Nature
  Methods, 17, 261

\bibitem[{{Wilson}(2019)}]{Wilson2019_harmonic_mean_pval}
{Wilson}, D.~J. 2019, Proceedings of the National Academy of Science, 116, 1195

\bibitem[{{Yang} {et~al.}(2025){Yang}, {Liu}, \&
  {Wang}}]{Yang2025_neut_from_disk_of_TXS0506}
{Yang}, Q.-R., {Liu}, R.-Y., \& {Wang}, X.-Y. 2025, \apj, 980, 255

\bibitem[{{Ye} {et~al.}(2023){Ye}, {Hu}, {Tian}, {Chang}, {Chang}, {Cheng},
  {Gao}, {Ge}, {Gong}, {Guo}, {Guo}, {He}, {Huang}, {Jiang}, {Jiang}, {Jing},
  {Li}, {Li}, {Li}, {Li}, {Li}, {Liao}, {Lin}, {Lin}, {Liu}, {Liu}, {Liu},
  {Miao}, {Mo}, {Morton-Blake}, {Peng}, {Sun}, {Tang}, {Tang}, {Tao}, {Tian},
  {Wang}, {Wang}, {Wang}, {Wei}, {Wei}, {Wu}, {Xian}, {Xiang}, {Xu}, {Xue},
  {Yang}, {Yang}, {Yu}, {Zeng}, {Zhang}, {Zhang}, {Zhang}, {Zhang}, {Zhi},
  {Zhong}, {Zhou}, {Zhu}, \& {Zhuang}}]{Ye2023_TRIDENT}
{Ye}, Z.~P., {Hu}, F., {Tian}, W., {et~al.} 2023, Nature Astronomy, 7, 1497

\bibitem[{{Yoshida} {et~al.}(2023){Yoshida}, {Petropoulou}, {Murase}, \&
  {Oikonomou}}]{Yoshida2023_low_blz_contr_to_neut}
{Yoshida}, K., {Petropoulou}, M., {Murase}, K., \& {Oikonomou}, F. 2023, \apj,
  954, 194

\bibitem[{{Zegarelli} {et~al.}(2025){Zegarelli}, {Franckowiak}, {Sommani},
  {Valtonen-Mattila}, \& {Yuan}}]{Zegarelli2025_IceCat2}
{Zegarelli}, A., {Franckowiak}, A., {Sommani}, G., {Valtonen-Mattila}, N., \&
  {Yuan}, T. 2025, arXiv e-prints, arXiv:2507.06176

\bibitem[{{Zhou} {et~al.}(2021){Zhou}, {Kamionkowski}, \& {Liang}}]{Zhou2021}
{Zhou}, B., {Kamionkowski}, M., \& {Liang}, Y.-f. 2021, \prd, 103, 123018

\end{thebibliography}

\FloatBarrier
\clearpage
\onecolumn

\begin{appendix}

\section{Low weight spatiotemporal associations} \label{appendix_additional_LCs}
Here, we give and describe the light curve and sky map of three low-weight spatiotemporal associations arising from the BB95 at the peak of a BBHOP metric, similar to Sect. \ref{sec_results_individual_assoc} where we did the same for high-weight spatiotemporal associations arising from the BB95 at the peak of a BBHOP metric.

\begin{figure*}[h!]
    \centering
    \includegraphics[width=18cm]{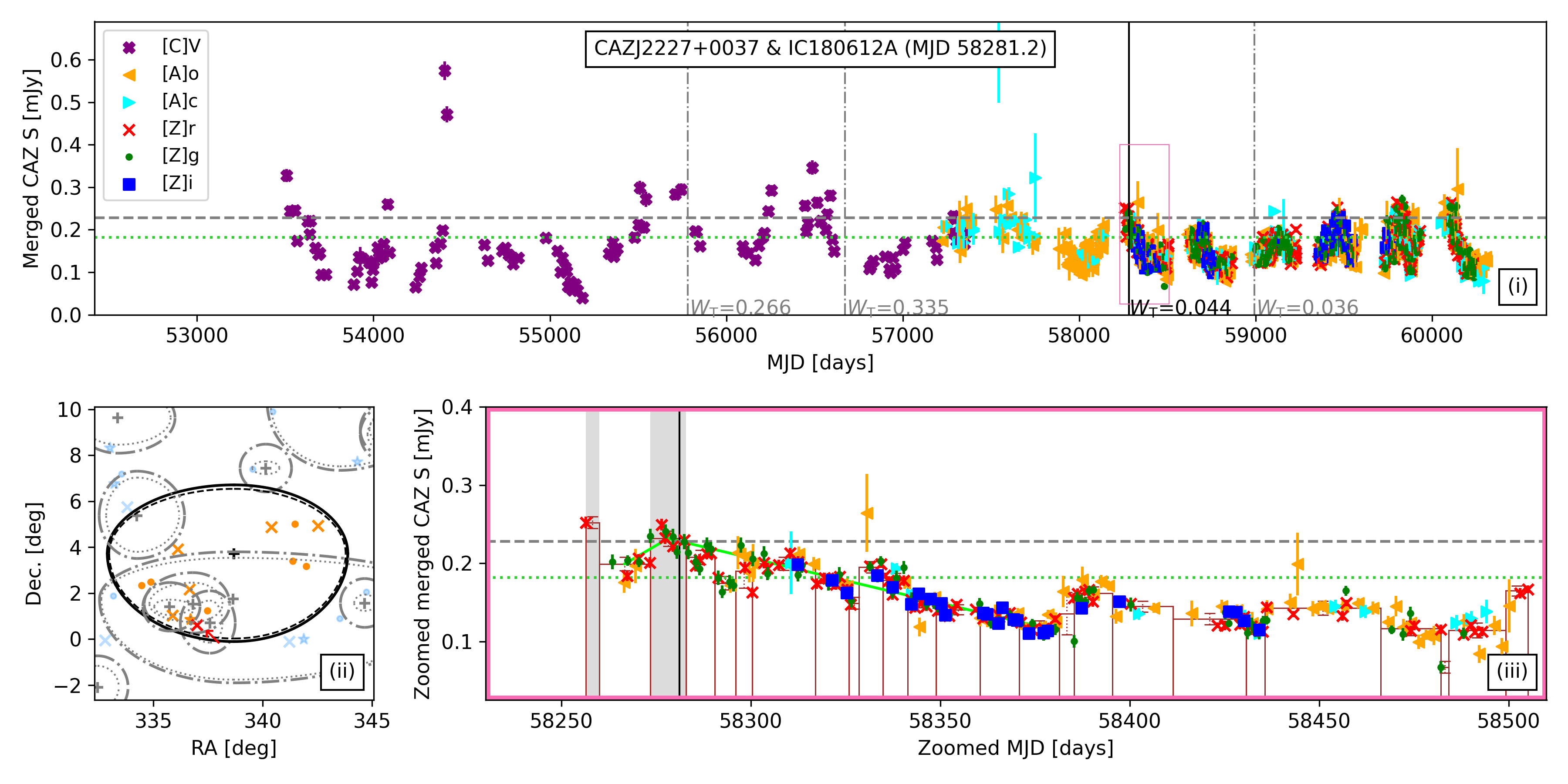}
    \caption{Light curve and sky map of the IBL CAZJ2227+0037 which is spatiotemporally associated with the neutrino IC180612A ($W_\mathrm{T}=0.044$). For plot details, see description of Fig. \ref{fig_CAZJ0211+1051_neut_MJD56579.909}. We note that the y-axis in plot (i) is truncated to omit a few outliers. The grey dash-dotted vertical lines in plot (i) show spatial associations with three other neutrinos (IC110807A, IC140114A, and IC200523A, which have weights of 0.266, 0.355, and 0.036, respectively).}
    \label{fig_CAZJ2227+0037_neut_MJD58281.190}
\end{figure*}

The first of the low-weight spatiotemporal associations arising from the BB95 at the peak of a BBHOP metric is of the IBL CAZJ2227+0037 (4FGL~J2227.9+0036) with IC180612A ($W_\mathrm{T}$=$0.044$), whose light curve and sky map are given in Fig. \ref{fig_CAZJ2227+0037_neut_MJD58281.190}. The neutrino IC180612A arrived at the peak of an outburst period. Intriguingly, CAZJ2227+0037 is associated with three other neutrinos: IC110807A, IC140114A, and IC200523A. These have weights of 0.266, 0.355, and 0.036, respectively. IC110807A arrived very close to the peak of another outburst period, but it unfortunately falls in a seasonal gap. This inadvertently excludes it from the spatiotemporal analysis as its temporal signal cannot be determined (for the adverse effects of seasonal gaps on our analysis, see Sect. \ref{sec_results_observational_gaps}). IC140114A also falls in a seasonal gap, although it likely did not arrive during a flare. Thus, this IBL has one definite, another likely, and another potential spatiotemporal association as well as one spatial association, which together make it an interesting neutrino-emitting candidate to follow up on. On top of this, its IBL nature fits the profile of the other spatiotemporally neutrino-associated sources well.

\begin{figure*}
    \centering
    \includegraphics[width=18cm]{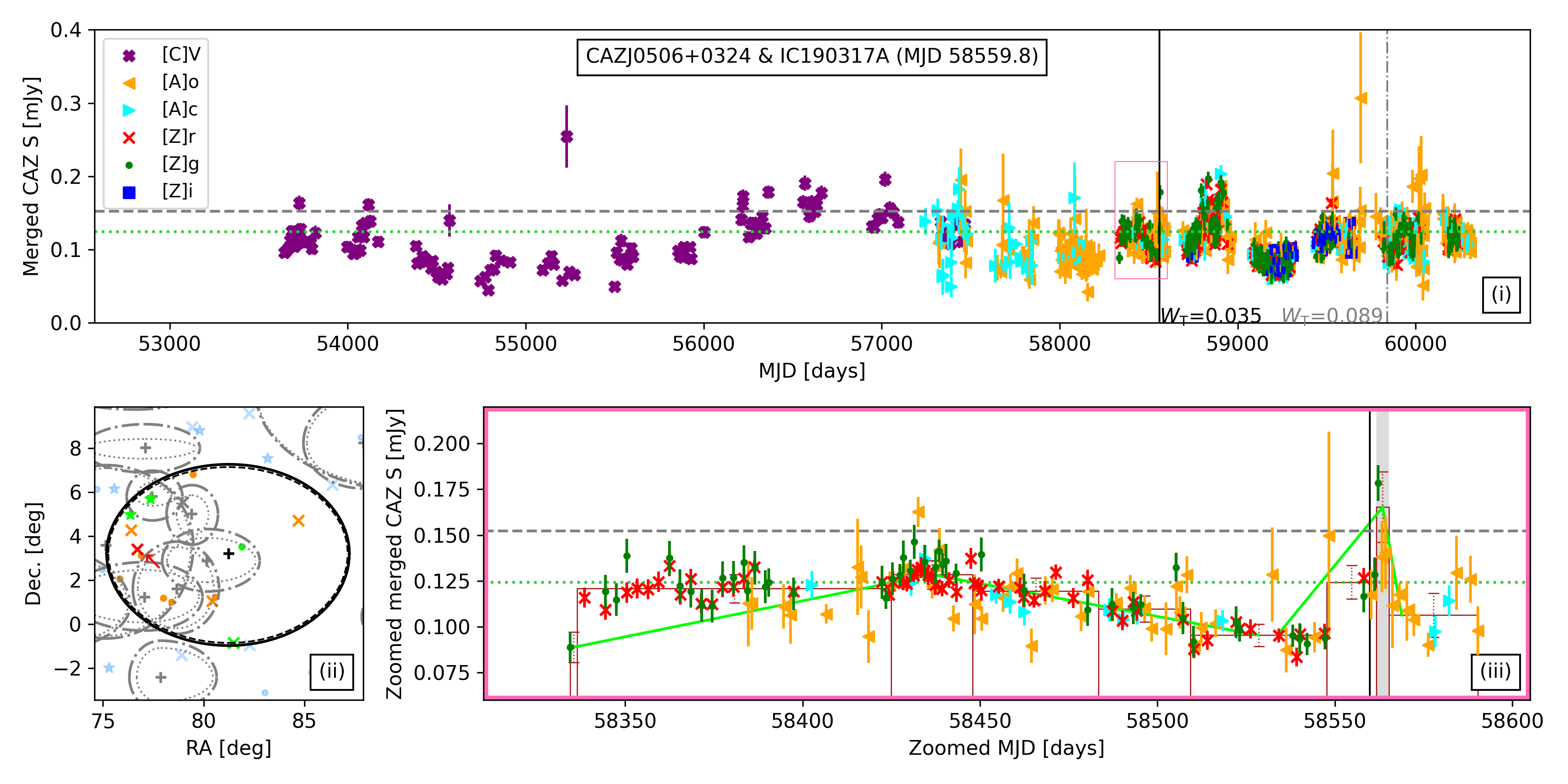}
    \caption{Light curve and sky map of the HBL CAZJ0506+0324 which is spatiotemporally associated with the neutrino IC190317A ($W_\mathrm{T}=0.035$). For plot details, see description of Fig. \ref{fig_CAZJ0211+1051_neut_MJD56579.909}. We note that the y-axis in plot (i) is truncated to omit a few outliers. The grey dash-dotted vertical line in plot(i) shows a spatial association with another neutrino (IC220918A with $W_\mathrm{T}=0.089$).}
    \label{fig_CAZJ0506+0324_neut_MJD58559.832}
\end{figure*}

The second such low-weight association is of the HBL CAZJ0506+0324 (4FGL~J0506.9+0323) with IC190317A ($W_\mathrm{T}$=$0.035$), whose light curve and sky map are given in Fig. \ref{fig_CAZJ0506+0324_neut_MJD58559.832}. This neutrino precedes the peak of a relatively notable flare by $\sim$2~d. CAZJ0506+0324 has a second neutrino association, IC220918A ($W_\mathrm{T}$=$0.089$), which is also associated with CAZJ0509+0541. There is no sign of increased optical activity when this neutrino arrived.

\begin{figure*}
    \centering
    \includegraphics[width=18cm]{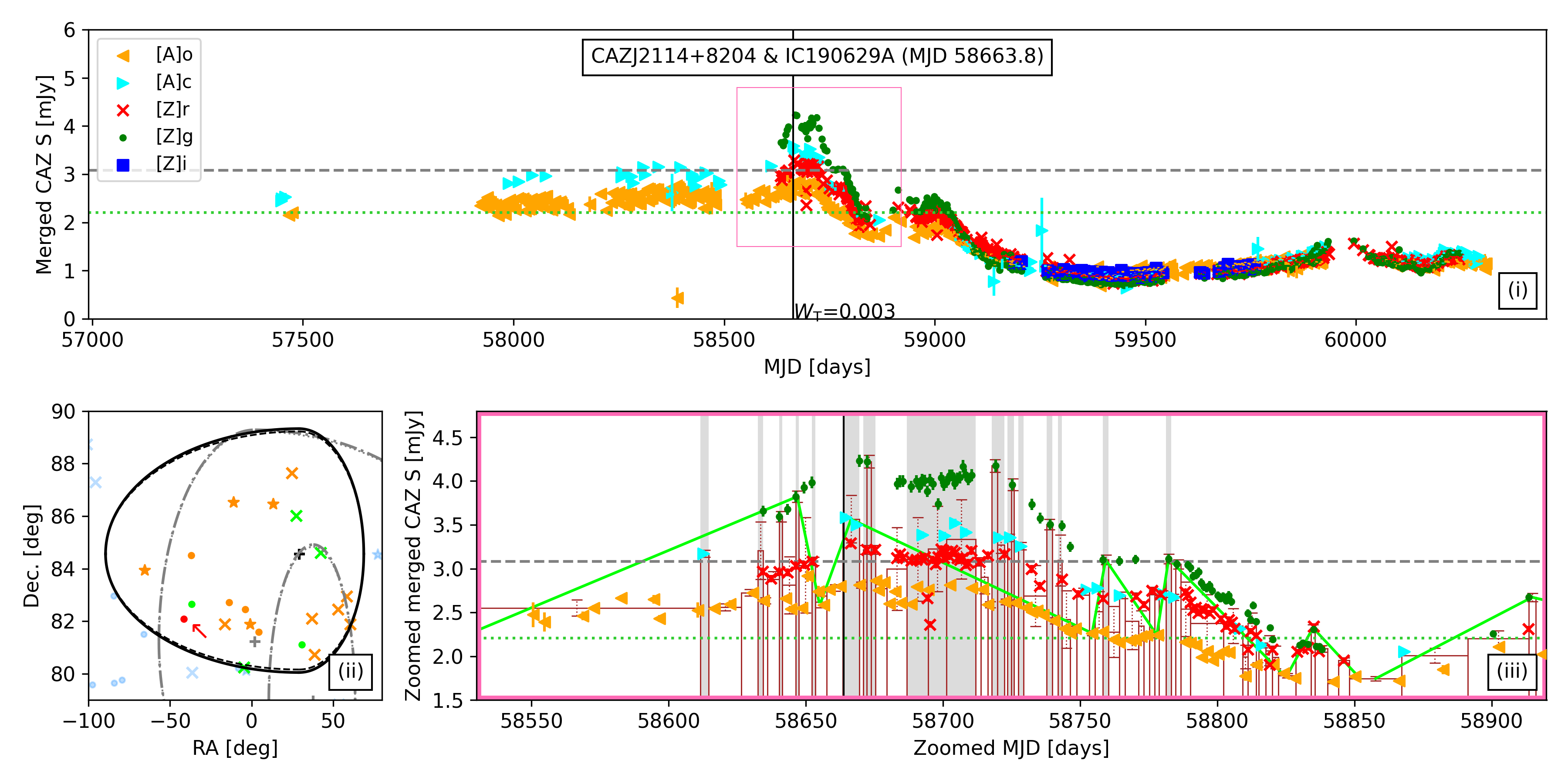}
    \caption{Light curve and sky map of the AGN CAZJ2114+8204 which is spatiotemporally associated with the neutrino IC190629A ($W_\mathrm{T}=0.003$). For plot details, see description of Fig. \ref{fig_CAZJ0211+1051_neut_MJD56579.909}. We note that around the neutrino arrival time, the quality of the merged CAZ light curve is poor (i.e., the filters exhibit substantial discrepancies in contemporaneous flux densities). However, it appears that the identified BBHOP, whose peak is the BB95 which coincides with the neutrino arrival, reliably traces the overall behavior of optical emission. We caution that, unlike all other such plots, the x- and y-axis of the sky map are not equal in scale. This blazar has another neutrino association (IC140410A; $W_\mathrm{T}$=$0.003$) which arrives before the CAZ light curve starts.}
    \label{fig_CAZJ2114+8204_neut_MJD58663.809}
\end{figure*}

Lastly, the third such low-weight association is of the AGN CAZJ2114+8204 (2MASS~J21140112+8204483) with IC190629A ($W_\mathrm{T}$=$0.003$), whose light curve and sky map are given in Fig. \ref{fig_CAZJ2114+8204_neut_MJD58663.809}. While the quality of the merged CAZ light curve is low at the time of the neutrino arrival, the neutrino clearly arrived at the peak of the all-time highest flare. This source is spatially associated with a second low-weight neutrino (IC140410A; $W_\mathrm{T}$=$0.003$) which unfortunately arrived before the CAZ light curve starts.

\end{appendix}

\end{document}